\documentclass[twoside,11pt]{article}
\usepackage{jair, theapa, rawfonts}

\jairheading{82}{2025}{1017-1076}{03/2024}{02/2025}
\ShortHeadings{Empirical Game-Theoretic Analysis: A Survey}
{Wellman, Tuyls \& Greenwald}
\firstpageno{1017}

\usepackage[utf8]{inputenc}
\usepackage{tabularx}
\usepackage{natbib}
\usepackage{graphicx}
\usepackage{amsmath,amssymb}
\usepackage{bbm}
\usepackage{xspace}
\usepackage{kbordermatrix}
\usepackage{appendix}
\usepackage{times}
\usepackage{amsthm}
\usepackage{amscd}
\usepackage{parskip}
\usepackage{enumitem}
\usepackage{verbatim}
\usepackage[labelfont=bf,labelformat=simple,singlelinecheck=off,justification=raggedright]{subcaption}
\usepackage{tikz-cd}
\usepackage{mathrsfs}
\usepackage[capitalise]{cleveref} % Should be last of all packages imported
\crefname{equation}{}{} % Do not show 'eq.' for \crefs
\usepackage{multirow}
\usepackage[normalem]{ulem}
\usepackage{nicefrac}
\usepackage{bm}
\usepackage{url}
\usepackage{algorithm2e}

\title{Empirical Game Theoretic Analysis: A Survey}

\author{\name Michael P. Wellman \email wellman@umich.edu  \\
       \addr University of Michigan\\
       Ann Arbor, Michigan, USA
        \AND
      \name Karl Tuyls \email karltuyls@meta.com \\
       \addr Meta AI\\
       Paris, France
       \AND
       \name Amy Greenwald \email amy@brown.edu \\
       \addr Brown University\\
       Providence, Rhode Island, USA}

\newcommand{\Samples}{\bm{Y}}

\newcommand{\NumberOfSamples}{m}

\newcommand{\UtilityRange}{c}
\newcommand{\UtilityVariance}{\bm{v}}

\newcommand{\ConditionDistribution}{\rand{D}}

\newcommand{\GameTuple}{\game}
\newcommand{\InducedGame}[1]{\GameTuple_{#1}}        % Typical use: \InducedGame{\ConditionDistribution}
\newcommand{\EmpiricalGame}[1]{\hat{\GameTuple}_{#1}} % Typical use: \EmpiricalGame{\Samples}

\newcommand{\SetOfPlayers}{N} % updated
\newcommand{\PlayerIndex}{i} % updated
\newcommand{\StrategySet}{S}

\newcommand{\StratProfile}{s} % deleted \bm
\newcommand{\StratProfileSpace}{\StrategySet} % deleted \bm

\newcommand{\Utility}{u} % deleted \bm

\DeclareMathOperator{\Nash}{\mathcal{E}}

\newcommand{\GS}{\ensuremath{\operatorname{GS}}}

\newcommand{\PSP}{\ensuremath{\operatorname{PSP}}}

\newcommand{\LandauTheta}{\Theta}

\newcommand{\abs}[1]{\left\lvert{} #1 \right\rvert{}}
\newcommand{\norm}[1]{\left\lVert{} #1 \right\rVert{}}

\newcommand{\rand}[1]{\mathscr{#1}}

\newcommand{\term}[1]{\textbf{\textit{#1}}}
\providecommand{\abs}[1]{\lvert#1\rvert}
\newcommand{\E}{\mathbb{E}}
\newcommand{\game}{\Gamma}
\newcommand{\Players}{N}
\newcommand{\numPlayers}{n}
\newcommand{\regret}{\epsilon}
\newcommand{\numRoles}{K}
\newcommand{\numReps}{p}
\newcommand{\numStrats}{\abs{S}}
\newcommand{\support}{\mathit{supp}}
\newcommand{\solcand}{\Psi}
\newcommand{\Goods}{G}
\newcommand{\numGoods}{g}
\newcommand{\val}{v}
\newcommand{\Val}{V}
\newcommand{\price}{p}
\newcommand{\Ind}[2]{\mathbbm{1}_{#1\ \text{wins}\ #2}}
\newcommand{\hist}{h}
\newcommand{\Hist}{H}
\newcommand{\BR}{\mathit{BR}}
\newcommand{\Counts}{\mathcal{N}}
\newcommand{\Util}{\mathcal{U}}
\newcommand{\hpt}{\mathcal{H}}
\newcommand{\alpharank}{$\alpha$-Rank}
\newcommand{\numSamp}{m}

\newcommand{\gamematrix}[6]{
  \begin{tabular}{c} 
    $\begin{array}{c} \\ #1 \\ #2 \end{array} \hspace{0mm}
    \begin{array}{r@{}c@{}c@{}l}
      & #1 & #2 & \\
      \bigg( & \begin{array}{c} #3 \\ #5 \end{array} & \begin{array}{c} #4 \\ #6 \end{array} & \bigg)
    \end{array}$
  \end{tabular}}

\newcommand{\kt}[1]{{\color{purple}[KT: #1]}}
\newcommand{\amy}[1]{{\color{red}[\textsc{Amy}: \emph{#1}]}}

\newcommand{\samy}[2]{{\color{pink}\sout{#1}\color{blue}#2}}{}
{}

\begin{document}

\maketitle
% \tableofcontents
\begin{abstract}
In the empirical approach to game-theoretic analysis (EGTA), the model of the game comes not from declarative representation, but is derived by interrogation of a procedural description of the game environment.
The motivation for developing this approach was to enable game-theoretic reasoning about strategic situations too complex for analytic specification and solution.
Since its introduction over twenty years ago, EGTA has been applied to a wide range of multiagent domains, from auctions and markets to recreational games to cyber-security.
We survey the extensive methodology developed for EGTA over the years, organized by the elemental subproblems comprising the EGTA process.
We describe key EGTA concepts and techniques, and the questions at the frontier of EGTA research.
Recent advances in machine learning are accelerating progress in EGTA, and promise to significantly expand our capacities for reasoning about complex game situations.
\end{abstract}

\section{Introduction}

When agents make decisions that interact with decisions of other rational entities, they play a \term{game}.
Understanding how to play games effectively is a central concern of AI, as is anticipating the outcomes of games played by agents using AI methods.
\term{Game theory} offers a rich conceptual and mathematical framework for describing and analyzing game situations \citep{Leyton-Brown08}.
As such, game-theoretic ideas and methods are ubiquitous in AI research, employed in theoretical treatments as well as computational approaches to reasoning about games.
AI today is thus a major consumer of game theory, and also a significant contributor---to concepts, representations, algorithms, and more.

At the heart of game-theoretic analysis is a formal representation of the game situation: a \term{game model}.
Classically (e.g., in game theory or AI textbooks), such models involve tables or trees, spelling out the agents' actions, information, and values, and the outcomes of the various decisions they might take.
Modern developments, often under the label of \term{computational} 
% (e.g., \cite{DaskalakisGP06})
or \term{algorithmic game theory} \citep{algorithmic-gt}, have significantly augmented these constructs in clever ways, for example by supporting compact expression of complex strategic environments.
Given a game model, the logic of game theory can be applied to identify or characterize game-theoretic \textit{solutions}: configurations of agent behavior satisfying well-defined conditions representing criteria for rational behavior.

Requiring a game model in analytic form, however, poses an impediment to the application of game theory in many settings of interest for AI.
In practice, formally specifying a multiagent scenario has proven feasible only for games that are artificially defined (e.g., recreational games in worlds of boards, cards, and dice) or are highly stylized representations of realistic situations.
Applied game theorists can be quite adept at stylization: isolating the salient features of a strategic situation and capturing them in an analytic form that is tractable for game-theoretic representation and reasoning.
Vast literatures document deep strategic insights about markets, conflicts, organizations, ecologies, and many other social domains---obtained through application of game-theoretic concepts to stylized models of such settings.
Inevitably, however, game-theoretic conclusions obtained in this way hinge on simplifying assumptions, adopted by necessity, and thus their application to real-world instances entails judgment about the relevance of the complications abstracted away.
Moreover, the necessities of stylization may be systematic in the kinds of actually relevant features that can be accommodated, and those that cannot.

An alternative is to express strategic environments \textit{procedurally}, in the form of a \term{simulation model}.
Simulation can readily accommodate important forms of complexity: for example, agent heterogeneity, and nonlinear dynamics generated through complicated state and information structures.
The methodology of \term{empirical game-theoretic analysis} (EGTA) aims to combine the flexibility of simulation with the strategic logic of game theory.
Simulating interacting agent decisions allows consideration of complex environments that would be difficult to express in an analytic game form, or that would be intractable for reasoning.

This motivation is similar to that of \term{agent-based modeling} (ABM), which emerged from the social sciences as a simulation-based alternative to mainstream economic theories \citep{Miller07,Tesfatsion06a}.
Pioneering advocates of ABM tended to eschew strict rationality and equilibrium assumptions, instead embracing tools from evolution and adaptive systems.
However, ABM alone is in a sense too flexible, as the possible outcomes achievable through \textit{some} agent behaviors can be extremely open-ended.
The outcomes we more specifically care about are those that follow from \textit{what rational agents would do}, according to some concept of rationality (including approximate or boundedly rational concepts) judged appropriate for our setting \citep{Wellman16}.
This is exactly what game theory provides and what EGTA inherits: a mathematical framework with representations for strategic situations and concepts for characterizing and identifying implications of rational choice. 

\begin{figure}[htb]
  \centering
\includegraphics[width=12cm]{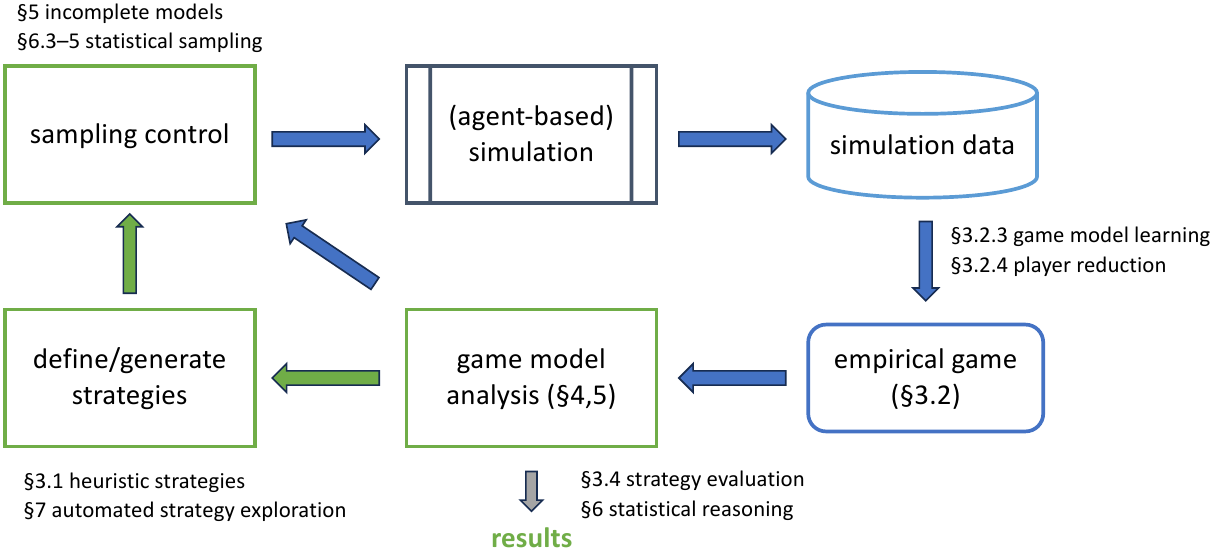}
  \caption{EGTA: A high-level view.
The underlying game is represented by an agent-based simulator, depicted in the top row, center.
Proceeding clockwise, data from the simulation is used to induce an empirical game model.
The green-box modules show how we develop the empirical game and ultimately obtain results, through a repeated process of game model analysis, extension, and refinement.
  Section references indicate where key issues and techniques are surveyed herein.}
  \label{fig:egta-diagram}
\end{figure}

The core idea of EGTA is to employ agent-based simulation to generate data from which we can induce a game model, which we call the \term{empirical game}. 
A high-level diagram of the EGTA process is shown in Figure~\ref{fig:egta-diagram}.
The process of inducing an empirical game from agent-based simulation is shown in the upper right.
That the fundamental game specification is in the form of a simulator affords a high degree of expressivity in a convenient manner.
% \amy{my overleaf comments are not working. what do you mean by convenience? maybe expressivity is enough?}
That the method produces a game model, tailored to the question at hand, in turn affords exploitation of a comprehensive algorithmic toolbox of computational game-theoretic techniques.
% \amy{is this the first mention of comp'l GT? if so, i think we need a reference. where can one find this toolbox?}
The result is an expansion of the scope of game-theoretic reasoning beyond domains where analytic game models can be plausibly and usefully constructed.

The empirical game is extended and refined through an iterative process, where analysis of the game model drives identification of new strategies to consider and determination of which strategy profiles to simulate.
The techniques by which these decisions are made based on interim analyses comprise the core of EGTA methodology, as described in this survey. 

It is important to keep in mind that the empirical game is a model, and as such it departs in important ways from the fundamental game situation of interest, sometimes referred to as the \term{underlying game}. 
First, the simulator itself may only approximate this underlying game, as after all, a simulation model is still a model.
Second, the empirical game typically covers only a strict subset of the possible agent strategies (i.e., those admissible by the simulator). 
It may also incorporate other abstractions or even structural modifications relative to the game defined by the simulator.
For instance, an empirical game may be expressed in normal form, even though the strategies themselves (i.e., as implemented in the simulation) are typically highly sequential and conditional
on implicit observations.
Finally, the empirical game's payoffs are induced from noisy and/or sparse simulation data, and so are subject to approximation error compared to the true payoffs of the underlying game.
As we survey here, much of the EGTA literature is devoted to characterizing these departures and developing techniques to promote the fidelity of EGTA results given the inherent imperfections of empirical modeling.

The EGTA approach was named ``empirical'' because it embraces empirical methods: simulation, sampling, search, machine learning (ML), etc. \citep{Wellman06}.
This contrasts with the prevalent mode of game-theoretic analysis at the time, which was based on deductive inference.
Nowadays it is far more common to invoke techniques like machine learning in service of game-theoretic reasoning,
% \amy{need references! not sure what exactly you are referring to?} 
beyond what we are here labeling EGTA.
This trend is accelerating, for example as foundation models engender an entirely new form of procedural interrogation by which to construct game models.

This survey presents an organized view of EGTA methodology as developed in the first decades of this 21st century.
The next section starts with some historical context and a presentation of the technical background in game theory, including evolutionary game theory, underpinning EGTA.
%\amy{if evolutionary GT underpins EGTA, i think it should have been mentioned earlier, near the discussion of ABM. why is it also insufficient in and of itself?}
Section~\ref{sec:concepts} describes the core concepts defining the EGTA approach, and enabling implementation of EGTA.
Concepts and methods associated with the evolutionary game perspective are presented in Section~\ref{sec:evolution}.
Reasoning about incomplete game models is of particular relevance for EGTA, and the subject of Section~\ref{sec:inc-game-reasoning}.
Statistical techniques for EGTA and works addressing statistical questions about empirical games are surveyed in Section~\ref{sec:statistical}.
Section~\ref{sec:explore} introduces the idea of automated strategy generation, and discusses the PSRO framework which today is one of the most popular approaches to EGTA.
EGTA applications and approaches to mechanism design are the topics of Sections \ref{sec:applications} and~\ref{sec:md}, respectively.
We conclude with a reflection on EGTA, discussing some promising future directions and fundamental challenges.

\section{Background}
% In this section we provide the necessary background and technical preliminaries to give the reader a good understanding of the origins and foundations of the area. We start by sketching some of the historical work and discussing the most closely related areas. Then we provide the technical preliminaries necessary to understand the remainder of the paper. We end this section by laying out a running example that will serve to illustrate some concepts in the rest of the article.

\subsection{History}
\label{sec:history}

To our knowledge, the concept of an empirical game defined by a constrained strategy set was first articulated by \citet{Armantier00}.
The earliest published instance we can identify of an explicit game model estimated by simulation of heuristic strategies appears in the PhD dissertation of William \citet[Chapter~6]{Walsh01}.
In that work, Walsh analyzed a game of supply chain formation \citep{Walsh00wy}, where agents bid in a combinatorial auction to determine whether a chain forms and who participates.
He used simulation to estimate payoffs for combinations of bidding strategies selected from a parametric family, and from those estimates identified approximate Nash equilibria among the evaluated strategy instances.

Around the same time, a group of IBM researchers developed some intriguing agent-based models of the emerging digital economy \citep{Kephart98,Tesauro01}, including some expressly appealing to game-theoretic concepts \citep{Greenwald99k,Greenwald99}.
Walsh eventually joined this group, and together they studied pricing and bidding games with up to twenty agents and three heuristic strategies.
Their paper \citep{Walsh02dtk} was the first to explicitly advocate equilibrium-based analysis (both strategic and evolutionary) of game models empirically derived by simulation.
They introduced the concept of \textit{heuristic payoff table} as a representation of expected payoffs over a heuristic profile space.

Research following these early works branched off in two directions: the first focusing on strategic reasoning for simulation-based games, and the second focusing on evolutionary dynamical analysis of agent behavior inspired by evolutionary game theory.
The overall methodology was given the name ``EGTA'', and the strategic reasoning direction systematically developed in a program of sustained research at the University of Michigan \citep{Wellman06}, shortly following
% {in the spirit of} 
the \citet{Walsh02dtk} paper. 
This began with a study of heuristic strategies for simultaneous ascending auctions \citep{Wellman03mrs}, which derived constrained equilibria for empirical games over selected parametric instances. 
A series of PhD dissertations over the next two decades advanced the methodology in a variety of directions.
A significant thread of EGTA work from this group was driven by the Trading Agent Competition (TAC), a series of market games posed as challenges to the research community \citep{Collins10,Wellman07gs}.
In these competitions, AI researchers developed innovative trading strategies for a variety of complex market environments.
Since the strategies were typically developed independently by diverse groups focusing on different approaches, understanding the strategic interactions among them often required careful post-hoc analysis.
For example, an empirical game study of strategic procurement in the TAC supply chain game provided insight into why the 2003 tournament was prone to bouts of ruinous price cutting \citep{Wellman04esvks}.
EGTA was subsequently employed with some regularity in analyses of TAC tournaments \citep{Jordan07kw,jordan10wb}, as well as other research competitions \citep{Baarslag13}.
% The same game served for a case study in \textit{empirical mechanism design} (EMD) \citep{Vorobeychik06kw}, where an incentive engineer searches over mechanism candidates using EGTA to assess strategic response.
% A recent extension of EMD methodology by \citet{Viqueira19cmg} also employed a TAC game \amy{add citation to AdX!} to exercise the authors' proposed EGTA methodology.

The second line of work was inspired by the discovery that canonical reinforcement learning (RL) algorithms, including learning automata \citep{Narendra} and Q-learning \citep{WatkinsD92}, are formally related to the process of replicator dynamics from evolutionary game theory (discussed in depth in Section~\ref{sec:evolution}).
%\amy{i'm not understanding why this connection matters? RD is a solver, and RL is a strategy?} \amy{answer from Karl: RL and RD are very similar. both reinforce successful actions / both are survival of the fittest. so, if an RD strategy proves successful (e.g., a regularization term that more readily leads to convergence to Nash), that insight may translate to RL, specifically to the exploration term of an RL algorithm. and perhaps vice versa as well (although this direction seems less obvious to me, since it would be going from single- to multi-agent)?}
The connections were illustrated by various authors, including \citet{Borgers97}, \citet{Sato_2003}, and \citet{Tuyls02,Tuyls03}.%
\footnote{Not coincidentally, some of the early EGTA works noted above also prominently employed replicator dynamics in their analyses \citep{Walsh02dtk}.} 
Agent-based research in the vein of evolutionary dynamics took off with the organization of the GTDT and EGTMAS workshops at AAMAS conferences starting in 1999 and 2003, respectively.
For instance, \citet{Phelps05} employ the evolutionary dynamics approach in an EGTA study comparing two alternative double-auction mechanisms.
%\amy{maybe this Phelps paper belongs in the EMD section? where you moved \citet{Vorobeychik06kw} and \citet{Viqueira19cmg}? but maybe not, since the next paragraph also mentions Karl's work on CDAs.}
In follow-on work, this team showed how to derive a new strategy through genetic search over a parametric strategy space, optimizing performance against the equilibrium derived from an empirical game model \citep{Phelps06}.
They further built on these ideas to consider basin size under an evolutionary dynamic as a fitness measure, and proposed an  algorithm for extending the strategy space by repeated iteration \citep{Phelps10}.
% \amy{all of this automated strategy generation sounds like a precursor to PSRO} \mike{indeed it is}

At the same time, researchers from Maastricht University, building on the links between RL methods and evolutionary game theory \citep{Tuyls02,Tuyls03}, studied the evolutionary dynamics of heuristic strategies for Texas Hold'em Poker \citep{PonsenTKR09} and continuous double auctions \citep{TuylsP07,KaisersTTP08}. 
Later, these works were extended by Tuyls and colleagues at DeepMind, as they appealed to EGTA to analyze and evaluate the RL breakthroughs achieved in Go, Capture the Flag, StarCraft, and other games \citep{Balduzzi18,TuylsPLLG18,TuylsSym,TuylsPLHELSG20}. 
%\kt{to be further completed}
By providing a flexible template for combining game-theoretic reasoning with deep RL, the PSRO framework (see Section~\ref{sec:psro}) developed by DeepMind researchers \citep{Lanctot17} spurred a significant expansion of interest in EGTA. 

%\kt{perhaps here we can sort of indicate that after the paper of Walsh EGTA branched of somehow (not unrelated) in two directions, one focusing on the strategic interactions and one on the evolutionary dynamics means of analysis. The first would be the research program developed at Michigan, and the second one would be the branch that developed somehow around my research, starting with my 2002 BNAIC paper and 2003 AAMAS paper.}
%\mw{sounds good to me}
%\kt{Also the EGTMAS workshop at AAMAS'03 was important, this is where Simon Parsons and Steve Phelps got interested in the topic (I merged the workshop at that time with the GTDT workshop of Simon and Piotr)}
%\mw{check out what I said about the Phelps/Parsons thread in paragraph above}
%\kt{later that year I gave a talk about this in Liverpool, discussing the details with Steve and Peter (on how to apply this to auctions). My early works were part of my PhD and postdoc, and later this thread of research moved with me to Maastricht (and Liverpool) where several more PhDs worked on this topic. With my move to DeepMind in 2017, we started to apply and extend these ideas further in combination with Deep RL.}

%\kt{I can edit this section in this sense and add all these works in}

%\textit{Add further background on the evolutionary thread, and the entry of DeepMind in this area.}

\subsection{Technical Preliminaries}

\subsubsection{Basic Terminology and Notation}
\label{sec:notation}

Formally, a \term{normal-form game} $\game = \langle \Players, (S_i), (u_i) \rangle $ consists of a finite set of players, $\Players$,  indexed by $i$; a nonempty set of (pure) strategies $S_i$ for player $i$; and a utility function $u_i : \prod_{j\in \Players} S_j\rightarrow \mathbb{R}$ for each player.
Let $\numPlayers = \abs{\Players}$ be the number of players.

Player~$i$ may play a \term{pure strategy}, $s_i\in S_i$, or a \term{mixed strategy}, $\sigma_i\in\Delta(S_i)$, which is a probability distribution over the pure strategies in $S_i$\@.
Let $\sigma_i(s_j)$ denote the probability that strategy $s_j$ is played in~$\sigma_i$.
The \term{support}, $\support(\sigma_i)$, of a mixed strategy $\sigma_i$ is the set of pure strategies played with positive probability: $\support(\sigma_i)=\{s_j\in S_i\mid \sigma_i(s_j)>0\}$.
A \term{strategy profile} $\sigma=(\sigma_1,\dotsc,\sigma_\numPlayers)$, assigns a (generally mixed) strategy to each player.
If the assignments are to pure strategies, $s=(s_1,\dotsc,s_\numPlayers)$ is a \term{pure profile}.
We use notation $\sigma_{-i}=(\sigma_1,\dotsc,\sigma_{i-1},\sigma_{i+1},\dotsc,\sigma_\numPlayers)$ to denote a strategy profile for all players excluding~$i$.

Thus, $u_i(s_i,s_{-i})$ denotes player~$i$'s \term{utility} for playing strategy $s_i$ when the other players play $s_{-i}$, also termed $i$'s \term{payoff} for that situation.
We extend $u_i$ to mixed profiles by taking expectations over the realization of mixtures in pure profiles,
\begin{displaymath}
u_i(\sigma_i,\sigma_{-i})\equiv \E_{s\sim \sigma_i, s_{-i}\sim \sigma_{-i}} [u_i(s_i,s_{-i})].
\end{displaymath}

The aim of game-theoretic analysis is typically to characterize and identify \term{solutions} of a given game, according to a specified \term{solution concept}.
Commonly employed solution concepts are based on notions of \term{equilibrium} among strategies.
The classic concept of strategic equilibrium, due to Nash, selects profiles such that no player can benefit from a unilateral deviation.
Define $\BR_i(\sigma_{-i})$ as player~$i$'s \term{best response} when player~$-i$ plays strategy~$\sigma_{-i}$:
\begin{displaymath}
 \BR_i(\sigma_{-i})\equiv
    \arg\max_{\sigma_i\in\Delta(S_i)} u_i(\sigma_i,\sigma_{-i}).
\end{displaymath}
Formally, profile $\sigma^*$ is a \term{Nash equilibrium} (NE) if and only if (iff) for all $i$, $\sigma^*_i\in\BR(\sigma_{-i}^*)$.
Equivalently, $\sigma^*$ satisfies
\begin{displaymath}
\forall s'_i\in S_i.\ u_i(\sigma_i^*,\sigma_{-i}^*) \geq u_i(s'_i,\sigma_{-i}^*).
\end{displaymath}

For a non-NE profile $\sigma$, at least one player can benefit by \term{deviating} to an alternative strategy.
Player~$i$'s potential gain to deviation from $\sigma$ in game~$\game$ is termed their \term{regret}, $\regret_i^\game(\sigma)$, and is given by 
\begin{displaymath}
    \regret_i^\game(\sigma) = \max_{s_i'\in S_i}u_i(s_i', \sigma_{-i})-u_i(\sigma_i, \sigma_{-i}).
\end{displaymath}
The \term{(game) regret} for a profile is the maximum over player regrets: 
\begin{equation}\label{eq:regret}
\regret^\game(\sigma)\equiv\max_{i\in\Players} \regret_i^\game(\sigma).
\end{equation}
We drop the superscript $\game$ when the game is clear from context.
% \amy{in footnote 6, you define $\regret_\game$, which you later use in the main body. if we ever need that notation, let's introduce it here, and then say something like, ``we drop the subscript $\game$, when the game is clear from context.}

We can use the notion of regret to define approximate solution concepts.
In particular, $\sigma^*$ is an \term{$\epsilon$-Nash equilibrium} iff no player can gain more than $\epsilon$ in payoff by deviating:
\begin{displaymath}
\forall i\ \forall s'_i\in S_i.\ u_i(\sigma_i^*,\sigma_{-i}^*) + \epsilon \geq u_i(s'_i,\sigma_{-i}^*).
\end{displaymath}
Equivalently, a profile $\sigma$ is an $\regret(\sigma)$-NE\@.
Exact NE have zero regret.

Additional solution concepts are discussed in Section~\ref{sec:soln-concepts}.

\subsubsection{Symmetric Games}

A game is \term{anonymous} if all players have the same strategy set ($\forall i,j.\ S_i=S_j=S$) and for all~$i$, $u_i(s_i,s_{-i})$ is invariant to permutations of the other players ($-i$).
In other words, the utility function depends only on how many of the others play each strategy---not which ones.
An anonymous game is further called \term{symmetric} if the utility functions are the same for every player.
For symmetric games, we may drop the subscript on $u$ and write $\E [u(\sigma,\sigma')]$ for the expected utility of (any player) playing strategy $\sigma$ when the remainder are playing according to other-player profile $\sigma'$.

A game is \term{role symmetric} if the players can be partitioned into roles $R_1,\dotsc,R_\numRoles$, such that symmetry applies within roles.
That is, players within each role have the same strategy sets and utility functions, and the utility functions depend only on how many (other) players in each role play each strategy.
Role symmetry is without loss of generality, with $\numRoles=\numPlayers$;
with $\numRoles=1$, the game is fully symmetric.
Thus, role symmetry interpolates between these two extremes.

\subsubsection{Bimatrix Games}\label{sec:bimatrix}

In the special case of two-player finite-strategy normal-form games, payoffs can be specified by a pair of matrices.
Formally, a \term{bimatrix game} has $\numPlayers=2$, and is given by $\game = \langle \Players, (S_1,S_2), (u_1,u_2) \rangle$, with the utility functions $(u_1,u_2)$  represented by a pair of matrices $(A, B)$ giving the payoffs for the respective players.

Figure~\ref{fig:bimatrix} illustrates a two-strategy example, in which one player (dubbed the \term{row player}) chooses one of the two rows (each corresponding to a pure strategy), and the \term{column player} chooses a column (each corresponding to a pure strategy), with the combination determining their joint payoff.

\begin{figure}[htb]
  \centering
  \gamematrix{}{}{a_{11}, b_{11}}{a_{12},b_{12}}{a_{21}, b_{21}}{a_{22}, b_{22}}
  \caption{General payoff $(A, B)$ for a two-action bimatrix game.}
  \label{fig:bimatrix}
\end{figure}

In case $S_1=S_2$ and $A=B^T$, the players are interchangeable and we have a symmetric game.

An example of an asymmetric bimatrix game is ``Bach or Stravinsky'' (BoS), illustrated in Figure~\ref{fig:BoSbimatrix}.
In this game, both players have the same strategy set ($S_1=S_2=\{B,S\}$), choosing whether to go to a Bach ($B$) or Stravinsky ($S$) concert.
They prefer to attend the same concert, however their payoffs differ, expressing divergent preferences between the two options.

\begin{figure}[h]
  \centering
  \gamematrix{B}{S}{3,2}{0,0}{0,0}{2,3}
  \caption{Payoff matrix for the BoS game. 
  Strategies $B$ and $S$ correspond to attending concerts of Bach or Stravinsky music, respectively.}
  \label{fig:BoSbimatrix}
\end{figure}

\subsubsection{Additional Solution Concepts}
\label{sec:soln-concepts}

% \amy{we also need to say somewhere that by far the most common empirical game model is the normal-form game. although there are numerous other game forms, most empirical analyses to date transform whatsoever sort of game being analyzed into a NFG. and often a relatively small NFG (is this true?), for which a Nash oracle may be available.}
% \\ \kt{agreed, and that seems correct to me}

By far the most commonly employed solution concept in EGTA (and game-theoretic analysis more broadly) is Nash equilibrium, defined in Section~\ref{sec:notation}.
Game theorists have introduced numerous refinements of and alternatives to NE to address a variety of considerations, such as special game structure 
or bounded rationality \citep{Gintis09}.
For example, \term{correlated equilibrium} generalizes Nash to allow strategy profiles with dependencies among player choices. 
\term{Quantal response equilibrium} is another generalization, designed to model approximately rational behavior.
Such concepts are likewise applicable to empirical game models, as the choice of solution concept is generally orthogonal to whether the model is based on simulation data, or any other knowledge source.

Some solution concepts are defined with respect to structured game forms.
For example, \term{subgame perfection} is a property of solutions applicable to \term{extensive-form games}.
Considering such properties in EGTA would require that the empirical game model express that structure. 
Our survey focuses on normal-form representations, as that covers the majority of EGTA literature to date.%
\footnote{We note where relevant some emerging works that extend EGTA beyond normal form.}

% \mw{Perhaps mention CE, QRE, BNE, Nash refinements, including concepts specific to extensive form.
% Probably do not need or want full definitions or deep discussions of these.}\kt{Agreed if we want to provide full definitions here I think we probably need to restructure a bit and not put them in one subsection}\amy{i would only mention solution concepts that come up in later discussions. so maybe Bayes Nash? anything else?}

The evolutionary perspective is also a rich source of solution concepts. 
These have played a salient role in EGTA methodology, including proposals for new solution concepts motivated in part by EGTA (\alpharank, discussed in Section~\ref{subsubsec:a-rank}).
We introduce technical background on evolutionary stability within a broader discussion of evolutionary game theory in the next section.

% \subsubsection{Imperfect Information}

% \mw{Explain concept, but avoid going deep on formalism here. 
% Mention extensive form and Bayesian games.
% Note that almost all games of interest for EGTA play out sequentially and have imperfect information, which is what makes them complex.
% Treating them in normal form as a kind of abstraction.}
% \kt{I guess we were thinking to move this elsewhere, right?}

\subsection{Evolutionary Game Theory}
\label{sec:egt}

%\amy{Karl, i have drafted an attempt at a new introductory paragraph to the EGT section, which would better articulate its goals. please feel free to revise as you see fit. thanks!}

Canonical game-theoretic solution concepts, starting with the classic equilibrium condition proposed by \citet{nash1950equilibrium}, tend to be defined by a \textit{static} relationship among player strategies.
But from the earliest days, game theorists have sought to produce more \textit{dynamic} accounts of how such equilibrium configurations might arise through player interactions.
A notable example is the method of \term{fictitious play} (FP), introduced by \citet{brown1951iterative}, which defines an iterative process where each player's strategy at a given time step is a best response to its belief about the play of others, based on past iterations.
In the standard version, the belief is that each strategy is played with probability equal to its frequency in past play.
Various analyses over the years have identified conditions for which FP and variants are guaranteed to converge (and to what), and studied FP's performance as a game-solving heuristic.

% Traditional game theory is primarily concerned with the study of static solution concepts, such as Nash equilibrium.
% But dating back to the 1950s, when Nash introduced his canonical solution concept, how players might learn to play a game was already a pressing question, e.g., Brown's and Robinson's analysis of fictitious play \kt{add citation}.
% Likewise, in addition to their focus on static equilibria, many EGTA researchers also analyze learning dynamics in empirical games, often applying techniques from dynamical systems.
% Such researchers are concerned with the convergence properties of various learning dynamics: e.g., whether they converge (i.e., land at a fixed point) and if so, whether that fixed point is also a Nash equilibrium; or, when they don't converge, whether they end in a limit cycle or in chaos; and the basins of attraction that lead to these outcomes. Subsequently, such analysis can then serve as a basis for devising new learning algorithms with desired convergence properties. %\amy{is there anything else they might care about?}

Biological evolution has been a particularly rich source of ideas about dynamically adapting behavior.
The field of \term{evolutionary game theory} (EGT) applies such ideas to strategic interaction, building dynamic accounts of adaptive game play \citep{Borgers97,Tuyls03,TuylsP07}, based on biological operators such as natural selection and mutation \citep{Smith73,Zeeman80,Zeeman81,Weibull97,Hofbauer98}.
%Concepts based on dynamical systems have been shown to be well suited to describing learning processes in multiagent systems \citep{Borgers97,Tuyls03,TuylsP07}.
The simplicity and concreteness of these operators provides a constructive basis for determination of joint behavior in complex strategic environments.
As evolutionary computation meshes well with agent-based simulation, a simulation-based approach to game theory (i.e., EGTA) is naturally suited to incorporate EGT principles and techniques. 
We lay out some of these techniques in Section~\ref{sec:evolution}, demonstrating the role of
evolutionary concepts in EGTA methodology.

\subsection{Running Example}

% \kt{is this effectively becoming our running example throughout the paper?}

To illustrate the concepts and methods of EGTA, we present an example game that can be studied with this approach.
The domain is \textit{bidding in sequential auctions}.
Like many auction games that have been tackled with EGTA, the setting is descriptively simple but strategically complex---beyond the realm of analytic tractability except in well-crafted special cases.
In its general form, the game comprises a series of one-shot sealed-bid auctions, each for a single good. 
In each auction, the players submit bids, representing positive amounts of a standard currency.
The good is awarded to the highest bidder (i.e., \term{winner} of the auction), who pays an amount that is a function of their own and the other players' bids, as dictated by specified auction rules.
The auction then reveals information about the result (e.g., the identity of the winner and price paid), which the players may use in determining their bids in subsequent auctions.
At the end of the series, players receive a payoff, namely the difference between their value for the combination of goods won and their total payment for these goods.

\citet{Greenwald12} provide a formal specification of this game along with some insightful theoretical observations and a constructive computational approach.
For our purposes, we may make do with the following notation:
\begin{itemize}
    \item $\Goods$ is the set of goods, indexed according to the auction sequence, $\numGoods=\abs{\Goods}$. 
    \item $\Ind{i}{j}$ is an indicator function, 1 if player~$i$ wins auction~$j$ and 0 otherwise.
    \item $\price_j$ is the price (i.e., payment) outcome from auction~$j$.
    \item $Y_{ij} = \{k\mid \Ind{i}{k}=1\}$ is the set of goods won by player~$i$ in the first $j$ auctions.
    \item \term{valuation} $\val_i:2^\Goods\to\Re$ is a function that describes player~$i$'s value for obtaining any set of goods.
    \item $\Val_i$ is the set of possible valuations $\val_i$.
    \item $\hist_{ij}$ is the history of observations received by player~$i$ from auctions $1,\dotsc,j$.
    \item $\Hist_{ij}$ is the set of possible histories $\hist_{ij}$.
\end{itemize}

The payoff to player~$i$ is determined by auction outcomes: goods won $Y_{i \numGoods}$ and their prices.
To describe the utility function in terms of strategies, we must account for the uncertainty due to incomplete and imperfect information.
Auctions are Bayesian games, in that players know their own valuations but have only probabilistic information about the others'. 
Sequential auctions also feature imperfect information, as players do not fully observe the others' actions. 
As the sequence of auctions proceeds, the players receive partial evidence about the other bids, summarized in auction results.
A player's bid in any given situation, therefore, depends on the player's valuation as well as their observations up to that point. 
Formally, the strategy set for player~$i$ is given by $S_i:\Val_i\times\Hist_{ij}\to \Re$\@.

A complete game specification would include probability distributions over valuations (if these distributions are the same for every player, the game is symmetric).
Given such distributions, we can write the utility function as an expectation over the auction outcomes:
\begin{displaymath}
u_i(s_i,\sigma_{-i}) %\samy{}{; \val}
= \E_{\price_j,\Ind{i}{j} \mid s_i, \sigma_{-i}} \left[ \val_i (Y_{i \numGoods}) - \sum_j \price_j \Ind{i}{j} \right].
\end{displaymath}

\section{EGTA: Key Concepts}
\label{sec:concepts}

% \amy{SECTION needs an INTRO !!} MPW: why?

We are now ready to introduce EGTA's defining concepts, most importantly, that of an empirical game model induced by simulation over a restricted strategy space. 
We illustrate the definition of a space of heuristic strategies (Section~\ref{sec:heuristic-strategies}), and consider a range of issues for the construction of empirical games (Section~\ref{sec:empirical-model}).
The remaining subsections raise issues for game solving, evaluating strategies, and game visualization that have been addressed within the EGTA approach.

\subsection{Heuristic Strategies and Restricted Games}
\label{sec:heuristic-strategies}

All game analysis is with respect to some set of included strategies $X_i$ for each player~$i$.
This set is typically a strict subset of all possible strategies, since these are generally infeasible to cover.
We use the symbol $\game_{\downarrow X}$ to denote a \term{restricted game} with respect to the \term{base game} 
% \amy{underlying? true?} 
% \mw{We've gone back and forth on terminology here, still open for discussion. 
% For the specific issue of restricting strategy sets, let's stay with "base" and be consistent about it for now.
% Perhaps we can introduce conceptually (informally?) in the intro a notion of ``underlying'' for the reality or true game that we are trying to capture with game modeling.}
$\game$, where each player~$i$ in $\game_{\downarrow X}$ is restricted to playing strategies in $X_i \subseteq S_i$, with $X = \prod_{i \in N} X_i$.

% \amy{let's use the word meta-strategy somewhere around here, if we plan to use it at all.}
% \mw{Let's not use ``meta-strategy'' to mean ``strategy''.
% I believe there is a useful interpretation for that term, but this is not the right place.}
% \amy{also, do we have any theory about what a game among meta-strategies might tell us about the underlying game?}
% \mw{Rewording: what a \textit{restricted game} might tell us about the \textit{base game}.}
% \amy{maybe not. we should be up front about this, and say that that is why this whole area is an empirical one, because we tend to invent meta-strategies, and build empirical game models around them. Mike is getting at this at the end of section 4.1.}
% \mw{Definitely worth discussion, but probably better at the end after we have presented how it all works.}

Restricted strategy sets may be formed in a variety of ways.
One basic approach is to design \term{heuristic strategies}, based on domain knowledge and intuitive simplifications, or perhaps based on known benchmarks.
We often specify a parameterized space of such strategies by exposing some controllable features of the heuristics.
Another approach that yields a parametric strategy space is to specify a more generic representation, such as a neural network policy implementation.
This parametric space is itself a restriction, as not all strategies in the base game may be expressible as parameter settings.
The actual set of included strategies may be further restricted with respect to this space, by either manual or automated selection.

For example, consider a heuristic strategy for bidding in sequential auctions based on \term{myopic marginal value}.
The strategy must specify how to bid for good $j+1$, given one's valuation and the results from the first $j$ auctions.
The myopic marginal value for this good is the increase in value it provides, given current winnings and assuming no further winnings: $\mu_{i,h_{ij}} \equiv \val_i(Y_{ij}\cup \{j+1\}) - \val_i(Y_{ij})$, for $j\in\{0,\dotsc,\numGoods-1\}$.
One possible strategy is to bid the myopic marginal value in every auction.
A simple parameterized extension is to bid a constant fraction of this value, that is, bid $\rho\mu_{i,h_{ij}}$ in each auction $j+1$, for some $0 < \rho \le 1$.
We refer to $\rho$ as a strategy parameter---in this case, a shading factor---and note that a parameterized heuristic strategy plus a set of allowed parameter settings defines a set of strategies.
Let $s^\rho$ denote the strategy of shading myopic marginal value by fraction~$\rho$ as described above. 
$X^{P}= \{s^\rho\mid \rho\in P\}$ would then comprise a set of strategies corresponding to a set of possible parameter settings~$P$.
Such strategy sets could, in turn, define a restricted game---for instance, if $\game$ is a game of sequential auctions, then a corresponding game restricted to myopic marginal value bidding could be written $\game_{\downarrow \{X^{P_i}\}}$, where $P_i$ is the set of shading factors allowed for player~$i$. 

Though myopic marginal value is limited as a strategy, analyzing a restricted game among these strategies could provide useful strategic insights about sequential auctions.
Such an analysis is directly relevant in cases where the restricted strategies are representative of real-world behavior.
More generally, it provides a baseline for gauging the benefit of incorporating more sophisticated reasoning into the bidding, most naturally transcending myopia by accounting for the opportunities of future auctions. 
By incrementally extending the restricted strategy set, we can refine the strategic analysis, ultimately approaching the base game.

\subsection{Empirical Game Models}
\label{sec:empirical-model}

% \amy{what about the language ``true'' game? does/will the meaning of a true game differ from an underlying one? if not, let's be careful to use only the one term.} \amy{the true game is the empirical game with accurate payoffs. the underlying game is more: everything about the real-world game}

% \amy{my feeling is that this paragraph should be moved to the intro to this whole section. iow, i think the empirical game model is paramount, and should be mentioned before restricted games/strategies, even if you can apply restrictions to base/underlying games. we are really only interested in applying those restrictions to empirical game models.}
The hallmark of EGTA is that the source of strategic information comes in the form of a simulation model of the environment, rather than an analytic game form.
In lieu of directly specified utility functions, the analyst must induce a utility model from payoff samples generated by simulation. 
The simulator generally produces a noisy sample of the payoff vector on each run, reflecting stochastic factors in the agent strategies or game environment.
Sample information is thus accompanied by some error, and so we add a hat to notationally distinguish an \term{empirical game model} 
% \amy{can we just say empirical game, instead of empirical game *model* everywhere, since an empirical game is a model of an underlying, true game} 
% \mw{I think it worth emphasizing that it's a model. 
% It makes sense generally to distinguish a game and a model of a game, and an empirical game model is a kind of model. But an empirical game is not a kind of game.}
$\hat{\game}$ induced from simulation data from the game $\game$ itself, and similarly for the empirical utility functions $\hat{u}_i$ defining the empirical game.

As illustrated in Figure~\ref{fig:egta-diagram}, the empirical game evolves throughout the EGTA process, continually refined as new simulation data accrues over an expanding strategy space based on results from intermediate analysis.
Iterative development of a sequence of models is common in game-theoretic reasoning, in standard analytic approaches as well as EGTA \citep{Wellman24m}.
Here we consider the empirical game at a particular point in the process, induced from a fixed body of simulation data collected to that point.

The most straightforward way to construct an empirical game model from data is through \term{direct estimation}.
In this approach, the empirical payoff for agent~$i$ in profile~$s$, $\hat{u}_i(s)$, is estimated as the sample average over a set of observations taken in simulation runs of profile~$s$.
More sophisticated sampling methods may weight observations non-uniformly, or employ other statistical techniques to sharpen estimates for a given body of data.
Statistical reasoning for EGTA is discussed further in Section~\ref{sec:statistical}.

Let us illustrate game estimation with the running example game of sequential auctions with myopic marginal value bidding. 
Consider a small version with $\numPlayers = \numGoods = 3$, and three strategies, defined by shading factors: $\rho\in\{0.3,0.5,0.7\}$.
Valuations are drawn as in the homogeneous-good model employed by \citet{Wellman17sg}, and the auction in each round is first-price.
We ran 10,000 simulations of all distinct profiles in this game, yielding the empirical game displayed in Table~\ref{tab:sec-auc3}.

\begin{table}[ht]
    \centering
\begin{tabular}{c|ccc}
     $\rho_3=0.3$   & 0.3 & 0.5 & 0.7  \\ \hline
    0.3 &  51.5 &  44.7 & 39.8 \\
    0.5 &  44.2 &  40.9  & 37.6 \\
    0.7 &  28.9 &  27.0  & 25.6 \\
    \end{tabular}
    \quad
 \begin{tabular}{c|ccc}
     $\rho_3=0.5$   & 0.3 & 0.5 & 0.7  \\ \hline
    0.3 & 44.7  & 36.9 & 32.0 \\
    0.5 & 40.9  & 36.8 & 33.3 \\
    0.7 & 27.0  & 25.6 & 23.9 \\
    \end{tabular}
    \quad
 \begin{tabular}{c|ccc}
     $\rho_3=0.7$   & 0.3 & 0.5 & 0.7  \\ \hline
    0.3 & 39.8  & 32.0 & 28.0 \\
    0.5 & 37.6  & 33.3 & 29.9 \\
    0.7 & 25.6  & 23.9 & 22.0 \\
    \end{tabular}
    \caption{Normal forms for three-player sequential auctions with three heuristic strategies. 
    Each $3\times 3$ table presents estimated payoffs (sample average rounded to tenths) 
% \amy{it seems you are indeed sampling valuations in these experiments, so that this is a Bayesian game?} 
%\mw{The underlying or base game could be modeled as a Bayesian game. The draw of valuations is in the simulation model. But once we have restricted to the heuristic strategies, all this is implicit and we have an NFG model.}
    for the row player, given the combination of row and column player, with a third player fixed at $\rho_3$.
}
    \label{tab:sec-auc3}
\end{table}

Inspection of Table~\ref{tab:sec-auc3} reveals that $(0.3,0.3,0.3)$ is a PSNE in this simple empirical game. 
Indeed, it is uniquely so. 
The setting $\rho=0.3$ is not quite dominant, as the strategy $\rho=0.5$ is a best response when the other-player profile is $(0.5,0.7)$ or $(0.7,0.7)$.
However, $\rho=0.7$ is dominated in this game, and after eliminating $\rho=0.7$, $\rho=0.5$ becomes dominated as well.
By iterated dominance, only $\rho=0.3$ survives.
Another way to visualize this game and identify the unique PSNE is depicted in Figure~\ref{fig:devgraph}.

Constructing an entire empirical utility function by direct estimation takes simulation time proportional to the number of strategy profiles.
For $\numPlayers$ players and $\numStrats$ available strategies per player, there are $\numStrats^\numPlayers$ possible profiles.
For instance, in the 3-player 3-strategy game of Table~\ref{tab:sec-auc3} there are $3^3 = 27$ table entries, one for each profile.
With symmetry, the number of distinct profiles is somewhat smaller: indeed, the payoffs in Table~\ref{tab:sec-auc3} can be derived from simulations of only ten profiles.
The savings, however, are limited, as a symmetric game comprises $\binom{\numPlayers+\numStrats -1}{\numPlayers}$ distinct profiles, 
which is still exponential in the smaller of $\numPlayers$ and $\numStrats$.
So we cannot expect simulation to be performed exhaustively for games with many players and strategies (or even large numbers of one and modest of the other).
The alternatives are to reason about an incompletely specified game model (i.e., where only a subset of profiles are evaluated; see Section~\ref{sec:inc-game-reasoning}), or to extend the game model to unevaluated profiles through some kind of generalization (i.e., machine learning; see Section~\ref{sec:learning}) process.

\subsubsection{Heuristic Payoff Tables}
\label{sec:hpt}

The natural representation of a finite $\numPlayers$-player normal-form game is as an $\numPlayers$-dimensional matrix, with cell $(j_1,\dotsc,j_\numPlayers)$ containing the \term{payoff vector}
\begin{displaymath}
(u_{1}(s_{j_1},s_{-1}),\dotsc,u_{\numPlayers}(s_{j_\numPlayers},s_{-\numPlayers})),\ \text{where}\ s_{-i}=(s_{j_1},\dotsc,s_{j_{i-1}},s_{j_{i+1}},\dotsc,s_{j_\numPlayers}).
\end{displaymath} 
An equivalent way to express these vectors is as a set of $\numPlayers$ $\numPlayers$-dimensional matrices, with cell $(j_1,\dotsc,j_\numPlayers)$ of matrix~$i$ containing the payoff scalar $u_{i}(s_{j_i},s_{-i})$.
The special case of $\numPlayers=2$ is the bimatrix game representation presented in Section~\ref{sec:bimatrix}.

Such a matrix representation of game $\game$ has size $\numPlayers\prod_{i\in \Players}\abs{S_i}$.
This size is required to fully represent $\game$, but an empirical game model is typically incomplete in the sense of including payoffs for only a subset of strategy profiles (Section~\ref{sec:incomplete}), and moreover may incorporate special structure such as symmetry that would afford more compact representation.
% \amy{well, i wouldn't say that the empirical game exhibits symmetry. maybe the underlying game itself exhibits symmetry, and the empirical game exploits that symmetry. or, maybe we just \emph{*impose*\/} symmetry on the empirical game, to make the analysis more tractable.}
% \mw{changed ``exhibit'' to ``incorporate''}
We therefore commonly employ a sparse-matrix representation that requires space proportional to the distinct
%\amy{why italics? why even mention the word distinct? is this a deterministic model, again?} 
%\mw{removed italics. The "distinct" qualifier is relevant to point out that we do not have to include redundant profiles, whose payoffs can be determined by payoffs of others -- for example based on symmetry.}
strategy profiles for which payoffs are evaluated.
This representation has been termed a \term{heuristic payoff table} (HPT)  \citep{Walsh02dtk}, and special-purpose infrastructure for data management of large HPTs has been developed \citep{EGTAOnline}.

Consider a symmetric normal-form game with $\numPlayers$ players and $\numStrats$ strategies.
For a symmetric game, what matters is not \textit{which} players choose a given strategy, but just \textit{how many}.
Therefore we can represent a strategy profile by a vector of \textit{counts}: for each strategy, the number of players who choose to play it.
As noted in Section~\ref{sec:empirical-model}, there are $\binom{\numPlayers+\numStrats -1}{\numPlayers}$ distinct count vectors for a symmetric game with $\numPlayers$ players and $\numStrats$ strategies, compared with the $\numStrats^{\numPlayers}$ profiles
%\mw{the extra factor of n gives the overall rep'n size, which is n per profile.}
ignoring symmetry.

Formally, let the HPT $\hpt=(\Counts,\Util)$, where $\Counts$ is a $\binom{\numPlayers+\numStrats -1}{\numPlayers}\times \numStrats$ matrix of counts, and $\Util$ is a matrix of utilities of the same dimension.
Each row represents a profile, such that entry $\Counts_{k,j}\in\{0,\dotsc,\numPlayers\}$ indicates the number of players choosing strategy $s_j$ in the $k^\text{th}$ profile.
The rows of $\Counts$ are all distinct, and satisfy $\sum_j \Counts_{k,j} = \numPlayers$ for all~$k$.
Entry $\Util_{k,j}$ is undefined if $\Counts_{k,j}=0$, and otherwise represents the payoff to playing strategy $s_j$ in the $k^\text{th}$ profile.
To connect $\Util$ to the standard utility function, let $s^k=(s_{j_1},\dotsc,s_{j_\numPlayers})$ be a profile in the standard strategy vector representation consistent with the $k^\text{th}$ HPT profile, that is, $\forall j.\ \abs{\{ i\mid j_i = j \}} = \Counts_{k,j}$.
Then $\forall i.\ \Util_{k,s_{j_i}} = u_i(s^k)$.

Table~\ref{tab:hpt-sa3} presents the HPT representation for the $\numPlayers=\numStrats=3$ sequential auction game specified in standard matrix form in Table~\ref{tab:sec-auc3}.
A partial HPT for a five-player version is presented in Table~\ref{tab:hpt-sa5}. 

% Let $N$ be a matrix, where each row $N_i$ is a vector of counts $(n_1,\dots,n_k)$ where $\sum_j n_j=p$, and $n_j$ indicates how many of the $p$ replicators play strategy $j$. The number of such distinct count vectors (which we also view as a discrete distribution) can be shown to be $m=\binom{p+k-1}{p}$, which is the number of rows of $N$ in our EHPT. Each distribution over strategies (rows) can be simulated (or derived from data), returning a vector of expected rewards $u(N_i)$ (one for each of the $k$ strategies). Let $U$ be an $m\times k$ matrix which captures the payoffs corresponding to the rows in $N$, i.e., $U_i = u(N_i)$. We refer to an EHPT as $M = (N, U)$. 
% Normalizing the count vector  $(n_1,\dots,n_k)$ by dividing it by $p$ gives us the probability vector $\mathbf{x}=(n_1/p,\dots,n_k/p)$, which we call a discrete strategy distribution. 

% Suppose we have an empirical game with $3$ meta-strategies ($k=3$) and $6$ players ($p=6$), this leads to a  payoff table of $28$ entries. 
% Table \ref{table:hpt} provides an example for three strategies and three players. The left-hand side shows the counts and gives the matrix $N$, while the right-hand side gives the payoffs for playing any of the strategies given the discrete profile and corresponds to matrix $U$.

\begin{table}[!ht]
\footnotesize
\centering
    \begin{subtable}{0.5\textwidth}
    \centering
		$\left( \begin{array}{ccccccc}
		& \Counts &  & \vline &  & \Util &  \\ 
% 		\Counts_{k,1}& \Counts_{k,2} & \Counts_{k,3} & \vline & \Util_{k,1} & \Util_{k,2} & \Util_{k,3} \\ 
		\hline
		3 & 0 & 0 & \vline & 51.5 & - & - \\
		2 & 1 & 0 & \vline & 44.7 & 44.2 & - \\
		2 & 0 & 1 & \vline & 39.8 & - & 28.9 \\
		1 & 2 & 0 & \vline & 36.9 & 40.9 & - \\
		1 & 1 & 1 & \vline & 32.0 & 37.6 & 27.0 \\
		1 & 0 & 2 & \vline & 28.0 & - & 25.6 \\
		0 & 3 & 0 & \vline & - & 36.8 & - \\
		0 & 2 & 1 & \vline & - & 33.3 & 25.6 \\
		0 & 1 & 2 & \vline & - & 29.9 & 23.9 \\
		0 & 0 & 3 & \vline & - & - & 22.0 \\
		\end{array} \right)$ 
	\caption{3 players (cf.~Table~\ref{tab:sec-auc3})}
	\label{tab:hpt-sa3}
    \end{subtable}%
    \begin{subtable}{0.5\textwidth}
    \centering
		$\left( \begin{array}{ccccccc}
		& \Counts &  & \vline &  & \Util &  \\ 
% 		\Counts_{k,1}& \Counts_{k,2} & \Counts_{k,3} & \vline & \Util_{k,1} & \Util_{k,2} & \Util_{k,3} \\ 
		\hline
		5 & 0 & 0 & \vline & 37.6 & - & - \\
		4 & 1 & 0 & \vline & 32.0 & 39.2 & - \\
		4 & 0 & 1 & \vline & 28.5 & - & 26.7 \\
		3 & 2 & 0 & \vline & 26.1 & 36.3 & - \\
		3 & 1 & 1 & \vline & 22.5 & 33.8 & 25.4 \\
		3 & 0 & 2 & \vline & 19.2 & - & 24.0 \\
		& \cdots & & \vline & & \cdots & \\
		2 & 1 & 2 & \vline & 15.2 & 27.5 & 22.4 \\
		& \cdots & & \vline & & \cdots & \\
		0 & 0 & 5 & \vline & - & - & 16.1 \\
		\end{array} \right)$ 
	\caption{5 players (partial: 8 of 21 profiles shown)}
	\label{tab:hpt-sa5}
    \end{subtable}
    \caption{Example heuristic payoff tables for 3- and 5-player versions of the sequential auctions game, with three strategies.
    Each row represents a profile, with counts and payoffs for strategies $\rho=0.3,0.5,0.7$, respectively.
    % Columns $\Counts_{k,1}$, $\Counts_{k,2}$, and $\Counts_{k,3}$ represent strategies $\rho=0.3,0.5,0.7$, respectively.
%\amy{i stared at these tables for at least 15 minutes, before i figured out that it is the entries in the matrix that are the $\Counts_{k,j}$s. if i finally understand things correctly, i think the columns should not be labelled with $k$ subscripts. that really threw me off. the columns are vectors, with each row corresponding to the $k$th entry.}
    }
    \label{table:hpt}
   % \vspace{-1cm}
\end{table}

The HPT construct can be straightforwardly extended to role-symmetric games.
Each role has a fixed strategy set and number of players, with symmetry holding within the role. 
We define the HPT $\hpt=(\Counts^1\times\dotsm\times\Counts^\numRoles,\Util)$, where $\Counts^r$ is a counts matrix for role~$r$, and $\Util_{(k_1,\dotsc,k_\numRoles),j,r}$ gives the payoff for playing strategy $s_j$ in role~$r$ given the rest of the strategy profile.

\subsubsection{Incomplete Evaluation of Profile Space}
\label{sec:incomplete}

When evaluating all profiles by simulation is computationally infeasible, a natural approach is to infer what we can from whatever part of the restricted game we can feasibly evaluate.
In many cases we can certify solutions or approximate solutions far short of exhaustively evaluating the profile space.
Verifying a Nash equilibrium or its degree of approximation is a matter of calculating a profile's regret. 
This requires only the payoffs for that profile and all \term{deviation profiles}: profiles formed by unilateral deviations.
For a pure profile, there are $\sum_i \abs{S_i} - \numPlayers$ deviation profiles.
For a mixed profile $\sigma$, the number of deviation profiles is generally exponential in $\max_i\abs{\support(\sigma_i)}$, which may still be modest for small-support profiles.

Finding an approximate solution is generally more expensive than verifying one, but can also often be accomplished short of exhaustive evaluation of the profile space.
We survey EGTA techniques for reasoning about incomplete game models in Section~\ref{sec:inc-game-reasoning}.

\subsubsection{Game Model Learning}
\label{sec:learning}

The machine learning approach to empirical game modeling is essentially a form of regression where the input is a set of (profile, payoff-vector) pairs, and the output is the vector of empirical utility functions $\hat{u}_i$.
These techniques can be used to infer a complete empirical game model from an incomplete one.

% \mw{Include simple illustration of regression to running example? This could simply interpolate the model over $\rho$ in the empirical game example above.}
% \amy{Absolutely! As it is now, it is hard to take away much of substance from this section. It is mostly a laundry list of where else to look for (interesting!) results in this vein.}

For illustration, suppose we wish to extend the example three-player sequential auction above (Table~\ref{tab:sec-auc3}) to include a fourth strategy, $\rho=0.4$.
There are many possible ways to extend the payoff matrix to include profiles with the new strategy.
In this case, since the strategies are defined parametrically, we can apply a nearest-neighbor approach with linear interpolation.
For example, consider a profile of the form $(0.4,\rho_2,\rho_3)$, with $\rho_2,\rho_3\in\{0.3,0.5,0.7\}$, the set of strategies for which we already have estimates.
To estimate $u(0.4,\rho_2,\rho_3)$, we could simply average $\hat{u}(0.3,\rho_2,\rho_3)$ and $\hat{u}(0.5,\rho_2,\rho_3)$, for example $\hat{u}_1(0.4,0.3,0.3) = (51.5+44.2)/2 = 47.85$.
The actual value, based on 10,000 samples, is 50.48.
Similarly, the interpolated estimate $\hat{u}_2(0.4,0.3,0.3)=(51.5+44.7)/2=48.1$ (actual sampled estimate is 47.2).
More sophisticated approaches could fit the payoff function to the data using richer hypothesis spaces, for example based on neural networks.

The first work to apply regression to learn payoff functions from simulation data was that of \citet{Vorobeychik07}.
\citet{Ficici08} employed clustering to partition a large number of players into two roles, then regression to find a best-fit representative payoff function for each role.
\citet{Jordan09w} applied cross-validation methods to select the best empirical game model, for instance to decide whether some strategies are similar enough to merge samples.
\citet{Wiedenbeck18} introduced methods to learn large symmetric games, using a representation that encodes the number of players choosing each strategy, rather than vectors based on a player ordering.
\citet{Sokota19} propose an approach that learns deviation payoffs (defined in Section~\ref{sec:hpt-RD}) with respect to role-symmetric mixed strategies, an alternative to directly learning payoff functions, which provides some advantages for equilibrium computation.
\citet{Li21w} employ this representation for game learning in an EGTA algorithm for symmetric Bayesian games, evaluated in a simultaneous-auction setting.
\citet{Shao25} propose a matrix completion approach, adopting a conservative bias to steer toward equilibria that are relatively well supported by the available data.

Some recent research in game representation has developed structural model forms that support succinct representation of games exhibiting particular regularities. 
An example is graphical games \citep{Kearns07}, which compactly capture situations where agents are affected by others' actions only in a local neighborhood.
There has also been some work on learning such graphical game models from payoff data \citep{Duong09,Fearnley15}.
\citet{Li20w} propose an approach that interleaves structure learning and payoff regression to induce tractable game models with many players.

\citet{Gatchel23w} broaden the problem to learn a model covering \textit{families} of games, defined by specified context features.
For example, a family of sequential auction games could encompass a range of scenarios parametrized by the number of goods, number of players, or features of the valuation distributions.
This approach supports reasoning about the relationship between environmental features and game solutions, with greater robustness and sample efficiency compared to learning separate models for an enumerated set of game instances.

%\amy{this paragraph doesn't seem relevant to me. maybe goes in related work, if anywhere?}
%\mw{Most of it is actually not, but seems necessary in this "game model learning" section to explain the relation of what is here to other things with the words "learning" and "games".}
%\mw{or in an "Unrelated Work" section. Seriously, though, here is where the point has some use.}
There is extensive additional literature at the intersection of machine learning and game theory, which is relevant but not specifically oriented toward learning game models from simulation data.
This includes a great deal of work on algorithms that learn to play games (i.e., converge to equilibria) through repeated interaction \citep{Fudenberg98}.
Some more recent work has exploited advances in deep neural networks to learn equilibrium behavior \citep{Bichler21,Gatti11r,Marchesi20tg} or optimal mechanisms \citep{Duetting24} directly from strategy simulations (i.e., without constructing a game model).
In the realm of game modeling, there is interesting research on learning to predict behavior of human players given a normal-form game specification \citep{Wright17}.
Yet another category is work on learning games from behavioral data \citep{Honorio15,Waugh11}, generally based on fitting structure and parameters of a game model under assumption that the behavior is generated rationally.
\citet{Gao10} suggest an approach that combines payoff data with behavioral data in this way.
In all of these categories we could cite many other works; satisfactory coverage would require a full-length survey treatment.

\subsubsection{Player Reduction}

Another technique developed to support scalability of empirical game modeling 
% \amy{using either of the above approaches} 
is \term{player reduction}.
The goal of this technique is to approximate a many-player game by one with considerably fewer~($\ll$) players.
The intuition is that in a large game, it may be approximately correct to reason coarsely about agent aggregates, expecting results not to be unduly sensitive to the exact number of agents adopting a particular strategy. 
%\amy{here is where we could mention mean-field games, if we want to say something about them. do you agree, Karl? can you possibly add a footnote noting the distinction/relationship?}
The approach inherently requires symmetry, otherwise aggregation would not be clearly meaningful.%
\footnote{We describe the techniques below for the case of full symmetry. 
Generalization to role symmetry is straightforward, limiting all aggregation to be applied within roles.}

Let $p < \numPlayers$; typically $p$ is a small fraction of $\numPlayers$.
Formally, a \term{reduced game} $\game^{(p)} = \langle P, (S_i), (u_i^{(p)}) \rangle $, $\abs{P} = p$, is a $p$-player game derived from an $\numPlayers$-player \term{full game} $\game = \langle \Players, (S_i), (u_i) \rangle$.
The reduced and full games have the same strategy sets.
The key to defining the reduction is specifying how the utility function $u_i^{(p)}$ for the reduced game can be derived from that of the full game.
Reductions of this sort mesh well with the simulation-based approach since payoff data for estimating the reduced game can be taken from simulations of the full game.

The idea of \term{hierarchical reduction} \citep{Wellman05rlcs} is to model the $p$-player game as if each player controlled $\numPlayers/p$ of the players in the full game.
Suppose for simplicity $p$ divides $\numPlayers$, so that $\numPlayers/p$ is integral.%
\footnote{The case where $\numPlayers/p$ is fractional can be handled by careful rounding or interpolation.}
Let $s_{-i}^{(p)}$ be an other-agent strategy profile in the reduced game, a vector of $p-1$ strategies.
Then we can define $u_i^{(p)}(s,s_{-i}^{(p)})=u_i(s,s_{-i}^{(\numPlayers)})$, where the full-game other-agent profile $s_{-i}^{(\numPlayers)}$ contains $\numPlayers/p$ copies of $s_{-i}^{(p)}$, plus $\numPlayers/p-1$ copies of $s$.
(Note that given symmetry, the ordering of strategies in $s_{-i}^{(\numPlayers)}$ does not matter.)

In \term{twins reduction} \citep{Ficici08}, an $\numPlayers$-player game is reduced to a 2-player game where each player views the other as representing an aggregate of all the rest playing the other strategy.
The derived utility function can be written simply as $u_i^{(2)}(s,s_{-i})=u_i(s,s_{-i}^{(\numPlayers)})$, where the full-game other-agent profile $s_{-i}^{(\numPlayers)}$ contains $\numPlayers-1$ copies of $s_{-i}$.
An interesting property 
% \amy{i feel like this should be stated up front as a goal of any reduction. a reduction for which this ``interesting property'' does not hold had better have some other compelling justification.} 
%\mw{The goal as stated above is "to approximate a many-player game by one with considerably fewer players". I don't think we should be restricting the nature of evidence for being an effective approximation.}
of the twins reduction is that it preserves symmetric PSNE, that is, $(s,s)$ is an NE of $\game^{(2)}$ twins-reduced from~$\game$ iff everyone playing $s$ is an NE of $\game$.
This is because the twins reduction (unlike hierarchical reduction) preserves the effect of single-player deviations.

\term{Deviation-preserving reduction} (DPR) \citep{Wiedenbeck12} generalizes twins reduction for $p>2$.
The idea of DPR is to consider each reduced player as controlling one player in the full game, but to treat the other reduced players as full-game aggregates.
For simplicity suppose $p-1$ divides $\numPlayers-1$.
The DPR mapping is given by $u_i^{(p)}(s,s_{-i}^{(p)})=u_i(s,s_{-i}^{(\numPlayers)})$, where the full-game other-agent profile $s_{-i}^{(\numPlayers)}$ contains $(\numPlayers-1)/(p-1)$ copies of $s_{-i}^{(p)}$.
For $p=2$, DPR is the same as twins, and for any $p$ it also preserves symmetric PSNE\@.

To apply player reduction in EGTA, one typically analyzes the reduced game as an approximation for the full game. 
Symmetric (mixed or pure) profiles can be interpreted as profiles over any number of players, so we can consider symmetric solutions of the reduced game as rough or candidate solutions of the full game.
If the support of the reduced-game solution is sufficiently small, evaluation of accuracy with respect to the full game by sampling can be quite tractable.
Although in the worst case an approximation by player reduction can be arbitrarily bad, DPR in particular has proved reasonably accurate for a range of natural games.
% \amy{reference?}
For example, \citet{Brinkman18} used DPR ($p=4$ or 6) to analyze financial market games with up to 216 agents, and verified statistically that regret due to the reduced-game approximation was small relative to the inherent variability of the models. 
% This has enabled scaling of EGTA studies to games defined by simulations of upwards of a hundred agents.
% \amy{does this idea work for symmetric Bayesian games? if so, i think that is worth mentioning. in the seq'l auction example, i am imagining that a bunch of bidders all use myopic bidding with some shading parameter, but that they all also draw different valuations? and that this idea allows us to analyze such games. maybe include an example, even, of a bigger version of the game in Table 1, using this reduction.}
%\mw{Confusing to draw this distinction. Bayesian games with identically drawn valuations are generally modeled here as symmetric NFGs.}

\subsection{Game Solving}

An empirical game model is fundamentally a model of a game, and so all the concepts and tools of computational game theory apply in principle to games induced from agent-based simulation or other data sources.
This includes game forms (e.g., normal form, extensive form), solution concepts, and algorithms for computing equilibria or other objects of game-theoretic analysis.
In practice, certain representations and methods have proven particularly relevant for EGTA studies.
An example is the HPT representation for symmetric games presented in Section~\ref{sec:hpt}.
This encoding is particularly convenient for applying \textit{replicator dynamics} (Section~\ref{sec:hpt-RD}), hence RD is commonly employed in EGTA.
EGTA-specific approaches have also been developed for special forms of learned game models, as well as representations that accommodate incompleteness in game models (Section~\ref{sec:subgame}).
% \amy{maybe first a sentence about backup methods, e.g., GAMBIT?}

\subsection{Strategy Evaluation}

By the very definition of a game environment, it is generally not possible to evaluate the quality of one player's strategy absent consideration of the strategies chosen by others. 
% \kt{maybe better: without considering the strategies chosen by others}
Thus, when we refer to solutions or solution concepts for games, we characterize these on the joint strategy space.
Nevertheless, we often are particularly interested in the perspective of a particular player, and would like some way to assess the efficacy of that player's strategy in absolute or relative terms (i.e., scores or rankings), without explicit reference to other-player choices. 
There is no escaping the need for making some assumptions about these choices, but it could be that some fairly generic assumptions about strategic context are sufficiently informative for strategy evaluation.

For example, one can invoke a worst-case assumption on other-agent play, and compare strategies on that basis.
The optimal strategy under this assumption is the \textit{maximin strategy}, and the  minimum payoff guaranteed by the maximin strategy is the player's \textit{safety value} for the game.

Except in zero-sum games, worst-case assumptions do not capture the interests of other players, and are thus not generally a realistic expectation for their choices.
An alternative is to assume rational play, which brings us back to the realm of game-theoretic solution concepts. 
For example, we might consider the performance of a player's  strategy when the others play according to a Nash equilibrium. 
\citet{Jordan07kw} defined the \term{NE-regret} of a strategy $s_i$ as $\regret_i(s_i,\sigma_{-i}^*)$, for $\sigma^*$ a NE profile. 
(In a game with multiple equilibria, there would be NE-regret measures with respect to each.)
\citet{Jordan10} showed how to use NE-regret for ranking strategies as part of an EGTA process.
% \amy{is there anything more to say here about his findings? an application of his ranking procedure, where we learned something about some game that we didn't o/w know?}

\citet{Balduzzi18} studied the question of empirical strategy evaluation in generalized form,
% \amy{what is ``generalized form''? (sorry if this was defined earlier.)} 
motivating the problem and pointing out strengths and limitations of various approaches.
% \kt{we can also mention not just AvA but also AvT, or maybe that is what you meant earlier with generalized form?}
Their proposal to rank strategies by \term{Nash-averaging} is technically equivalent to ranking by NE-regret, with an additional prescription to select the benchmark equilibrium that maximizes entropy.

Alternative approaches for ranking strategies appeal to evolutionary solution concepts.
In particular, \alpharank\ has been deployed to evaluate the strength of various learning strategies in the games of Go, soccer, and poker \citep{omidshafiei2019alpha}.
Amenability to a variety of measures that account for strategic context make EGTA a valuable tool for evaluating novel multiagent training algorithms \citep{Li24w}.

\subsection{Visualization}
% \kt{very nice section!}

\begin{figure}
    \centering
    \includegraphics[width=8cm]{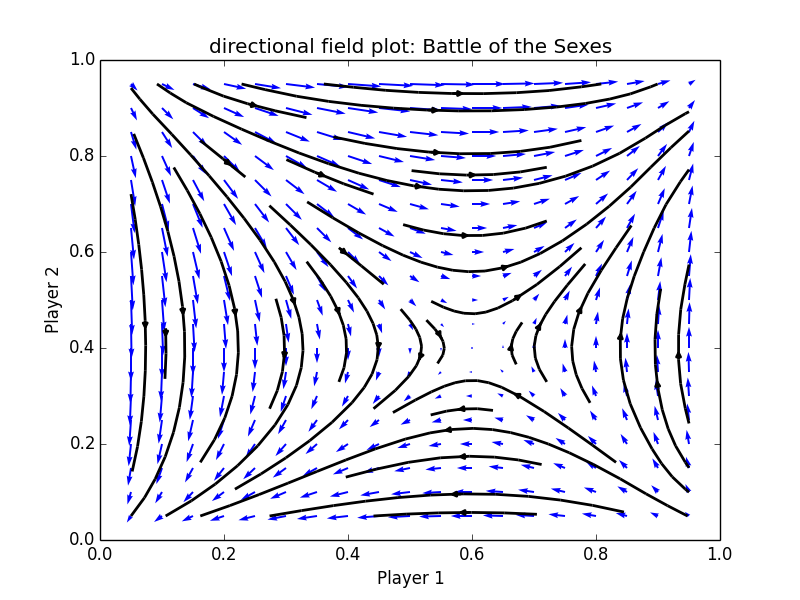}
    \caption{Directional field plot of the BoS game. 
    Flows point toward the two PSNE, $(B,B)$ and $(S,S)$, whereas the mixed Nash equilibrium at $(\frac23, \frac13)$ is unstable.}
    \label{fig:dfieldBoS}
\end{figure}

Game analysts are often motivated to draw strategic insights about a game environment, beyond identifying specific solutions.
Toward that end, researchers have devised various ways to visualize game analyses. 
An example is the use of directional field plots (e.g., Figures \ref{fig:PD} and~\ref{fig:dfieldBoS}) to illustrate evolutionary dynamics over a profile space.

\begin{figure}[ht!]
  \centering
 	\includegraphics[width=0.9\textwidth]{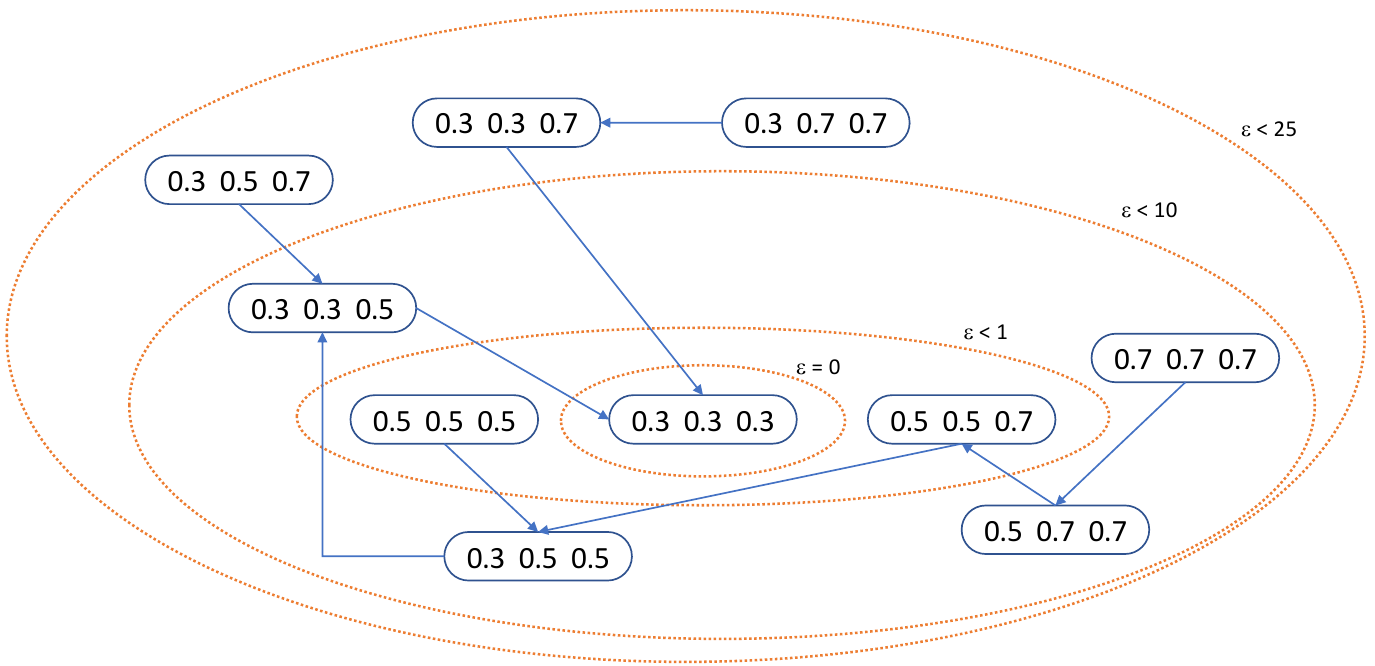}
 	\caption{Contour deviation graph for the sequential auction game model of Table~\ref{tab:sec-auc3}.
  Nodes are strategy profiles, and edges denote most beneficial one-player deviations.
  The dashed ellipses in orange delimit increasing levels of regret ($\regret$) as one moves outward in the plot.}
 	\label{fig:devgraph}
\end{figure}

Another approach developed in EGTA research graphs the relationships over an enumerated set of pure profiles.
A \term{deviation graph} connects profiles by edges indicating the most beneficial deviations.%
\footnote{These are related to response graphs, introduced in Section~\ref{subsubsec:a-rank} for analyzing learning dynamics.
Response graphs encode transitions for all alternative strategies, whereas here we focus on best deviations.}
% \amy{Mike, FYI, we use the term response graph for perhaps the exact same thing in the Section 2.3.3, when discussing stochastic stability.}
A plot of such a graph organized to display levels of profile regret is called a \term{contour deviation graph} \citep[Chapter~3]{Jordan10}.
Figure~\ref{fig:devgraph} presents a contour deviation graph corresponding to the $\numPlayers = \numGoods = 3$ auction empirical game model given by Table~\ref{tab:sec-auc3}.
The ten nodes of the graph represent the pure profiles of the game. 
In this visualization, an outgoing edge represents the most beneficial one-player deviation.
(One could also include edges for all deviations, weighted by the associated gain.)
For example, from the profile $(0.3,0.7,0.7)$ near the top right, a player could gain 14.2 by switching from strategy $\rho=0.7$ to $\rho=0.3$.
The sole node without an outgoing edge is $(0.3,0.3,0.3)$, the game's unique PSNE.

% Deviation graphs are closely related to the response graphs that encode learning dynamics, analyzed in the computation of stochastically stable distributions \citep{young:evolution, wicks:ssd} and \alpharank\ \citep{omidshafiei2019alpha}.
% In that context, the graph represents a Markov chain, with edges labeled by transition probabilities reflective of deviation gain.

Scaling these visualizations to high-dimensional or otherwise large profile spaces is a challenge.
New and creative ideas are required to support the extraction of strategic insights in game analysis. 
One interesting concept proposed by \citet{Czarnecki20} is to view strategic relationships on two dimensions: \textit{transitive} and \textit{intransitive}.
The upright axis represents transitive relationships, where strategies can be ordered relatively unambiguously in strength or competence. 
% \kt{I guess we could include the visualisation itself, WDYT?}
The radial axis represents intransitive relationships, like the canonical rock-paper-scissors interaction. 
Through an extensive EGTA exercise over popular two-player zero-sum games, these authors find a pervasive ``spinning top'' structure, in which the degree of non-transitivity tapers off as strategies improve along the upright dimension toward the game's NE solutions.
\citet{ShayeganOmidshafiei_11_2020} present a more comprehensive set of tools, based on response graphs and other constructs, for categorizing and visualizing the strategic landscapes of a wide variety of games.

Another type of strategic question of interest in EGTA is how solutions vary as a function of the game environment. 
Such questions are typically addressed by conducting EGTA over a space of parametric instances, and plotting features of the solution.
For example, \citet{Wang21hyw} used EGTA to derive empirical equilibria among trading strategies for 36 instances of a financial market game: with and without the presence of a market manipulator, for 18 configurations of trader population size and stochastic parameters of the market.
Plotting features of the strategies adopted in equilibrium, as well as outcome features (in this case, spoofing effectiveness, price accuracy, and welfare) yields insights about how spoofing manipulation operates across a space of market situations.
These kinds of visualizations are now quite common in EGTA studies.
% \kt{can we give an example?}

\section{Evolutionary Game Analysis}
\label{sec:evolution}

As noted in Section~\ref{sec:egt}, evolutionary game theory is a natural fit for games defined by agent-based simulation.
In this section, we elaborate on some of the basic concepts underpinning the evolutionary perspective on game theory, and discuss EGTA methods developed based on these concepts.

\subsection{Evolutionarily Stable Strategies}\label{sec:ESS}

%\kt{Finetune and adjust notation}
%\mw{note this all assumes symmetry}

%Imagine a single population $P$ (the symmetric case) of simple agents (aka replicators) playing the same strategy (from set $S_i = S$).

Imagine a large population of agents, each playing a pure strategy, from which two are repeatedly selected at random to play a game.
Imagine further that these agents reproduce according to their success in these interactions, so that successful strategies multiply, while unsuccessful ones die off.
In evolutionary game theory, we often view such populations as conceptually infinite, and interpret the distribution of strategies adopted as a symmetric mixed-strategy profile.
If one strategy attempts to invade another (i.e., if a small part of the population mutates), but if the reproductive success of the new one lags behind that of the original, then the new strategy will eventually disappear. 
In other words, the original strategy is \textit{evolutionarily stable} against this invading strategy.
More generally, an \term{evolutionarily stable strategy} (ESS) is a strategy that is robust against evolutionary pressure from any mutant strategy not yet present in the population, or present only as a very small fraction.
%\kt{write 'f' in function of 'y'}

Formally, suppose the population can be described by the state vector (or mixed strategy) $\sigma = (\sigma(s_{1}), \dotsc, \sigma(s_{\numStrats}))$, with $\forall j.\ 0 \leq \sigma(s_{j}) \leq 1$ and $\sum_j \sigma(s_{j}) = 1$, representing the fractions of the population playing pure strategies $1,\ldots,\numStrats$, respectively.
A strategy $\sigma$ is an ESS if it is immune to invasion by other strategies that initially occupy only a small fraction of the population. 
% \amy{i want to resolve my old comment about this sentence. but i cannot see it!}
%\amy{Karl, where does the def'n insist that $\pi$ is small relative to $\sigma$? i'm missing why that is required.}\kt{this comes from the original definition of ESS, see e.g. Herb Gintis' book. Suppose the state of a population is $\sigma$. If we now introduce a small fraction $\epsilon$ playing $\pi$, the mutant strategy, then the population state is $(1-\epsilon)\sigma + \epsilon\pi$, then the expected payoff to the nonmutant is $(1-\epsilon)f(\sigma,\sigma)$ + $\epsilon f(\sigma,\pi)$ and to the mutant is $(1-\epsilon)f(\pi,\sigma)+\epsilon f(\pi,\pi)$. So the mutant can invade the population for a sufficiently small $\epsilon$, if it holds that $f(\pi,\sigma) \geq f(\sigma,\sigma)$. and 2 can be derived as well.}
Let $f(\sigma, \pi)$ be the (expected) fitness of strategy $\sigma$ against strategy $\pi$.
Formally, then, strategy $\sigma$ is an ESS iff, for any mutant strategy $\pi$, the following hold:
\begin{enumerate}
  \item $f(\sigma, \sigma) \geq f(\pi, \sigma)$, and
  \item if $f(\sigma, \sigma) = f(\pi, \sigma)$, then $f(\sigma, \pi) > f(\pi, \pi)$.
\end{enumerate}
The first condition states that an ESS is also a Nash Equilibrium of the original game, which implies that ESS is a refinement of the Nash solution concept.
The second condition states that if the invading strategy does as well against the original strategy as the original strategy does against itself, then the original strategy must do better against the invader than the invader does against itself.
It turns out that every ESS is an asymptotically stable fixed point of the replicator dynamics process \citep{Weibull97}, which we define next.

\subsection{Replicator Dynamics}
\label{sec:replicator}
%\kt{Need to adjust this to Michael's notation from previous sections}

\term{Replicator dynamics} (RD), introduced by \citet{Taylor78} and developed further by \citet{Schuster1983533}, is another key concept from EGT. 
As a dynamical system, RD describes mathematically how a population can evolve over time, either computationally or based on biological operators such as selection, mutation, and crossover. 

In their most basic form, the RD equations express the canonical biological selection mechanism: survival of the fittest.
Suppose the fitness of pure strategy $i$ is given by a fitness function $f_i (\sigma)$, typically defined as $i$'s expected payoff, with the average population fitness $\bar{f}(\sigma) =\sum_j \sigma_jf_j(\sigma)$.
In the single population case, which is applicable to symmetric games, 
%\amy{how/why is this single pop'n? what would be different if there were multiple pop'ns?}\kt{you get a coupled system of diff eqns, see below - asymmetric games} 
the RD equations are:
\begin{align}\label{eq:singleRD1}
    &\dot{\sigma}_{i} = \sigma_{i} \left(f_i(\sigma) - \bar{f}(\sigma) \right),
\end{align}
%
%hese dynamics can be deployed for symmetric bimatrix games, assuming both agents' strategies are sampled from the same profile $\sigma$.}{}
%More specifically, 
If $A$ denotes a symmetric bimatrix payoff matrix, as in Section~\ref{sec:bimatrix},
%, defining fitness as expected payoff,
%Usually we work directly with $2$-player payoff matrices $A$ as defined in Section~\ref{sec:bimatrix}. Using now $x$ to denote our mixed strategy $\sigma_i$ (or population state or distribution from which replicators are sampled), 
the RD equations simplify as:
\begin{equation}\label{eq:singleRD2}
    \dot{\sigma}_{i} = \sigma_{i}((A\sigma)_{i} - \sigma^TA\sigma).
\end{equation}

\begin{figure}[ht!]
\centering
\begin{subfigure}{.28\textwidth}
  \centering
  \gamematrix{C}{D}{3,3}{0,5}{5,0}{1,1}
\end{subfigure}%
\begin{subfigure}{.55\textwidth}
  \centering
    \includegraphics[width=8cm]{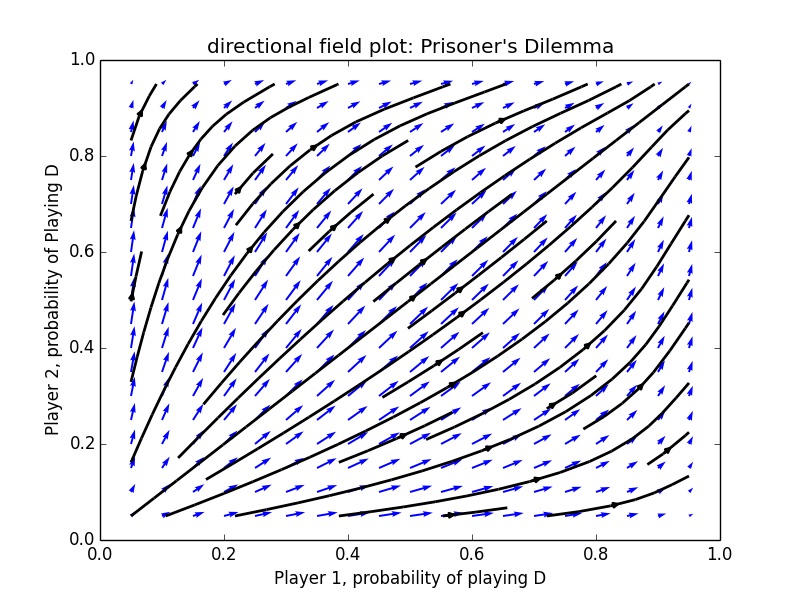}
\end{subfigure}
\caption{Payoff matrix (left) and directional field plot (right) for the Prisoners' Dilemma game.
  Strategies $D$ and $C$ represent \textit{Defect} and \textit{Cooperate}.
  Flows point towards full probability on the unique Nash equilibrium, $(D,D)$, which is a strong attractor.}
\label{fig:PD}
\end{figure}

We illustrate these dynamics on a classic example, the Prisoners' Dilemma, whose payoff table is shown in Figure~\ref{fig:PD}.
The axes of the field plot correspond to the probability of the respective players playing action $D$ (Defect). 
The gradient flow indicates that all probability mass flows to coordinates $(1,1)$, which represents the pure Nash equilibrium $(D,D)$. 
One can also observe from this plot that the Nash equilibrium $(D,D)$ is evolutionarily stable: injecting any number of cooperators into the population would not lead the dynamics to stray from the $(D,D)$ state.

%\kt{add an example here, and also refer to ESS}

% \begin{figure}[ht]
%   \centering
%   \gamematrix{C}{D}{3,3}{0,5}{5,0}{1,1}
%   \caption{Payoff matrix for the Prisoners' Dilemma game.
%   $D$ and $C$ correspond to the strategies \textit{Defect} and \textit{Cooperate}.}
%   \label{fig:PD}
% \end{figure}

% \begin{figure}[ht!]
%     \centering
%     \includegraphics[width=8cm]{figs/PD_directionfield.jpg}
%     \caption{Directional field plot of the Prisoners' Dilemma game. Flows point towards full probability on the unique Nash equilibrium, state $(D,D)$, which is a strong attractor.}
%     \label{fig:dfieldPD}
% \end{figure}

We can also extend the replicator equations to asymmetric games, but we then need multiple populations. 
In a two-player asymmetric bimatrix game with payoff matrices $(A,B)$, and population states $x$ and $y$, respectively, for populations 1 and~2 (i.e., corresponding to players 1 and~2), evolution 
%of players' strategy profiles
under RD is now given by
\begin{equation}\label{eq:asymRD}
    \dot{x}_{i} = x_{i}((Ay)_{i} - x^TAy) \qquad     \dot{y}_{j} = y_{j}((x^TB)_{j} - x^TBy) \qquad \forall (i, j)\in S_1 \times S_2.
\end{equation}
%where $x_{i}$ and $y_{j}$ are, respectively, the proportions of strategies $i \in S_1$ and $j \in S_2$ in two infinitely-sized populations.
This system of coupled differential equations models the temporal dynamics of the interactions among agents in these two %infinitely-sized
populations.

We illustrate the asymmetric replicator dynamics in the BoS game in Figure~\ref{fig:BoSbimatrix}, which depicts the gradient dynamics of player 1 (row, $x$-player) and player 2 (column, $y$-player), respectively. 
This game has three equilibrium fixed points: two pure-strategy NE (PSNE)---at the origin and $(1,1)$, and a third (evolutionarily unstable) mixed NE at coordinates $(\frac{2}{3},\frac{1}{3})$.
This example thus illustrates how the ESS solution concept is a refinement of Nash.

%\amy{what else do these examples illustrate? when we answer this question, let's do so by referencing the goals of EGT laid out at the very start of 2.4. let's discussion the basins of attraction, that there are no limit cycles or chaos, etc.}\kt{we could, but maybe that is too detailed for what we want to achieve in this paper?}

%\kt{quickly describe figure here}

\subsection{Evolutionarily Dynamic Solution Concepts}
\label{subsubsec:a-rank}

%\kt{you integrated SSD in this section - I was somehow expecting a separate subsection before this one, but this makes sense too.}\amy{oops. should have discussed first. can easily change. but -- there seems to be SOOO much overlap b/n the two ideas, that i think it makes more sense to include both in one section. in fact, much of what i wrote about SSDs i stole from what you had written about $\alpha$-rank.}

%\kt{or: from replicator dynamics to alpha-rank}
Whereas Nash equilibria are guaranteed to exist in all finite games \citep{nash1950equilibrium}, their computation is believed to be intractable in general-sum games \citep{daskalakis2009complexity, goldberg2013complexity}, and their non-uniqueness leads to an equilibrium selection problem \citep{harsanyi1988general}. 
% From this point of view, their applicability as an underlying principle for analyzing large empirical games is limited.
Such indeterminacy motivates the consideration of behavior under  learning dynamics (e.g., fictitious play), viewed as stochastic processes.

%The main advantages of this new concept, based on discrete-time evolutionary dynamics \citep{Gintis09}, 
%%\kt{we could also explain the discrete version of the RD in the previous section}
%concerns its uniqueness and efficient computation in many-player and general-sum games, making it a promising foundation for empirical game-theoretic analysis and new multiagent learning algorithms.

One simple way to apply this approach, dating back to \citet{young:evolution}, is to represent the learning dynamics of interest by a Markov chain.
% \amy{new text follows. could not add overleaf comment, sorry:}
One such possible Markov chain is a \term{response graph}, with states as strategy profiles, and transitions/edges indicating the extent to which a player has an incentive to deviate unilaterally from one state/profile to another.
One can then consider a stationary distribution of this Markov chain as a solution concept, but the stationary distribution is not necessarily unique.
Instead, one often introduces small perturbations to this chain, which render it irreducible and its stationary distribution unique \citep{LevinPeresWilmer2006}.
Note that such a solution concept captures agent interactions regardless of initial conditions (i.e., the agents' initial strategies).
\if 0
\samy{}{, unlike, say, ELO ratings, viewed as a solution concept}.
\amy{what does this mean?}
\kt{the point that is being made relates to for example Elo rating, where it is not possible to capture strength of strategies under intransitive behaviors; e.g. in an RPS setting, it would choose one of the three strategies as strongest, depending on the initial conditions of where you start in the strategy simplex.}
\fi

The \term{stochastically stable distribution} (SSD) of a Markov chain is the limit as $\epsilon \to 0$ of such a perturbed Markov chain,%
\footnote{This limit is guaranteed to exist when the perturbed Markov chain is \emph{regular} \citep{young:evolution}.}
and the \term{stochastically stable states} (SSS) are those in the support of the SSD\@.
The SSS yield an alternative solution concept, possibly ranked according to the corresponding SSD probabilities.
%, and the stochastically stable states (SSS) are those in the support of the SSD.
Unlike Nash equilibrium, the SSD (and hence the set of SSS) is unique, and computing it is tractable \citep{wicks:ssd}.
\citet{young:evolution} thus proposed it as a solution to equilibrium selection in his work on the evolution of conventions.

More recently, a related solution concept called \term{\alpharank} was proposed by \citet{omidshafiei2019alpha}.
Not only was it applied as described above to 
%solve equilibrium selection by ranking 
rank strategy profiles, it was also applied to rank agents, by building a Markov chain whose states are agents instead.
Encoded in the Markov chain's probabilities are evolutionary dynamics that model a selection-mutation process over a set of interacting populations.
Individuals sampled from each population play an $n$-player game, and then the strongest individuals either reproduce, or with a very small probability mutate.

\begin{figure}[ht!]
    \centering
    \includegraphics[width=12cm]{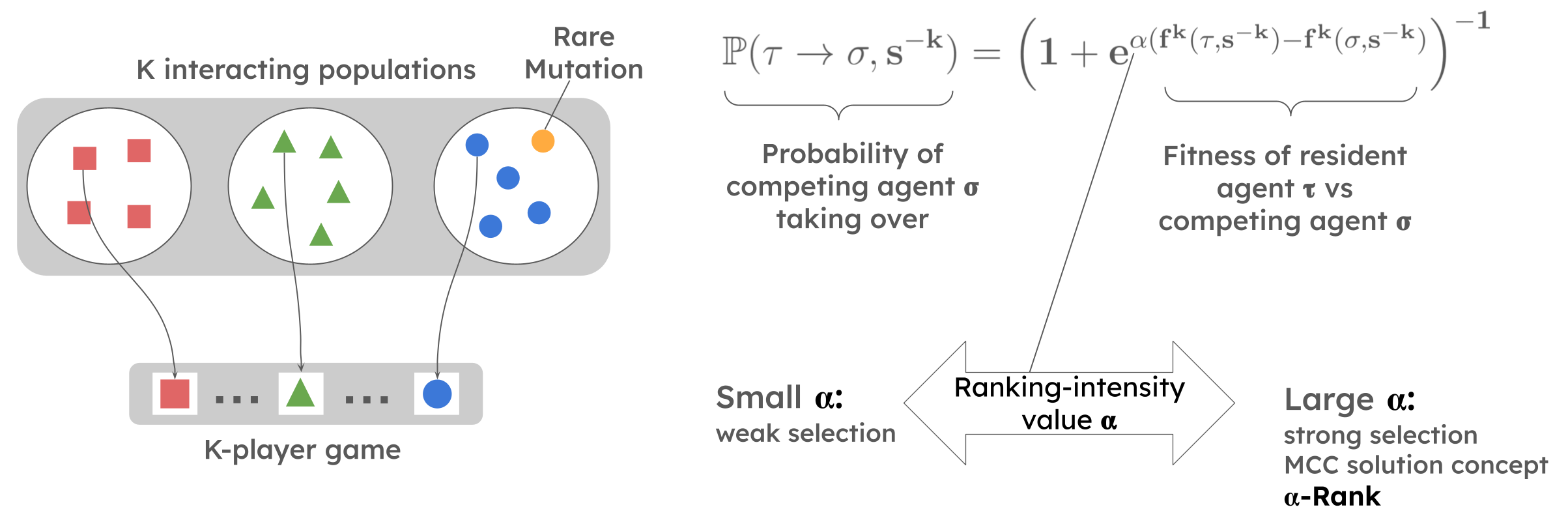}
    \caption{A discrete-time selection-mutation evolutionary process described as a Markov chain.}
    \label{fig:evoARank}
\end{figure}

Specifically, the propagation of strong agents is driven by a selection function that compares the fitness of a resident agent $\tau$ with a competing agent $\sigma$, as shown on the bottom right of Figure~\ref{fig:evoARank}.
The \term{ranking-intensity} of selection,
%, or the so-called ranking-intensity value
which is given by a parameter $\alpha$, influences the probability that a mutant overtakes a population.
A low $\alpha$ corresponds to weak selection, while a large $\alpha$ ensures that only the strongest mutants survive.
This selection-mutation model is encoded as the transition matrix of the \alpharank{} Markov chain.
(See the formula at the top right of Figure \ref{fig:evoARank}.)
%depicts the formula for the transition probabilities of the \alpharank{} Markov chain.

%this evolutionary model converges exactly to the perturbed MCC model \amy{didn't get this?} we outlined previously, and yields the \term{\alpharank} ranking method.

% \begin{figure}
%     \centering
%     \includegraphics[width=8cm]{figs/alpharankmethod.png}
%     \caption{The $\alpha$-Rank algorithm summarized.}
%     \label{fig:ARank}
% \end{figure}

The $\alpha$-Rank algorithm proceeds as follows:
% is illustrated in Figure~\ref{fig:ARank}.
\begin{enumerate}
    \item Estimate the empirical game.
    \item Define a Markov chain where states are the agents being evaluated.
    \item Compute the transition matrix according to a selection-mutation model parameterized by $\alpha$.
    \item Compute the stationary distribution of this Markov chain.
    \item Rank the agents via ordered masses of this distribution.
\end{enumerate}
% As this Markov chain is irreducible for $\alpha > 0$, we can compute its stationary distribution, and rank the agents accordingly.
The final steps are repeated by sweeping over $\alpha$ values,  until the resulting rankings stabilize over a few consecutive iterations.
%\amy{how is eventual stability ensured?}

%For additional details, we refer the reader to \cite{omidshafiei2019alpha}.

\begin{figure}[ht!]
    \centering
    \includegraphics[width=7cm]{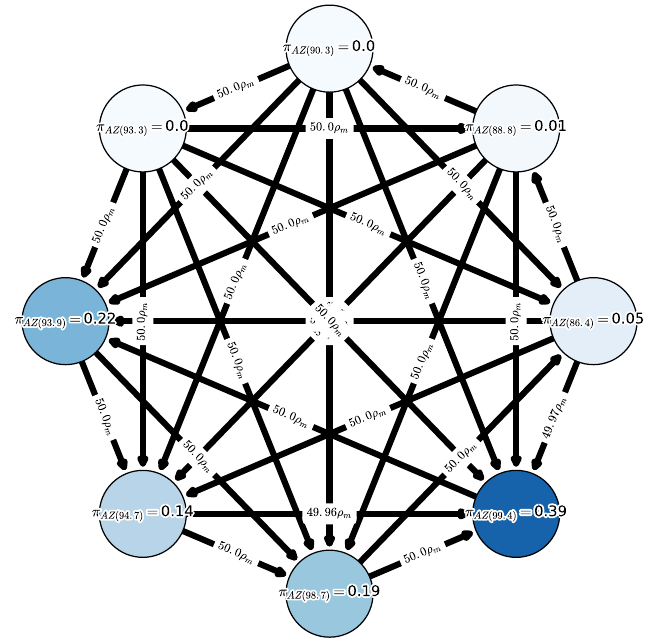}
   \hspace{9mm}
    \includegraphics[width=4cm]{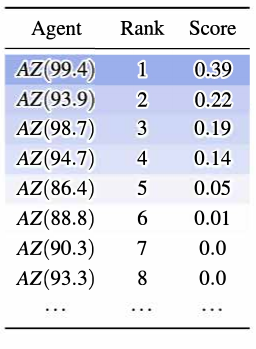}
    \caption{(LHS) Graph representation of the discrete-time dynamics applied to the AlphaZero dataset, also referred to as the response graph. Nodes represent the various agents or checkpoints during training of the Chess agent. (RHS) The outcome of $\alpha$-Rank applied to AlphaZero Chess, considering 56 agents and ranking the top 8 (listed).}
    \label{fig:ARankAlphaGo}
\end{figure}

Figure~\ref{fig:ARankAlphaGo} presents an example. 
The empirical game here is a two-player symmetric game with 56 AlphaZero Chess checkpoints, ranging from beginning to end of training. 
For example, AZ(99.4) has completed 99.4$\%$ of training. One note to make here is that these interactions are evaluated using the entire 56-agent dataset, though we show only the top-8 ranked agents for clarity.

The payoffs in the empirical game are the win ratios of pairwise match-ups between these agents.
On the left we depict the ensuing discrete-time Markov chain, 
%where each node is a distinct agent,
with node color indicating the mass of the stationary distribution on that agent.
The majority of top-ranked agents indeed correspond to snapshots taken near the end of training.
There are some more interesting outcomes, however. For example, the agent ranked 5th is AZ(86.4), which places higher than several agents with longer training time. 
The key point is that by using this type of evaluation, we can narrow the focus to agents of interest during training, and conduct a more refined analysis of their interactions.

\subsection{Replicator Dynamics on HPTs}
\label{sec:hpt-RD}

% \amy{my new understanding is the following: given the HPT rep'n, there are two things we might study: mixed strategies and infinite pop'ns. for each, there is a corresponding definition of $U_{\sigma, j}$. but then -- i don't see why we wouldn't run replicator dynamics on either. iow, i don't think the discussion of replicator dynamics should be wedded to eq'n 5, as if it is not also relevant to the second def'n of $U_{\sigma, j}$, meaning the one in the infinite pop'n section. i think there are two ways to define these utilities, and that the discussion of RD should be orthogonal to these two def'ns.}

The HPT representation can be leveraged for use with any game-solving method, but is particularly well-suited for implementing algorithms that exploit symmetry, such as the method of replicator dynamics described in Section~\ref{sec:replicator}.
That section presented RD as a dynamical system in terms of differential equations.
Here we describe it as an equilibrium search algorithm, which traverses a simplex as does the RD dynamical system, though in discrete steps.
Points in the simplex represent mixed strategies, which can also be interpreted as symmetric mixed profiles.%
\footnote{As discussed in Section~\ref{sec:evolution}, it is common in evolutionary game theory to interpret the points as fractions of an infinite population of agents employing various pure strategies.
We present the use of RD for studying evolutionary dynamics under this interpretation in Section~\ref{sec:infinite-pop}.}
%Viewed in search terms, 
As a search algorithm,
RD is an iterative improvement method that evaluates all strategies in the current solution, and updates their probabilities based on relative fitness.
If all strategies are equally fit,
%(i.e., have same expected payoff)
% \kt{if the rate of change of all strategies are equal to 0; not sure I understand 'if all strategies are equal'}
RD is at a fixed point.
RD may cycle or otherwise fail to converge, but if it does converge to a stable fixed point (e.g., an ESS), the result also constitutes a Nash equilibrium \citep{Gintis09}.

Fitness in this context is expected payoff, under the assumption that all other players choose according to the current mixed strategy.
Let $\sigma$ be a mixed strategy, with $\sigma(s_j)$ the probability that strategy~$s_j$ is played in~$\sigma$.
Given $\sigma$ and HPT $\hpt=(\Counts,\Util)$, the \term{deviation payoff} $\Util_{\sigma,j}$ for strategy~$s_j$
represents the expected payoff to that strategy conditional on other players following the mixture~$\sigma$.
To express this expectation, define $\Counts_{k,\ell}^{-j}$ as a vector of other-agent counts given that the designated player plays $s_j$ in the $k^\text{th}$ profile.
That is, $\Counts_{k,j}^{-j} = \Counts_{k,j}-1$, and $\Counts_{k,\ell}^{-j} = \Counts_{k,\ell}$ for $\ell\neq j$.
Then
\begin{equation}
\label{eq:expected-payoff}
\Util_{\sigma,j}= 
   \sum_{k\mid \Counts_{k,j} > 0} 
   \binom{\numPlayers -1}{\Counts_{k,1}^{-j},\dotsc,\Counts_{k,\numPlayers}^{-j}} \prod_{\ell=1}^{\numStrats} \sigma(s_\ell)^{\Counts_{k,\ell}^{-j}}
   \,\Util_{k,j},
\end{equation}
where $\binom{\numPlayers-1}{\Counts_{k,1}^{-j}, \dotsc, \Counts_{k,\numPlayers}^{-j}}$ is a multinomial coefficient, which describes the number of possible partitions of $\numPlayers-1$ objects into groups of sizes $\Counts_{k,1}^{-j}, \ldots, \Counts_{k,\numPlayers}^{-j}$.

\citet{Wiedenbeck23b} present a series of data structure improvements, building on the HPT representation, that facilitate representation and computations over deviation payoffs. 
The cumulative effect of these is a $10^4$-fold speedup in RD for games with many players, compared to the baseline matrix game representation. 

Given a specification of deviation payoffs, the RD algorithm starts from an initial mixed strategy $\sigma_{0}$ and updates the mixture at each step~$t$ using a discrete-time version of the replicator equation~\eqref{eq:singleRD1}:
\begin{equation}\label{eq:discrete-rd}
  \sigma_{t+1}(s_j)\gets 
     \sigma_{t}(s_j)\left[1 + \alpha 
        \left( \Util_{\sigma_t,j}
            - \sum_{\ell} \sigma_{t}(s_\ell)\,\Util_{\sigma_t,\ell} \right) \right],
\end{equation}
where $\alpha>0$ is a learning-rate parameter.

A typical implementation of RD for game-solving will start from a uniform or random initial mixed strategy, and iterate~\eqref{eq:discrete-rd} until convergence (e.g., change in probabilities fall below a numeric threshold), or until a maximum number of steps have been reached. 
At termination, the result can be evaluated for distance from Nash equilibrium, as measured by regret.
Multiple runs from different starting points is also common practice, to deal with non-convergence and to identify multiple equilibria if they exist.
Though experience with RD has confirmed it is usually effective on empirical games encountered in practice, it is not guaranteed to find solutions and therefore having backup solution algorithms is recommended.
For example, \citet{Wiedenbeck23b} recommend running both RD and gradient descent (also supported by HPT-based representation of deviation payoffs) from a diverse set of starting points in the strategy simplex.

\subsection{Approximating Infinite Populations for Evolutionary Dynamics}
\label{sec:infinite-pop}

The previous section described how the HPT representation of empirical games can facilitate RD computations, assuming the dynamic state variable $\sigma$ is interpreted as a symmetric mixed strategy.
Evolutionary game theory more commonly appeals to an alternate interpretation, where $\sigma$ is viewed as proportions of an infinite population.
RD using the HPT representation can apply under this interpretation as well \citep{Bloembergen15}, where the population is approximated by a large, but finite, population of $\numReps$ \term{replicators}.

An evolutionary game played by a population of $\numReps\ge\numPlayers$ replicators is derived from, but not the same as,
% \amy{Karl, if possible, it would be good to succinctly describe why these games are different: e.g., is one an approximation of the other?}\kt{yes, good point} 
an underlying game among a fixed set of $\numPlayers$ players.
In the evolutionary game among replicators, $\numPlayers$ instances are drawn from the population, inheriting their strategies, with or without replacement. 
The selected replicators then play their strategies and receive payoffs per the underlying $\numPlayers$-player game.
The payoff to each replicator in this game is their deviation payoff conditional on their selection in this sampling process, given the population distribution.
The larger the population $\numReps$, the better the game among replicators approximates an infinite-population evolutionary game.

Using the HPT representation, we can express an evolutionary game with $\numReps$ replicators as if it were a $\numReps$-player game.
The idea is simply to define the payoffs by the result of the sampling process.
For example, consider the $2\times 2$ Prisoner's Dilemma game (Figure~\ref{fig:PD}).
Payoffs for the game with two replicators and no replacement (Table~\ref{table:hpt2nr}) are exactly the same as for the underlying normal-form game, since $\numReps=\numPlayers$.
With replacement (Table~\ref{table:hpt2wr}), we see a difference in the profile where the population comprises one $C$ and one~$D$ ($\Util_C$ in red).
Conditional on a $C$ replicator being chosen for one player, the probability of the other being~$D$ is one under no-replacement (payoff 0), but 0.5 if the sampling is with replacement (payoff $\nicefrac12(3) + \nicefrac12(0)$).

\begin{table}[ht]
    \centering
    \caption{\small Prisoner's Dilemma HPTs with two replicators.} 
    \label{table:hpt2reps}
    \begin{subtable}[b]{0.4\textwidth}
%\footnotesize
\caption{No replacement}
    \label{table:hpt2nr}
%		\begin{center}
		$\left( \begin{array}{ccccc}
		\Counts_{C}& \Counts_{D} &  \vline & \Util_{C} & \Util_{D}  \\ 
		\hline
		2 & 0 & \vline & 3 & -  \\
		1 & 1 &  \vline & {\color{red}0} & 5   \\
		0 & 2 &  \vline & - & 1  \\
		\end{array} \right)$ 
	%	\end{center}
    
    \end{subtable}
    \begin{subtable}[b]{0.4\textwidth}
    \caption{\small With replacement.}
    \label{table:hpt2wr}
%\footnotesize
	%	\begin{center}
		$\left( \begin{array}{ccccc}
		\Counts_{C}& \Counts_{D} &  \vline & \Util_{C} & \Util_{D}  \\ 
		\hline
		2 & 0 & \vline & 3 & -  \\
		1 & 1 &  \vline & {\color{red}1.5} & 3  \\
		0 & 2 &  \vline & - & 1  \\
		\end{array} \right)$ 
	%	\end{center}
   % \vspace{-1cm}
\end{subtable}
\par \bigskip
\end{table}

For the infinite population case, one may consider the probability that each profile~$k$ is realized under the given $\sigma$:
\begin{displaymath}
\Pr(k\mid \sigma) = \binom{\numPlayers}{\Counts_{k,1}, \dotsc, \Counts_{k,\numPlayers}} \prod_{j=1}^{\numStrats} \sigma(s_j)^{\Counts_{k,j}}.
\end{displaymath}
The deviation payoff $\Util_{\sigma,j}$ can then be computed (in alternative to~\eqref{eq:expected-payoff}) as the normalized weighted combination of the profile payoffs:
\begin{displaymath}
\Util_{\sigma,j}= \frac{\sum_{k\mid \Counts_{k,j} > 0} \Pr(k\mid \sigma)\,\Util_{k,j}}{{\sum_{k\mid \Counts_{k,j} > 0} \Pr(k\mid \sigma)}}.
\end{displaymath}

% \amy{Karl, i think we need to explain the relationship b/n \eqref{eq:expected-payoff} and this new equation for $\Util_{\sigma,j}$ in the infinite case. my suspicion is it is something like we are treating replicators as if they were non-atomic here, b/c there are so many of them. so maybe this new equation is the limit of \eqref{eq:expected-payoff}, as the number of replicators goes to infinity. once we understand the relationship b/n these equations, we can better understand how to relate the different outcomes that might arise after applying RD to each.}

\begin{figure}[h!]
    \centering
    \includegraphics[width=8cm]{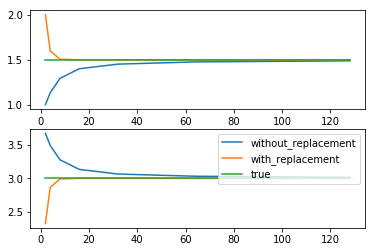}
    \caption{Expected payoffs approach the infinite-population value as we increase the number of replicators in the $2\times 2$ Prisoner's Dilemma game.}
    \label{fig:ehptreplicators}
\end{figure}

Tables \ref{table:hpt6reps} and~\ref{table:hpt10reps} show the HPTs for six and ten replicators, respectively.
As $\numReps$ increases, the replicator HPT provides a better approximation to the infinite game.
Suppose the population is half $C$ and half~$D$\@.
Figure~\ref{fig:ehptreplicators} shows how the expected payoffs for $C$ (top) and $D$ (bottom) approach the true infinite-population value (green line) as the number of replicators increases.
Convergence is relatively quick whether sampling is with (orange curve) or without (blue curve) replacement, but the former can provide a significantly better approximation for small numbers of replicators.

% \amy{no need to depict so many replicators in Figure 6. 60 would be enough. better to focus more on the area of the graph which is interesting, which is before about 20. also, i'm not sure how much we really need these graphs. i think the tables tell the story well enough.}

\begin{table}[t]
    \centering
    \caption{\small Prisoner's Dilemma HPTs with six replicators.} 
    \label{table:hpt6reps}
    \begin{subtable}[b]{0.4\textwidth}
%\footnotesize
\caption{No replacement}
    \label{table:hpt6nr}
    %	\begin{center}
		$\left( \begin{array}{ccccc}
		\Counts_{C}& \Counts_{D} &  \vline & \Util_{C} & \Util_{D}  \\ 
		\hline
		6 & 0 & \vline & 3 & -  \\
		5 & 1 &  \vline & 2.4 & 5   \\
		4 & 2 &  \vline & 1.8 & 4.2  \\
		3 & 3 &  \vline & {\color{red}1.2} & {\color{red}3.4} \\
		2 & 4 &  \vline & 0.6 & 2.6 \\
		1 & 5 &  \vline & 0 & 1.8 \\
		0 & 6 &  \vline & - & 1 \\
		\end{array} \right)$
	%	\end{center}
    
    \end{subtable}
    \begin{subtable}[b]{0.4\textwidth}
    \caption{\small With replacement.}
    \label{table:hpt6wr}
%\footnotesize
	%	\begin{center}
	$\left( \begin{array}{ccccc}
		\Counts_{C}& \Counts_{D} &  \vline & \Util_{C} & \Util_{D}  \\ 
		\hline
		6 & 0 & \vline & 3 & -  \\
		5 & 1 &  \vline & 2.5 & 4.33   \\
		4 & 2 &  \vline & 2 & 3.66  \\
		3 & 3 &  \vline & {\color{red}1.5} & {\color{red}3} \\
		2 & 4 &  \vline & 1 & 2.33 \\
		1 & 5 &  \vline & 0.5 & 1.66 \\
		0 & 6 &  \vline & - & 1 \\
		\end{array} \right)$ 
	%	\end{center}
   % \vspace{-1cm}
\end{subtable}
\par \bigskip
\end{table}

\begin{table}[t]
    \centering
    \caption{\small Prisoner's Dilemma HPTs with ten replicators.} 
    \label{table:hpt10reps}
    \begin{subtable}[b]{0.4\textwidth}
%\footnotesize
\caption{No replacement}
    \label{table:hpt10nr}
%		\begin{center}
	$\left( \begin{array}{ccccc}
		\Counts_{C}& \Counts_{D} &  \vline & \Util_{C} & \Util_{D}  \\ 
		\hline
		10 & 0 & \vline & 3 & -  \\
		9 & 1 &  \vline & 2.66 & 5   \\
		8 & 2 &  \vline & 2.33 & 4.55  \\
		7 & 3 &  \vline & 2 & 4.11 \\
		6 & 4 &  \vline & 1.66 & 3.66 \\
		5 & 5 &  \vline & {\color{red}1.33} & {\color{red}3.22} \\
		4 & 6 &  \vline & 1 & 2.77 \\
		3 & 7 &  \vline & 0.66 & 2.33 \\
		2 & 8 &  \vline & 0.33 & 1.88 \\
		1 & 9 &  \vline & 0 & 1.4 \\
		0 & 10 &  \vline & - & 1 \\
		\end{array} \right)$ 
	%	\end{center}
    
    \end{subtable}
    \begin{subtable}[b]{0.4\textwidth}
    \caption{\small With replacement.}
    \label{table:hpt10wr}
%\footnotesize
	%	\begin{center}
	$\left( \begin{array}{ccccc}
		\Counts_{C}& \Counts_{D} &  \vline & \Util_{C} & \Util_{D}  \\ 
		\hline
		10 & 0 & \vline & 3 & -  \\
		9 & 1 &  \vline & 2.7 & 4.6   \\
		8 & 2 &  \vline & 2.4 & 4.2  \\
		7 & 3 &  \vline & 2.1 & 3.8 \\
		6 & 4 &  \vline & 1.8 & 3.4 \\
		5 & 5 &  \vline & {\color{red}1.5} & {\color{red}3} \\
		4 & 6 &  \vline & 1.2 & 2.6 \\
		3 & 7 &  \vline & 0.9 & 2.2 \\
		2 & 8 &  \vline & 0.6 & 1.8 \\
		1 & 9 &  \vline & 0.3 & 1.4 \\
		0 & 10 &  \vline & - & 1 \\
		\end{array} \right)$ 
	%	\end{center}
   % \vspace{-1cm}
\end{subtable}
\par \bigskip
\end{table}

\section{Incomplete Game Models}
\label{sec:inc-game-reasoning}

As noted in Section~\ref{sec:incomplete} above, it is not always feasible to simulate all strategy profiles, even over a restricted strategy space.
Two approaches to deal with this are to generalize from the sampled profiles to fit a learned game model (Section~\ref{sec:learning}), or to reason over a model covering only part of the profile space.
Here we discuss EGTA methods developed for the latter option.

To accommodate incomplete specification, we allow that the estimated payoff function in an empirical game model % \samy{}{may} 
leave utility unassigned for some profiles.
Formally, some profiles may be mapped to a null value, $\hat{u}_i : \prod_{j\in \Players} S_j \rightarrow \mathbb{R} \cup \{\bot\}$.
A profile $s$ such that $\hat{u}_i(s) \in \mathbb{R}$ for all~$i$ is termed \term{evaluated}.
% \amy{shouldn't there be some notion of confidence here? is a profile considered ``evaluated'' if we have only seen one sample? i.e., if it has only been simulated once? i guess we need to say that we defer a statistical definition of ``evaluated'' until a later section.}
Otherwise (i.e., $\hat{u}_i(s)= \bot$ for some~$i$), we say $s$ is \term{unevaluated}.%
\footnote{This binary distinction is overly coarse, as profiles may also be evaluated to varying levels of accuracy or confidence.
The methods described in the current section make this simplification, operating as though the evaluations are exact and correct.
We present more statistical treatments of profile evaluation in Section~\ref{sec:statistical}.}
If all $s\in \prod_{j\in \Players} S_j$ are evaluated we say the empirical game $\hat{\game}= \langle \Players, (S_i), (\hat{u}_i) \rangle$ is \term{completely evaluated}.

% \amy{i am not sure this paragraph goes here. this looks like an eqm-finding algo, as opposed to an empirical game model. is it not? feels to me like material that should go closer to the Searching Subgames section.}
% \mw{I think it goes here. Do not want to postpone the demonstration that there is a lot we can do with incomplete game models. What is discussed here and in the next few paragraphs are some particular approaches and works, not a comprehensive algorithm.}
A core problem within incomplete game models is to search for a profile (pure or mixed) that can be established to have sufficiently small regret.
Several works have examined this problem from a search perspective.
The first was due to \citet{Sureka:2005uq}, who proposed an algorithm  to identify PSNE based on best-response dynamics combined with tabu search \citep{Glover89}.
In their formulation, the basic operation is evaluation of a strategy profile by simulation until it is labeled evaluated, and the search successively evaluates profiles until a PSNE is reached.
\citet{Jordan08vw} termed this problem formulation the \term{revealed payoff model},
% \amy{i think we need to emphasize t he took all evaluated profiles to be exactly evaluated. this binary defines the revealed payoff model.} 
and extended the approach to include approximate equilibria, since PSNE may not exist.
These authors also proposed an algorithm called \term{minimum-regret-first search} (MRFS)\@.
This algorithm maintains for each profile $s$ evaluated a lower bound $\bar{\regret}(s)$ on the profile's regret, defined as the maximum gain to deviating from $s$ considering the deviation profiles evaluated. 
If all deviations have been evaluated, then $\bar{\regret}(s)$ is the actual regret, so that $s$ constitutes an $\bar{\regret}(s)$-Nash equilibrium.
MRFS selects an evaluated profile with minimal $\bar{\regret}$, and then chooses an unevaluated deviation to simulate. 
This simple approach typically identifies and confirms low-regret profiles after exploring only a small fraction of the profile space. 
%\amy{add reference}
%\mw{reference is Jordan08vw already cited in this para.}

\subsection{Query Complexity}

\citet{Fearnley13} examined the problem from the algorithmic complexity perspective.
They define the \term{query complexity} of a game class with respect to some solution concept as the number of profile evaluations that are required (under the revealed-payoff model) to identify a solution, in the worst case.
For example, the solution concept may be $\regret$-NE, in which case the query complexity would generally depend on $\regret$ as well as parameters of the game class. 
The authors investigated several settings, producing interesting query complexity results for bimatrix games as well as structured game classes with compact payoff representations \citep{Fearnley15}.

For instance, one striking result for bimatrix games is that it is possible to guarantee finding an approximate NE with a linear number of queries.
Consider a two-player game where each player has strategy set $S$, and assume payoffs lie in the range $[0,1]$.
Recall $\BR_i(s)$ denotes player~$i$'s best response when player~$-i$ plays strategy~$s$.
The following algorithm identifies an $\regret$-NE for $\regret=0.5$ with $2\numStrats-1$ queries.
\begin{enumerate}
    \item Choose an arbitrary action $s^0\in S$.
    \item Compute $\BR_2(s^0)$. 
    This can be accomplished in $\numStrats$ queries; simply evaluate profiles $(s^0,s)$ for all $s\in S$ and select the $s$ with greatest payoff as $\BR_2(s^0)$.
    \item Compute $\BR_1(\BR_2(s^0))$. 
    This can be accomplished with an additional $\numStrats-1$ queries, of profiles $(s,\BR_2(s^0))$ for all $s\in S \setminus s^0$. 
    Note that profile $(s^0,\BR_2(s^0))$ was already evaluated by a query in the preceding step.
    \item Return the profile where player~1 plays $s^0$ or $\BR_1(\BR_2(s^0))$, each with probability 0.5, and player~2 plays $\BR_2(s^0)$.
\end{enumerate}
To see that the profile returned by this algorithm is a $0.5$-NE, we observe that each is playing a best-response to the other with probability 0.5. 
With the remaining probability they can be worse by at most one (the payoff range), thus 0.5 in expectation.

\citet{Fearnley15} further show how the regret bound for bimatrix games can be tightened somewhat with more queries.
% {clever querying algorithms}. 
Finding an exact equilibrium, however, requires exhaustive evaluation. 
On the other hand, games with known structure may enable alternative approaches, like generalization from queries across profiles (i.e., model learning, discussed in Section~\ref{sec:learning}).
If structure exists but is not known to the analyst, it still may be implicitly exploited by heuristic search methods, such as MRFS discussed above.
These methods may be expected to perform better than worst-case, and thus continue to provide a practical means of performing EGTA, in the usual case where exhaustive evaluation of the profile space is infeasible.
% \samy{More recent heuristic search methods, described in Section~\ref{sec:subgame}, aim to find symmetric mixed-strategy NE \citep{Brinkman16,Wellman13kd}, building on ideas from support-enumeration search for NE in normal-form games \citep{Porter08ns}.}{}

\subsection{Equilibrium Search over Subgames}
\label{sec:subgame}

% \amy{is this section using the HPT representation? if not -- if it more generic -- maybe it goes before discussing that representation? and maybe HPT fits into a section about RD, with subsections 1.~representation = HPT; 2.~replicator dynamics on HPTs; 3.~infinite-pop'n RD}
% \mw{representation is orthogonal to anything considered here. It is all fully agnostic to how equilibria are calculated for complete subgames.}

The heuristic methods (tabu search and MRFS) mentioned above focus on finding single profiles that represent approximate pure equilibria.
To verify that a mixed profile is in equilibrium (exact or approximate) from an incomplete game model, we require that the model contains full payoff information over the support of the profile, plus all one-player deviations from any pure profile in the support.
Given such information, we can employ any game-solving algorithm for finding equilibria within the support space, and then check whether any of the deviations outside the support are beneficial.
In this section, we define an analysis algorithm for incomplete game models based on searching over strategy subspaces for which full payoff information is specified.
As necessary, the search process proposes profiles to extend the evaluated space so that an equilibrium can be identified and confirmed.

Let $\game$ be symmetric with strategy set $S$,%
\footnote{The methods in this section can be generalized in a straightforward manner to handle role symmetry.}
and recall the notation $\game_{\downarrow X}$, $X\subseteq S$, for the restricted game over strategies $X$\@.
We say that $\game_{\downarrow X}$ is a \term{complete subgame} if the empirical game $\hat{\game}_{\downarrow X}$ is completely evaluated.
We typically focus attention on \term{maximal} complete subgames, where adding any strategy to $X$ would render the subgame incomplete.

The game analysis algorithm maintains a set $\solcand$ of candidate solution profiles of empirical game $\hat{\game}$.
For $\sigma$ to qualify as a candidate, we require that $\game_{\downarrow \support(\sigma)}$ be a complete subgame, and that there be some completion of $\hat{u}$ (i.e., assignment of payoff values to $\bot$ entries) that renders $\sigma$ a solution.
If $\sigma$ is a solution in \textit{all} completions of $\hat{u}$, we say that $\sigma$ is \term{confirmed}.
Otherwise $\sigma$ is \term{unconfirmed}.

For example, suppose the solution concept is symmetric mixed $\regret$-NE\@.
To assess profile $\sigma$ as a candidate solution, we would first verify that it is an $\regret$-NE of $\hat{\game}_{\downarrow \support(\sigma)}$.
We then consider deviation profiles of $\sigma$.
In order to determine whether strategy $s\in S\setminus \support(\sigma)$ is a beneficial deviation, we must evaluate all profiles of the form $(s,s')$, with $s'$ an other-agent profile over $\support(\sigma)$.
We then compare $\E [ \hat{u}(s,\sigma) ]$ with $\E [ \hat{u}(\sigma,\sigma) ]$, and if the difference exceeds $\regret$, then $\sigma$ is \term{refuted} by $s$ and no longer considered a candidate.
In other words, $\sigma$ is by definition a candidate solution iff it is an $\regret$-NE of $\hat{\game}_{\downarrow \support(\sigma)}$ and there is no $s$ that refutes~$\sigma$.
If $\sigma$ is a candidate solution with all deviation profiles evaluated, then $\sigma$ is confirmed. 

Note that if a profile \( \sigma \) has regret bounded by \( \regret \) in empirical game \( \hat{\game} \), then it will also have regret at most \( \regret \) in any complete subgame \( \hat{\game}_{\downarrow X} \) for which \( X \) contains \( \support(\sigma) \).%
\footnote{More generally, we observe that regret is monotone in strategy sets, in the following sense. 
Recall that $\regret^\game(\sigma)$ denotes the regret of profile $\sigma$ in game~$\game$.
Then $\forall \sigma.\ \support(\sigma)\subseteq X\subset X' \implies \regret^{\game_{\downarrow X}}(\sigma)\le \regret^{\game_{\downarrow X'}}(\sigma)$.
}
The game analysis algorithm therefore simply runs standard equilibrium-finding methods for each maximal complete subgame to identify potential candidates.
% \amy{i think ``standard equilibrium-finding methods'' should be mentioned much earlier! they are the backup methods, i suppose. but what exactly are they? e.g., running GAMBIT? or running RD?}
It then filters any that are refuted in the broader strategy space, and classifies the remaining as confirmed or unconfirmed: candidate sets \( \solcand_C \) and \( \solcand_U \) respectively.

\begin{figure}[ht!]
  \centering
 	\includegraphics[width=0.65\textwidth]{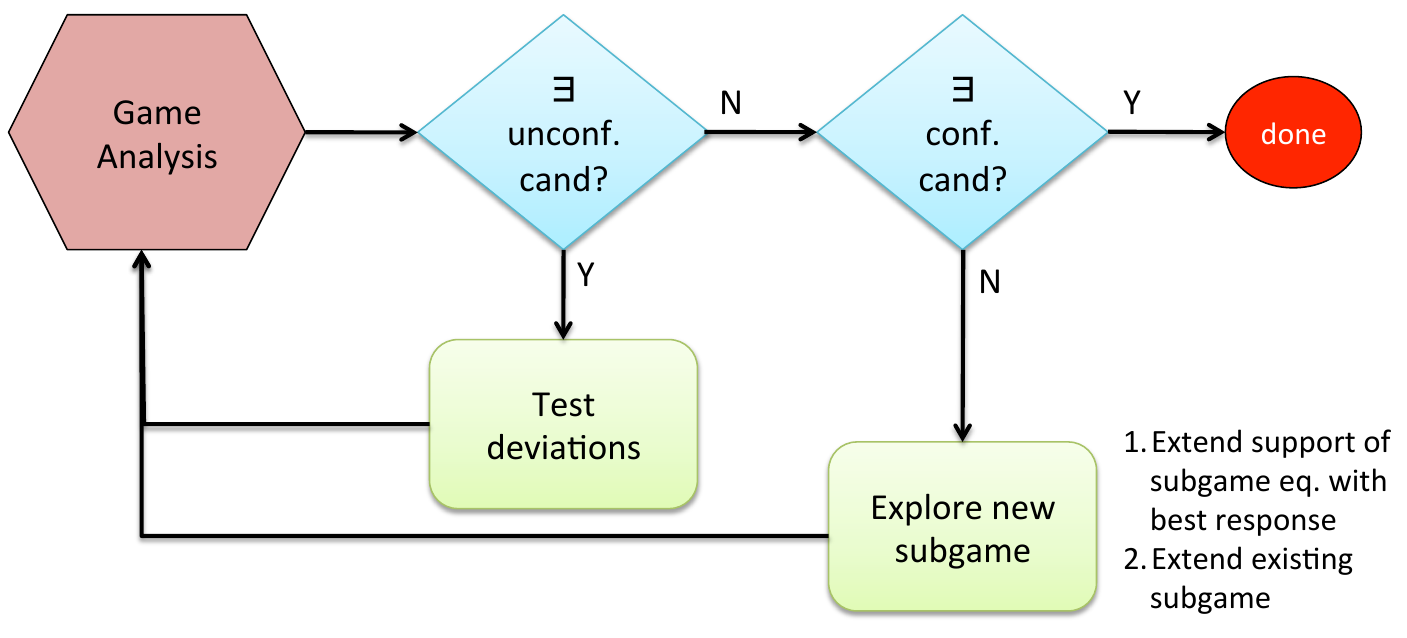}
 	\caption{Search over complete subgames to identify and confirm solution candidates.}
 	\label{fig:inner-loop}
\end{figure}

Figure~\ref{fig:inner-loop} shows how the algorithm sketched above (hexagon in the figure) can be incorporated within an iterative search for confirmed solutions.
If the game analysis reveals unconfirmed candidates, we attempt to confirm or refute them by testing deviations. 
Specifically, we evaluate by simulation the profiles $\bigcup_{\sigma\in\solcand_U} \mathit{UD}(\sigma)$, where $\mathit{UD}(\sigma)$ denotes the unevaluated deviation profiles of $\sigma$.
Once these are all evaluated, each of the unconfirmed candidates is either refuted or confirmed.

If at any point we have at least one confirmed candidate and no unconfirmed candidates, the procedure terminates and returns \( \solcand_C \).
If there are no candidate solutions, then we need to further explore the profile space. 
Evaluating additional profiles cannot affect results of the game analysis algorithm unless the additional profiles lead to completion of a new maximal subgame.
Figure~\ref{fig:inner-loop} lists two approaches for extending subgames.
The first aims to complete subgames that appear promising based on existing results.
Specifically, for any case where profile \( \sigma \) is an \( \regret \)-NE of a maximal subgame $\hat{\game}_{\downarrow S'}$, but the best evaluated response to \( \sigma \) is some strategy \( s\in S\setminus S' \), we consider the subgame defined by $\support(\sigma)\cup \{s\}$.
If all such subgames have already been evaluated yet still no confirmed equilibria have been found,%
\footnote{It may seem counterintuitive, but this can happen.
For example, consider a three-player symmetric game with three strategies \( \{A,B,C\} \) such that all profiles except \( (A,B,C) \) have been evaluated.
We could have a situation exhibiting a non-transitive response pattern reminiscent of rock-paper-scissors, such as the following: \( (A,A,A) \) is the equilibrium of subgame \{A,B\}, with best-response \( C \); \( (B,B,B) \) is the equilibrium of subgame \{B,C\}, with best-response \( A \); and \( (C,C,C) \) is the equilibrium of subgame \{A,C\}, with best-response \( B \).
All best-response-enhanced subgames have been evaluated, but to find the true equilibrium (which has full support), we need the missing profile.
} 
the second approach nondeterministically chooses to extend one of the current maximal subgames.
In the worst case, this process can lead to exhaustive exploration of the profile space, which necessarily contains an $\regret$-NE, which would be confirmed at that point.%
\footnote{To guarantee that the procedure returns a confirmed solution candidate requires that the equilibrium-finding procedure applied to subgames is itself assured to return a solution.
In particular, RD alone is not sufficient.}

% \mw{Mention Brinkman's more refined search algorithm?}

\section{Statistical Reasoning in Empirical Games}
\label{sec:statistical}

A defining feature of empirical game models is that they are estimated or induced from simulation data.
Simulating a complex game typically involves stochastic factors, embedded in the game environment or in the players' strategies (i.e., exogenous or endogenous randomization, respectively).
In our running example of sequential auctions, the main stochastic element in the environment is the initial random draw of player valuations.
Notationally, we mark empirical game models $\EmpiricalGame{}$ with a hat as a reminder that their associated utility functions $\{\hat{u}_i\}$ may differ from the \term{true game} utility functions $\{u_i\}$ due to error inherent in the process of inducing the model from a sample of data. 
Many EGTA techniques operate on the game model as if it were perfectly accurate.
For example, the revealed payoff model discussed in Section~\ref{sec:incomplete} treats each strategy profile as either completely unknown or exactly evaluated.
In this section, we discuss methods that recognize the noise inherent in actual payoff samples, and explicitly consider the statistical character of empirical game models.

The first question we address %(Section~\ref{sec:var-reduction}) 
is how to exploit statistical properties of simulator-generated data to improve the quality of payoff-function estimates. 
The next subsequent sections examine statistical questions about the game models themselves. 
That is, given a game model generated by simulation according to a sampling process, what kinds of statistical claims can we make about properties of the game and its solutions?
Further, how might we design a process that interleaves sampling and game reasoning to produce an empirical game model supporting the sharpest possible conclusions about the true game?

\subsection{Variance Reduction Methods}
\label{sec:var-reduction}

Estimating summary statistics from stochastic processes is a well-studied problem in simulation \citep{Ross02}, and techniques from that field are directly applicable to estimation tasks in EGTA\@.
In particular, methods that reduce variance in the sampling distribution for payoff estimates can substantially improve the reliability of game models from a given corpus of simulation data.

One variance-reduction technique that has been extensively applied in EGTA and other strategic analysis contexts is the method of \term{control variates} \citep{Lavenberg81,LEcuyer94,McGeoch12}.
The key idea of this technique is to exploit correlation between the random variable of interest (e.g., payoffs of a strategy profile) and other observable variables (the control variates).
In our sequential auction example, potential control variates might be some summary statistics on one's own and/or other-agent valuations. 
Payoffs for typical strategies are likely to be positively correlated with one's own valuation (representing potential for profit), and negatively correlated with others' (representing degree of competition). 
A given sample may be more or less favorable, and over a course of sampling it might be that some strategies tended to be simulated in more favorable instances than others. 
The control variates method effectively adjusts estimates for this kind of luck, based on the observed favorability of sampled instances compared to expectation.

An early example of the use of control variates in EGTA was in a study of strategic procurement in a supply chain game \citep{Wellman04esvks}.
In this game---part of the Trading Agent Competition series---agents representing computer manufacturers competed to buy parts and assemble computers for sale over the course of a simulated year.
Simulating this environment is expensive (seven CPU-hours per sample), and samples are quite noisy.
The noise is due to high variance in consumer demand, which played a significant role in payoffs for many strategies.
To apply demand level as a control variate, the simulation data was employed to estimate a linear model of payoff as a function of demand, reflecting the underlying correlation between these variables. 
This linear model was then combined with the known demand distribution to yield an improved estimate of expected payoff, decreasing the amount of simulation required by up to 50\%, compared to estimating payoffs directly.

Control variates and related variance reduction methods have also been applied in analyses of other TAC games \citep{jordan10wb,Sodomka:2007vn,Wellman07gs}. 
Similar techniques were developed to evaluate AI poker strategies \citep{Burch18}, which proved essential for deriving confident statistical comparisons among strategies that are quite close in strength.

\subsection{Statistical Characterization of Empirical Games}

Analysts may be interested in various characteristics of a game: its equilibria or other solution concepts, welfare properties such as price of anarchy, etc.
Since an empirical game model is induced from simulation data drawn from the true game, its accuracy is subject to sampling error, and results from analyzing the empirical model are related only probabilistically to properties of the true game.
Researchers have investigated this probabilistic relationship both theoretically and experimentally.
Theoretical studies have sought to derive probabilistic guarantees relating the results from empirical game analysis to properties of the true game, given characteristics of the sampling process or model accuracy.
Experimental studies have sought ways to measure statistical performance of EGTA techniques, or to estimate the reliability of model results deriving from statistical observations.

In one of the first statistical EGTA investigations, \citet{Vorobeychik10Probabilistic} showed that as we approach infinitely many i.i.d.\@ simulation queries, estimation by averaging sample payoffs converges to an empirical game reflective of the true game.
In particular, the set of equilibria of a true game and its empirical counterpart coincide.
He further noted that with only finitely many samples, spurious equilibria (i.e., false positives) can arise in the empirical game. 

To state this more precisely, let $\Nash(\game)$ denote the set of equilibria (according to some specified solution concept) of game $\game$\@. 
We say that a profile $\sigma$ is a \term{spurious equilibrium} of the empirical game $\EmpiricalGame{}$ if $\sigma\in\Nash(\EmpiricalGame{})$ but $\sigma\not\in\Nash(\game)$.
We can quantify the prevalence of spurious solutions using the language of \textit{precision} and \textit{recall}, in the sense of information retrieval \citep{salton1983introduction}:
if all equilibria of the true game are ``recalled'' as equilibria in the empirical game---that is, if $\Nash(\game)\subseteq \Nash(\EmpiricalGame{})$---this constitutes perfect recall.
More generally, the degree of recall is measured by the fraction 
\[
\frac{\abs{\Nash(\game)\cap \Nash(\EmpiricalGame{})}}{\abs{\Nash(\game)}}.
\]
Conversely, if $\Nash(\EmpiricalGame{})\subseteq \Nash(\game)$ (i.e., there are no spurious equilibria in the empirical game), precision is perfect, and the degree of precision can be measured as above, but with $\abs{\Nash(\EmpiricalGame{})}$ as the denominator.

To illustrate this phenomenon, we conducted a simple experiment using our running sequential-auctions example.
The game has three players and four goods, and three available bidding strategies: $\rho \in \{ 0, 0.5, 1 \}$.
Valuations are based on the homogeneous-good model of \citet{Wellman17sg}, with maximum values drawn from player-specific distributions.
We generated 100 random games, and for each game evaluated the corresponding empirical game after various numbers of samples (100, 200, 500, and 1000) per profile.

\begin{figure}[ht]
\centering
\includegraphics[width=\textwidth]{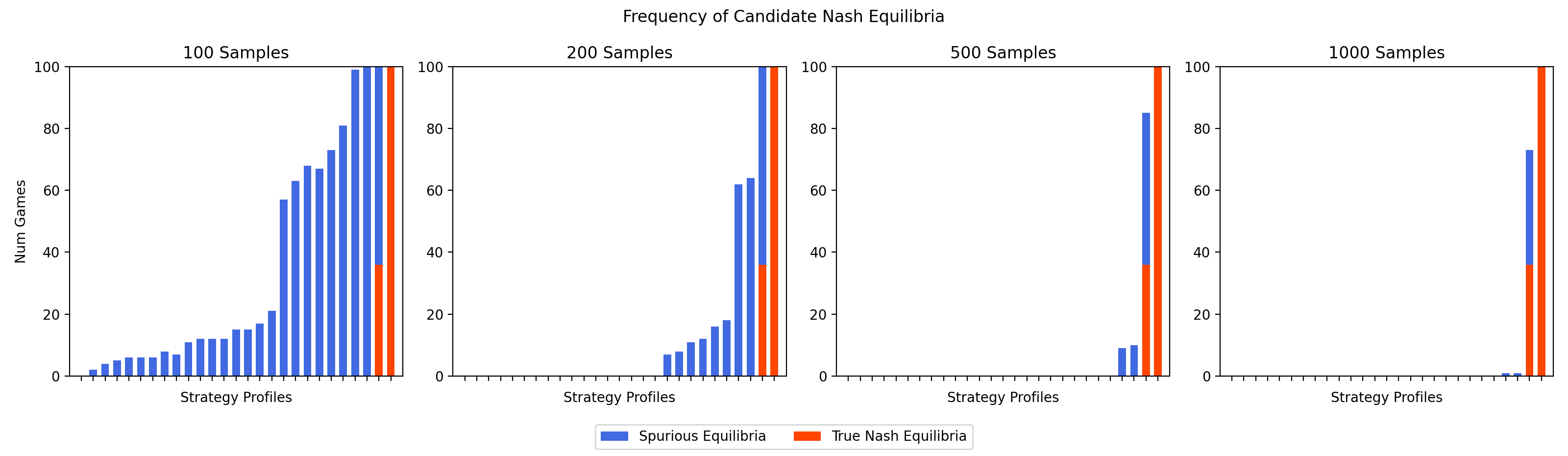}
%{expts/gs_frequency_bidding_game_with_green.png}
\caption{Spurious and true equilibria as a function of the number of samples.
Bar height represents frequency of identification as empirical $\regret$-PSNE in 100 random games, with blue indicating spurious and red, true equilibria.
Strategy profiles whose bars are both blue and red were spurious in some games, and true in others, in the proportions shown.
In each plot, the order of the 27 strategy profiles on the $x$-axes is fixed, from highest to lowest regret, computed using 10,000 samples.
``True'' equilibria were likewise found using 10,000 samples.}
\label{fig:nash_freq}
\end{figure}

The results are shown in Figure~\ref{fig:nash_freq}.
For each empirical game, we tested whether each of the 27 pure profiles was an approximate PSNE, with approximation threshold $\regret$ set based on an empirical version of Bennett's inequality \citep{cousins2022computational} for the given number of samples.
We then plot, for each sample level and profile, in how many of the 100 games that profile was deemed an $\regret$-PSNE in the empirical game after that many samples.
As we can see, at 100 samples, there are many spurious equilibria, which are steadily eliminated as the number of samples increases.

\subsection{Sampling and Approximation Bounds}

A \term{uniform approximation} of a game is one in which all utilities are estimated to within the same error, simultaneously.
At well-defined regret levels, uniform approximations of games support perfect recall and approximate precision: perfect recall, because the set of approximate equilibria of $\hat{\game}$ contains all true positives---that is, all $\game$'s equilibria; and approximate precision, because all false positives (i.e., spurious equilibria) are approximate equilibria in $\game$.
Stated more precisely, $\hat{\game}$ with utility $\hat{\Utility}$ is said to be an $\epsilon$-\term{uniform approximation} of $\GameTuple$ with utility $\Utility$ iff $\norm{\Utility - \hat{\Utility}}_{\!\infty} = \max_{\PlayerIndex \in \SetOfPlayers, \StratProfile \in \StratProfileSpace} \abs{ \Utility_{\PlayerIndex} (\StratProfile) - \hat{\Utility}_{\PlayerIndex} (\StratProfile) } \leq \epsilon$.
%When one game $\GameTuple'$ is a uniform approximation of another $\GameTuple''$, all equilibria in $\GameTuple'$ are \emph{approximate\/} equilibria in $\GameTuple''$.
% When an empirical game's payoffs $\hat{\Utility}$ uniformly approximate the true payoffs $\Utility$, it holds that all equilibria in $\GameTuple$, the game defined by $\Utility$, are approximate equilibria in $\hat{\GameTuple}$, $\GameTuple$'s counterpart defined by $\hat{\Utility}$ (perfect recall); conversely, all equilibria in $\GameTuple$ are approximate equilibria in $\hat{\GameTuple}$ (approximately perfect precision).
%
Then, letting $\Nash_\epsilon (\GameTuple)$ denote the set of $\epsilon$-Nash equilibria, if $\hat{\game}$ is an $\epsilon$-uniform approximation of $\GameTuple$, then $\Nash (\GameTuple) \equiv \Nash_0 (\GameTuple) \subseteq \Nash_{2\epsilon} (\hat{\game}) \subseteq \Nash_{4\epsilon} (\GameTuple)$ \citep{TuylsPLHELSG20,areyan2020improved}.

It follows that any algorithm that learns an $\epsilon$-uniform approximation of a game
%\footnote{A uniform approximation of a simulation-based game is one in which all utilities in the empirical game tend toward their expected counterparts.}
also learns the equilibria of that game, up to an accuracy that depends on $\epsilon$.
A standard approach to learning uniform approximations uses concentration inequalities to first establish high-probability confidence intervals around each individual parameter (in our case, a player's utility at a strategy profile), and to then apply 
a statistical correction (e.g., Bonferroni
or \v{S}id\'{a}k (\citeyear{Sidak67})) to bound the probability that all approximation guarantees hold simultaneously.
%, often incurring looseness in one or both steps.

\citet{TuylsPLHELSG20} were the first to try this approach, using a \term{global sampling} (\GS) algorithm where the empirical game model is estimated from batches of $\numSamp$ samples per profile.
%that learns an empirical game which uniformly approximates a simulation-based game with finite-sample guarantees.
%Taking an interactive perspective instead,
To derive their guarantee, the authors used Hoeffding's inequality, a sub-Gaussian tail bound for averages of $\numSamp$ $\nicefrac{\UtilityRange}{2}$-bounded random variables that yields a supremum deviation bound $\norm{\Utility - \hat\Utility}_{\!\infty} \le \epsilon \in \LandauTheta \left( c \sqrt{\frac{\ln(\nicefrac{2}{\delta})}{2\numSamp}} \right)$ with probability at least
$1-\delta$.
%confidence intervals of width $\LandauTheta \left( c \sqrt{\frac{\ln(\nicefrac{2}{\delta})}{2\numSamp}} \right)$.
For a constant failure probability $\delta$, rather than compute the accuracy $\epsilon$ given sample size~$\numSamp$, it is often useful to calculate the 
sample size 
$\numSamp$ required to achieve 
%accuracy
a target error $\epsilon$.
Applying this logic, and a Bonferroni correction, \GS\ requires $\numSamp \ge \frac{\UtilityRange^{2} \ln \left( \nicefrac{2 \abs{\GameTuple}}{\delta} \right)}{2\epsilon^{2}}$ samples per profile to guarantee an $\epsilon$-uniform approximation of the game with probability at least $1-\delta$, where $\abs{\GameTuple}$ is the number of game parameters (e.g., utility values). 
% \amy{N.B.\ $\abs{\GameTuple}$ can be much smaller than the total number of utilities, b/c of symmetries. maybe use a function: size of game.} \amy{introduce counting notation in incomplete models section; discuss savings there as well.}

As Hoeffding's inequality is sensitive only to the range $\UtilityRange$ of the random variables, not their variance $\sigma^2$, \citet{areyan2020improved} analyze \GS\ with an alternative concentration inequality.
Bennett's inequality [\citeyear{bennett1962probability}] relaxes the dependency on $\UtilityRange$, replacing it with a dependency on $\sigma^2 < \UtilityRange^2$, yielding
%$1 - \delta$ probability confidence intervals of width 
$\epsilon \in \LandauTheta \left( \frac{c \ln (\nicefrac{1}{\delta})}{\numSamp} + \sqrt{\frac{\sigma^2 \ln (\nicefrac{1}{\delta})}{\numSamp}} \right)$, 
%which can be interpreted as a subgamma tail bound, 
which in this case
%as above, by a union bound
implies that 
$\numSamp \ge 2 \ln \frac{2 \abs{\GameTuple}}{\delta} \left( \frac{\UtilityRange}{3\varepsilon} + \frac{\norm{\UtilityVariance}_{\infty}}{\varepsilon^{2}} \right)$ samples per profile are required to guarantee an $\epsilon$-uniform approximation of the game with probability at least $1-\delta$.
This bound depends on the so-called \emph{wimpy variance}, the maximum variance across all game parameters, denoted by $\norm{\UtilityVariance}_{\infty}$ \citep{boucheron2013concentration}.
Using an upper bound on the wimpy variance in terms of its empirical counterpart \citep{cousins2020sharp}, one can derive a supremum deviation bound that does not depend on any \emph{a priori\/} variance knowledge.
%~\citep{audibert2007tuning,audibert2007variance}.

\begin{figure}[t!]
\centering
\includegraphics[width=1.0\textwidth]{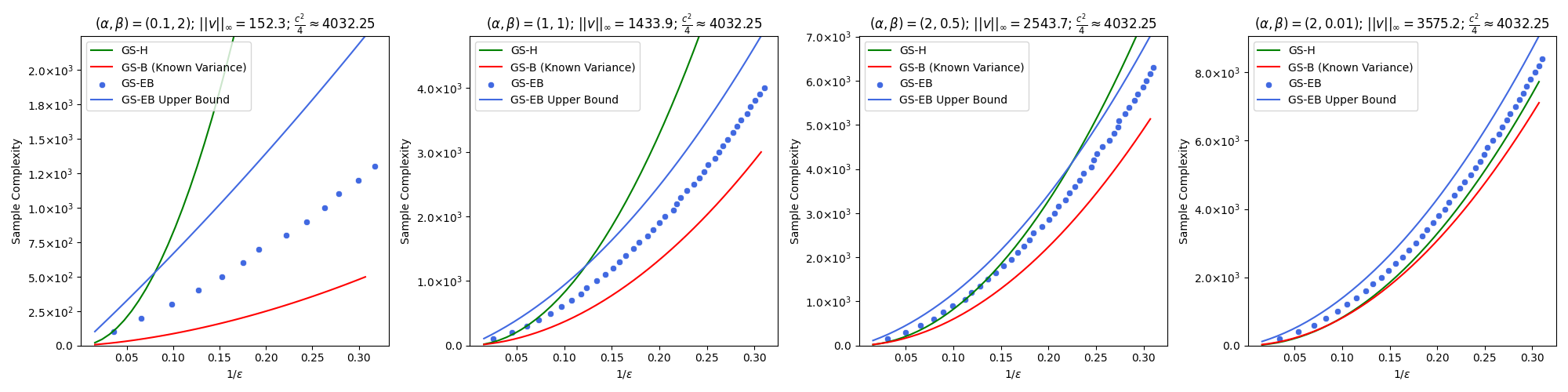}
\caption{Sample complexity $\numSamp$ of GS as a function of $\nicefrac{1}{\epsilon}$, calculated using four bounding methods.
Variance increases across the instances plotted moving left to right.
Algorithms based on Bennett's bound are most effective in the leftmost plot, and least in the rightmost. 
The algorithm based on Hoeffding's bound is independent of variance; the GS-H curve appears to shift due to compression of the y-axis scale from left to right.}
\label{fig:gs_across_bounds}
\end{figure}

Figure~\ref{fig:gs_across_bounds} depicts the sample complexity $m$ of \GS\ as a function of $\nicefrac{1}{\epsilon}$, for $\delta = 0.05$, calculated according to four methods.
The methods employ Hoeffding's bound (GS-H), as well as three variants of Bennett's bound:
Bennett's original bound assuming known wimpy variance (GS-B);
\citeauthor{areyan2020improved}{}'s empirical Bennett bound based on the actual empirical wimpy variance realized in the experiments (GS-EB);
% \footnote{In fact, these experiments use \cite{cousins2022computational}'s slightly improved version of the original bound.} 
and a third upper bound derived by \citet{cousins2022computational} (GS-EB Upper Bound).
% \footnote{This result depends on a ``reverse'' tail bound, which bounds empirical variance in terms of its true counterpart.}
The first and third of the aforementioned curves are smooth, because they are obtained by plugging a ``known'' variance (determined by 30,000 samples) and a target error $\epsilon$ into a formula that produces $\numSamp$, a sample complexity.
In contrast, in the GS-EB experiments, the independent variable is the number of samples, while the dependent variable is $\epsilon$, as per \citeauthor{areyan2020improved}{}'s empirical Bennett bound.

These sample complexities were calculated in four bidding games, each with three bidders, one good, and six shading factors ($\rho \in \{ 0, 0.2, 0.4, 0.6, 0.8, 1 \}$).
The games varied in that the bidders' values were drawn from four different beta distributions, which yield four different ``known'' wimpy variances, as indicated in the figures.
%The plots towards the left exhibit lower wimpy variance than those towards the right.
%We observe, as expected, that in game with low wimpy variance, GS-B and GS-EB drastically outperform GS-H, whereas in games with high wimpy variance, GS-H can perform at best a small constant factor better than GS-EB.
When wimpy variance is low (left), Bennett's bound significantly outperforms Hoeffding's;
when wimpy variance is high (right), Bennett's bound still performs within a small constant factor of Hoeffding's. 
%%%\amy{multiplicative or additive?}
% SANITY CHECK
%In all plots, the sample complexity of GS-EB falls between the known case and its upper bound.

\subsection{Measuring Uncertainty in Empirical Game Analysis}

Once an empirical game model is constructed and analyzed, we often wish to quantify the reliability of results, accounting for sampling error.
Even if the model was estimated based on \GS\ or other approaches that provide \textit{a priori} approximation bounds, an \textit{a posteriori} analysis may yield further precision about the uncertainty attached to game-theoretic conclusions.

\begin{figure}
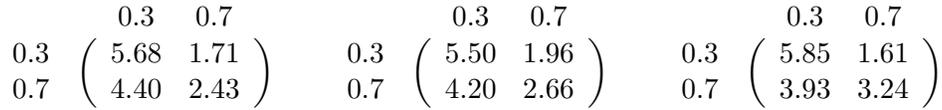

    \centering
  \gamematrix{0.3}{0.7}{5.68}{1.71}{4.40}{2.43}
  \gamematrix{0.3}{0.7}{5.50}{1.96}{4.20}{2.66}
  \gamematrix{0.3}{0.7}{5.85}{1.61}{3.93}{3.24}
    \caption{Statistical analysis of an empirical game for a simple sequential auction example.
    The game is symmetric, and payoffs are shown for the row player.
    Left: true game~$\game$; Middle: empirical game~$\hat{\game}$ ($\numSamp=20$); Right: resampled version~$\hat{\game}'$ of empirical game.}
    \label{fig:bootstrap-ex}
\end{figure}

For example, let us define a simple version of our sequential auction game, with $\numPlayers=\numGoods=2$, two available strategies ($\rho\in\{0.3,0.7\}$), and three possible valuations drawn uniformly and independently for each player.
Thanks to its simplicity, we can work out the exact normal-form game~$\game$, shown in Figure~\ref{fig:bootstrap-ex}(left).
The middle payoff matrix in Figure~\ref{fig:bootstrap-ex} presents an empirical game~$\hat{\game}$ for these settings, estimated from $\numSamp=20$ samples per profile.

This game has $c=28$ and a known wimpy variance of 13.7, so by the GS-B bounds of the previous section an empirical game generated with $\numSamp=20$ samples/profile has at least 0.95 probability of being an $\regret$-approximation with $\regret \approx 5.63$.
Such a bound is not very helpful given the payoff scale of the empirical game generated, and as we can see, it is actually a much better approximation than that.%
\footnote{Of course, in a real application we would not know the true game. 
In this example, where the number of samples $\numSamp$ is tiny, so we can expect weak bounds.
But the bounds are nonetheless quite conservative; so we may have been lucky, but not extraordinarily so.}
So after we build an empirical game, what can we say about the accuracy of its conclusions?

First, let's solve~$\hat{\game}$. 
It has a unique symmetric NE~$\sigma^*_{\hat{\game}}$, which plays $\rho=0.3$ with probability 0.35.
As discussed above, given sampling error, we do not expect empirical-game solutions to be exact solutions of the true game.
The question of interest, rather, is how likely is it to be an approximate equilibrium, at various levels of approximation?
In other words, what is the probability distribution over $\regret^\game (\sigma^*_{\hat{\game}})$, viewed as a random variable conditional on the samples that produced~$\hat{\game}$?

\citet{Wiedenbeck14} proposed a bootstrap approach for addressing such statistical questions about empirical games.
The basic idea of bootstrapping is to use the sampling data itself as the basis for modeling uncertainty in data generation.
This technique has been applied to a broad variety of statistical questions \citep{Davison97}, and is amenable to straightforward implementation \citep{Shasha11}.
To bootstrap the sampling distribution for regret, we \textit{resample} with replacement from the payoff data to construct new game models, a process that we assume captures the distribution of empirical games that would arise from random sampling.

\begin{figure}
    \centering
    \includegraphics[width=0.5\textwidth]{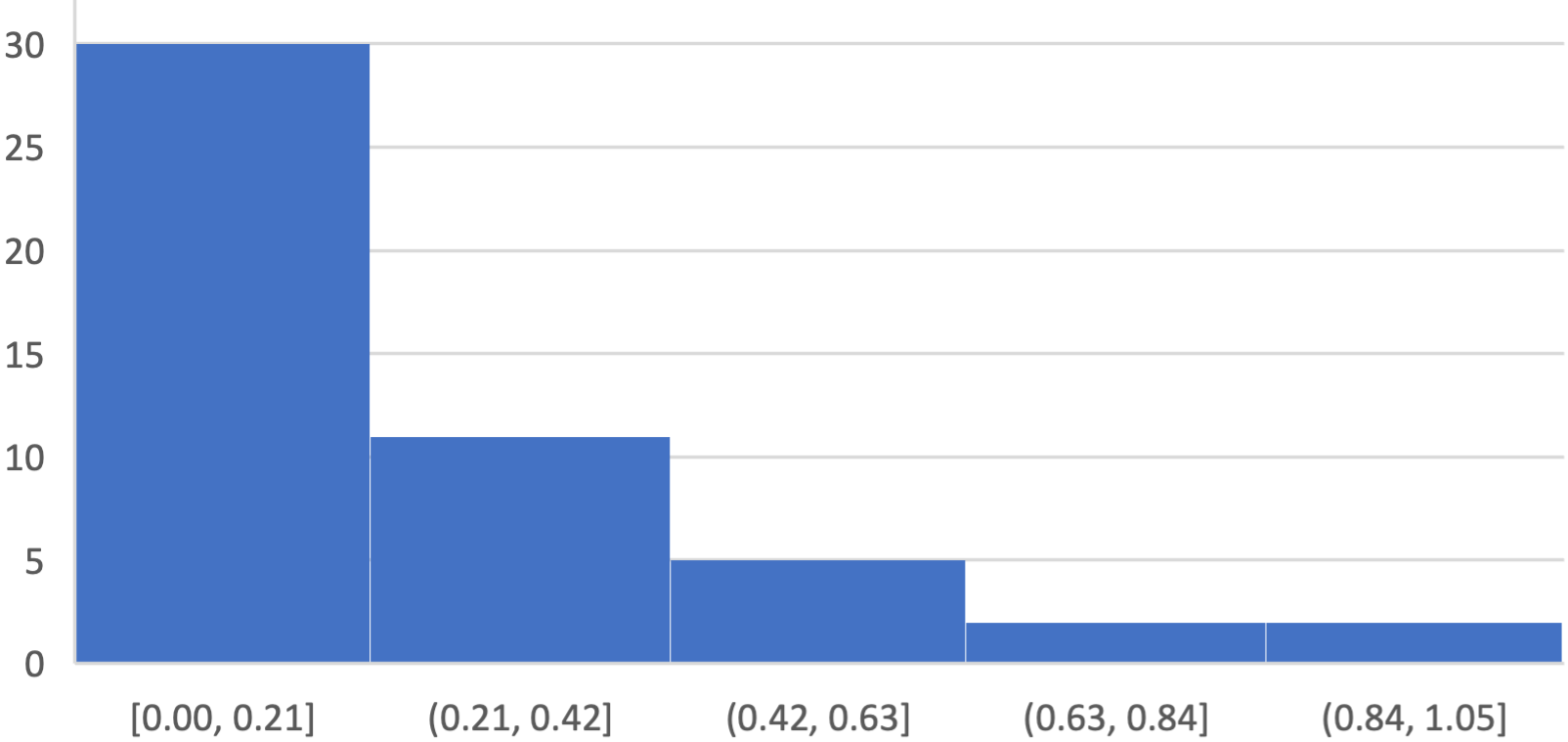}
    \caption{Bootstrapped regret distribution for the empirical game~$\hat{\game}$ of Figure~\ref{fig:bootstrap-ex}(middle).}
    \label{fig:bootstrap-regret}
\end{figure}

In our example, empirical game~$\hat{\game}$ was estimated from $\numSamp=20$ samples for each cell of the game matrix.
For the resampled version~$\hat{\game}'$ shown in Figure~\ref{fig:bootstrap-ex} (Right), each cell is the average of 20 payoffs resampled with replacement from the original data set. 
Next, we calculate the regret of the profile of interest with respect to the resampled game model.
In this instance, $\regret^{\hat{\game}'}(\sigma^*_{\hat{\game}})=0.14$.
This value represents one data point in the sampling distribution for regret.
We resampled 49 more times to build the histogram shown in Figure~\ref{fig:bootstrap-regret}.
According to this histogram, the expected (mean) regret is 0.24, the median is 0.16, and the 95\% confidence level is 0.72.
In this example, the true-game regret turns out to be at the optimistic extreme of the bootstrap distribution, $\regret^{\game}(\sigma^*_{\hat{\game}})=0.01$.
Nonetheless, \citet{Wiedenbeck14} found that bootstrapping regret from the empirical game was often well-calibrated with regret in the true game.
% \samy{In our instance, true-game regret $\regret_{\game}(\sigma^*_{\hat{\game}})=0.01$, so toward the optimistic extreme of the bootstrap distribution.}{}

The bootstrapping approach described above attempts to quantify uncertainty using the data already generated for estimating the empirical game.
It may also be valuable to perform additional sampling expressly for the purpose of assessing reliability in game-theoretic solutions. 
\citet{Jecmen20} propose a bandit-based algorithm for sampling profiles and deviations in order to bound regret estimates for profiles in an empirical game model.
\citet{rowland2019multiagent} design sampling strategies to achieve confidence guarantees in orderings produced by \alpharank.

\subsection{Dynamic Sampling Algorithms}
\label{sec:dynamic-sample}

Measurements of statistical confidence in EGTA can be useful not just for assessing conclusions, but also to guide the collection of samples during the EGTA process. 
The idea of statistical sample control for EGTA was first pursued by \citet{Walsh03}.
In that early study, where the goal was to learn Nash equilibria, the authors proposed selecting strategy profiles to sample based on \textit{expected confirmational value of information}, a prediction of the degree by which sampling would decrease estimated error in the current solution candidate.
\citet{Jordan08vw} framed the statistical sample control task within EGTA based on a \term{noisy payoff model}, by contrast with the revealed payoff model discussed in Section~\ref{sec:incomplete}.
These authors proposed an algorithm based on maximizing information gain, defined relative to a mapping from empirical games to beliefs about strategies played \citep{Vorobeychik06w}.
% They found this approach outperformed prior sample control methods, including \GS\ heuristics based on the revealed-payoff model.
Using the method for bootstrapping regret distributions described above, \citet{Wiedenbeck14} proposed using confidence intervals on these distributions for sample control.
In particular they introduced heuristics based on bootstrap estimates that guide how to allocate and when to stop sampling, and found that they outperformed more common rules of thumb.

Several works frame EGTA as a \term{black-box optimization} problem, where the objective function is given not analytically, but in the form of a simulator \citep{Audet17}.
More specifically, the \term{Bayesian optimization} approach to black-box optimization \citep{garnett2023bayesian} employs  an \term{acquisition function} to guide sampling from the simulator.
To apply Bayesian optimization to game-solving, a natural acquisition function would be based on regret~\eqref{eq:regret},
% $\min_{\sigma \in \Delta(S)} \max_{i} \max_{\tau_i \in \Delta(S_i)} u_i (\tau_i, \sigma_{-i}) - (\sigma_i, \sigma_{-i})$,
though as the regret function $\regret$ is not available analytically, a surrogate must be used.
\citet{AlDujaili18} use Gaussian process (GP) regression to build a probabilistic game model, from which they estimate $\regret_i(s)$ as the maximal difference across all strategies $s'_i$ between $i$'s mean utility $\bar{u}_i(s'_i, s_{-i})$ plus one standard deviation, and the 
%%% DOUBLE CHECK: mean
mean utility $\bar{u}_i(s)$ of $s$.
They then propose the minimum of the maximum of this regret across all players as the next sample with probability $1 - \eta$, for some $0 < \eta \ll 1$, choosing at random according to the uncertainty of the GP, otherwise.
\citet{Tay23} take a similar GP approach, also basing their acquisition function on maximum regret, but in a worst-case sense using confidence bounds: the difference between the lower limit of the confidence bound for a given strategy (a pessimistic outcome) and the upper limit across all alternatives (an optimistic one).
\citet{Picheny19} likewise employ GP, and propose two acquisition functions. 
Their first is designed to maximize the probability that each player's strategy is part of an equilibrium (i.e., yields no regret), while their second is designed to reduce uncertainty in this measure.

%\amy{hoping for a better segue. maybe can figure out how to compare sample complexities.}

Dynamic sampling methods like these can improve over global sampling (\GS) by allocating
%differential 
sampling effort to profiles based on relevance as determined through intermediate analysis.
% \amy{Naive?} Global Sampling simulates \emph{all\/} strategy profiles equally often, in the latter case producing a PAC-style guarantee on the quality of the equilibria.
% A natural alternative would be to somehow distribute those expensive simulation queries more wisely.
For example, if the goal is to find equilibria, then we can avoid sampling strategy profiles which we deem sufficiently unlikely to affect equilibrium determination.
More generally, if it can be established that a strategy profile's variance is sufficiently lower than another's, it may not be necessary to query the first as often as the second.
These observations, respectively, form the basis of \term{progressive sampling with pruning} (PSP) algorithms \citep{areyan2020improved,cousins2022computational}.

Using GS as a subroutine, a progressive sampling algorithm builds an empirical game iteratively, based on progressively larger samples, and hence more refined utility estimates, until a desired accuracy is achieved.
%, or the maximum number of iterations is reached.
\PSP{} builds on this idea by pruning game parameters between iterations (i.e., ceasing to estimate them) if a certain pruning criterion is met.
%the property of interest is
%all utilities are
%approximated to the desired accuracy, or when the sampling budget is exhausted.
\citeauthor{cousins2022computational}{}'s algorithm prunes low-variance parameters before high-variance parameters, as soon as it establishes that they have been sufficiently well estimated (PSP-WE).
With the goal of learning equilibria, \citeauthor{areyan2020improved}{}'s algorithm prunes the utilities of strategy profiles once it establishes that they are not likely an approximate equilibrium (PSP-REG).
% \amy{commented out the more precise explanation based on $\epsilon$ b/c i have not specified what $\epsilon$ is here, and it is complicated to do so. can i add it back w/o explaining $\epsilon$? we had an $\epsilon$ in the last section, which gives a hint, and the actual def'n is coming, in the next paragraph. the difficulty with saying what it is here is that it was different in A-V's 2020 AAMAS paper than it is in these exp'ts. in the former case it was an output (but the algo could fail); now it is an input (and the algo can't fail).}
%\amy{sufficiently high regret, and not needed to refute other strategy profiles}

%\amy{should we describe these pruning algos as stopping algos?}

%\PSP{} can potentially learn game properties using vastly fewer resources than \GS.

\begin{figure}[htbp]
    \centering
    \includegraphics[width=\textwidth]{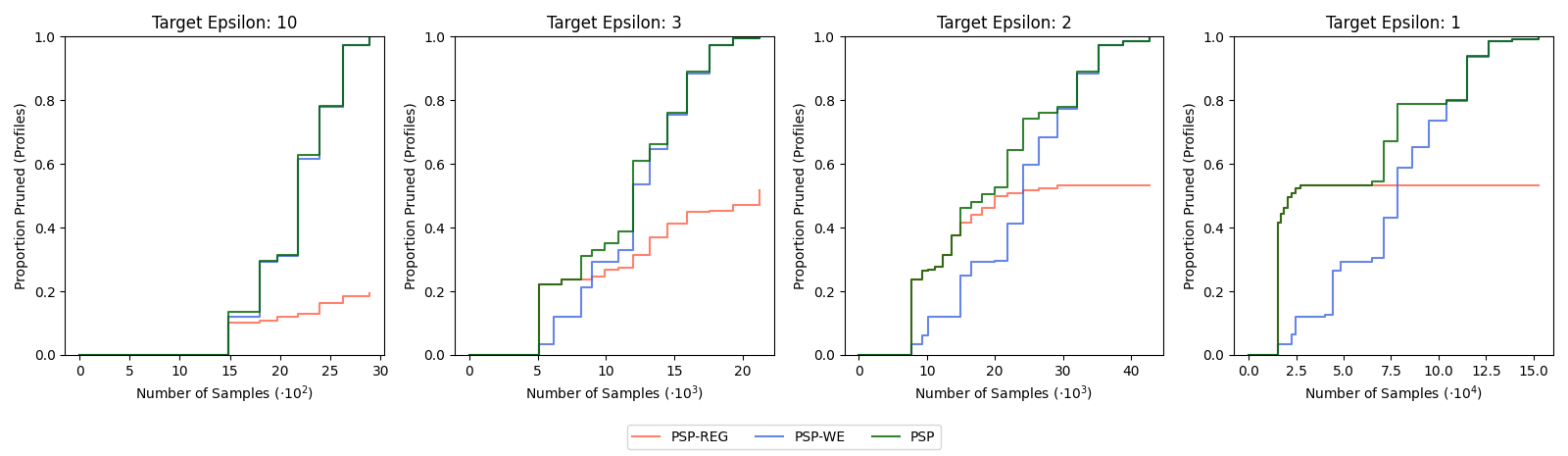}
    \caption{The proportion of strategy profiles pruned by each algorithm for each target $\epsilon$.
    The red curves correspond to PSP-REG; the blue curves, to PSP-WE; and the green curves, to PSP, which prunes using both criteria.
    The white space between the red and blue curves increases as the target error decreases, corresponding to the increase in significance of regret vs.\ well-estimated pruning.}
    \label{fig:PSP}
\end{figure}

Figure~\ref{fig:PSP} depicts the proportion of strategy profiles pruned, hence the work saved, for three PSP variants on an instance of the sequential auctions game.
Our setting employed homogeneous valuations  \citep{Wellman17sg} with three bidders, four goods, and six shading factors ($\rho \in \{ 0, 0.2, 0.4, 0.6, 0.8, 1 \}$).
%maximum values drawn uniformly in the range $[0, 127]$.
%We set αi = βi = 1 for each buyer type i.
We used \citeauthor{cousins2022computational}'s PSP-WE sampling schedule, with 
%$\delta = 0.05$ and 
$\beta = 1.1$, which is tailored to achieve a desired accuracy, for which we set four targets, namely $\epsilon \in \{ 10, 3, 2, 1 \}$.
We report the proportion of strategy profiles pruned by each algorithm for each target, averaged across five sample games.

We observe that as desired accuracy increases (i.e., as target $\epsilon$ decreases),
%(i.e., as stronger approximations are desired)
the PSP-REG curves eventually stabilize, whereas the PSP-WE curves maintain their basic shape, stretched across the number of samples.
These observations are in line with intuitions, as regret pruning is independent of the desired accuracy (it depends only on the number of samples), 
%it depends only on $\hat{\epsilon}$ (i.e., is a function of the number of samples), not $\epsilon$
whereas well-estimated pruning depends on the target.
Indeed, \citeauthor{cousins2022computational}{} provide a lower bound on number of samples required to prune a strategy profile, which is inversely proportional to the target. 
As a result, PSP-REG's contribution to PSP relative to PSP-WE's increases in significance
%the space b/n the blue and red curves
as the desired accuracy increases.

\subsection{Game Properties beyond Equilibrium}

% \amy{stopped here!} 

Whereas the property of greatest interest for EGTA has typically been identification of solutions (e.g., Nash equilibria), statistical reasoning is relevant for other game properties as well, such as social welfare of solution profiles.
A property $f$ is called $\lambda$-Lipschitz if $\norm{f(\cdot; \Utility) - f(\cdot; \Utility')}_{\infty} 
\doteq \sup_{x \in \mathcal{X}} \abs{f(x; \Utility) - f(x; \Utility')} \leq \lambda\norm{\Utility - \Utility'}_\infty$.%
\footnote{Here, $\mathcal{X}$ is usually the set of strategy profiles, but it need not be.}
% \lambda bounds how the property's error can expand from one game to another
For example, common variants of social welfare---utilitarian, egalitarian, and Gini---are all $\lambda$-Lipschitz with $\lambda = 1$ \citep{beliakov2009some,cousins2023learning}.
Regret
%, defined as $\max_{\PlayerIndex \in \SetOfPlayers} \sup_{\StratProfile' \in \Adjacent_{\PlayerIndex, \StratProfile}} \Utility_{\PlayerIndex} (\StratProfile') - \Utility_{\PlayerIndex} (\StratProfile)$, where $\Adjacent_{\PlayerIndex, \StratProfile} \doteq \{ \StratProfileAlt \in \StratProfileSpace \mid \StratProfileAlt_q = \StratProfile_q, \forall q \neq \PlayerIndex \}$, which in turn defines Nash equilibrium, 
is $\lambda$-Lipschitz with $\lambda = 2$~\citep{areyan2020improved}.

Observe the following: if $f$ is $\lambda$-Lipschitz and $\norm{\Utility - \Utility'}_{\infty} \leq \epsilon$, then
%it is immediate that
$\norm{f(\cdot; \Utility) - f(\cdot; \Utility')}_{\infty} \leq \lambda \epsilon$.
Equivalently, if $f$ is $\lambda$-Lipschitz and $\norm{\Utility - \Utility'}_{\infty} \leq \nicefrac{\epsilon}{\lambda}$, then $\norm{f(\cdot; \Utility) - f(\cdot; \Utility')}_{\infty} \leq \lambda (\nicefrac{\epsilon}{\lambda}) = \epsilon$.
Thus, as regret is 2-Lipschitz, it can be $\epsilon$-approximated from an $\nicefrac{\epsilon}{2}$-uniform approximation of a game.
Likewise, utilitarian, egalitarian, and Gini social welfare can all be $\epsilon$-approximated directly from an $\epsilon$-uniform approximation~\citep{cousins2023learning}.

%\amy{could say somewhere that xxx use this observation to prove that Nash eqm is PAC-learnable. indeed, all Lipschitz properties are likewise PAC-learnable.}

Based on this observation, \citet{cousins2023learning} derive two-sided approximation bounds on the \emph{extrema\/} of $\lambda$-Lipschitz game properties (e.g., maximum utilitarian welfare).
Specifically, they derive a two-sided bound on their values,
and a dual containment (recall and approximate precision) result characterizing their witnesses.
This theorem implies that global sampling can be used to learn any $\lambda$-Lipschitz properties of games beyond regret/Nash equilibrium.
Pruning algorithms based on game properties other than regret have not yet been fully explored in EGTA, although welfare-based pruning has been analyzed to learn competitive equilibria \citep{areyan2021learning}.

%More generally, any algorithm that learns an empirical game that is an $\nicefrac{\epsilon}{\lambda}$-uniform approximation of an underlying game can likewise learn any $\lambda$-Lipschitz property of that game.

In summary, statistical tools offer EGTA practitioners guidance in tackling questions about the sample complexity 
%(i.e., number of simulation queries)
required to achieve a desired accuracy when estimating a game's properties, and how to distribute the requisite number of queries across the game's various strategy profiles.

\section{Strategy Exploration}
\label{sec:explore}

The EGTA methods discussed in this survey thus far take the empirical game strategy sets, $X_i$, as given and fixed.
Techniques for dealing with incomplete game models, statistical (Section~\ref{sec:statistical}) or otherwise (Section~\ref{sec:incomplete}), allow for partial evaluation over the space of profiles induced by these strategy sets, iteratively extended through simulation queries.
The subgame search methods (Section~\ref{sec:subgame}) also iteratively extend the game model, reasoning across different strategy-set restrictions.
Nevertheless, all methods discussed to this point treat the base set of strategies defining the overall profile space as constant.

\subsection{Automated Strategy Generation}

\begin{figure}[ht!]
  \centering
 	\includegraphics[width=0.6\textwidth]{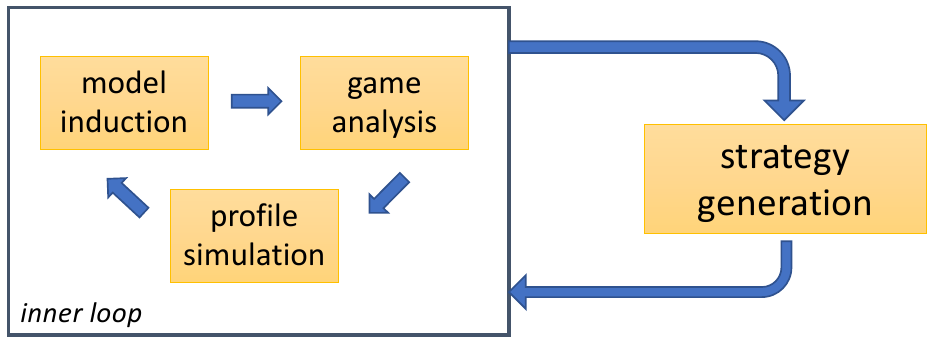}
 	\caption{EGTA with automated strategy generation.
 	The inner loop constructs and analyzes empirical game models over restricted strategy sets $X_i$.
 	The outer loop searches over $S_i$ to generate one or more new strategies to include, based on this analysis.
 	These new strategies are then added to the $X_i$, and the process iterates.}
 	\label{fig:outer-loop}
\end{figure}

Whereas manual specification of heuristic strategies often provides a good starting point for game-theoretic analysis, restricting attention to these sets a priori fundamentally limits the scope of the analysis.
One way to address this limitation is through \term{automated strategy generation}: a process of search through the unrestricted sets $S_i$ of possible strategies for new strategies to add to the restricted sets $X_i$ delimiting the empirical game model.
An iterative process for EGTA with automated strategy generation is illustrated in Figure~\ref{fig:outer-loop}.
The box on the left, termed \textit{inner loop} by \citet{Wellman13kd} (essentially corresponding to the blue-arrow cycle of Figure~\ref{fig:egta-diagram}), iteratively performs simulation, game model induction (estimation or learning), and game analysis, within a fixed and restricted space of strategies.
For example, the inner loop might implement the process of Figure~\ref{fig:inner-loop}, and might include the sample-control methods of Section~\ref{sec:statistical}.
An \textit{outer loop} (incorporating the green-arrow path of Figure~\ref{fig:egta-diagram}) employs results from analyzing the restricted empirical game $\hat{\game}_{\downarrow X}$ to generate new strategies from the base game for inclusion.

Ultimately, this iterative analysis is still limited by the restriction of the empirical game model to the set of strategies generated.
However, by invoking unrestricted search as part of the process, the analyst consults the full sets $S_i$ repeatedly.
This tends to strengthen the analysis, as the intermediate EGTA results can provide an informed basis for selecting strategies to evaluate explicitly.
% \amy{i feel like this is an important point that deserves more discussion. perhaps it comes later?}
We would generally expect
% {Assuming we employ reasonable heuristic generation strategies, we would hope to achieve} 
better coverage from strategy sets generated in an informed manner incrementally over multiple (up to $\abs{X}$) iterations,
% \amy{why $\abs{X}$ iterations? why not some arbitrary number $n$? not sure why we want to tie the number of iterations back to the size of $X$?} 
% \mw{The maximally informed case is when we add strategies one per iteration, interleaved with analysis}
compared to an a priori identification of $\abs{X}$ strategies from a parameterized heuristic space.

\textit{How} to incrementally extend a game model in an informed manner is what we call the \term{strategy exploration problem} \citep{Jordan10sw}.
To illustrate the problem, consider the simple $4\times 4$ game shown in Table~\ref{tab:game_with_increasing_regret}.
The strategy exploration problem asks in which order to introduce the strategies to our empirical game analysis.
Note that regardless of the ordering, once \( X = S \), equilibria in the restricted game and base game coincide, so regret is zero.
Thus, we might expect that regret would tend to start high, and decrease progressively until reaching zero in the last step.
This is not necessarily the case, however.
Introducing strategy~1 first, for example, would produce the solution profile \( (1,1) \) after the first iteration, which has a regret \( \regret(1,1) = 3 \).
If we then introduce additional strategies in the order (2,3,4), the additional sequence of regrets we observe would be (4,5,0), thus increasing monotonically until inevitably falling to zero at the very end.

\begin{table}[htb]
\centering
\begin{tabular}[htb]{cc|cccc}
\multicolumn{2}{c|}{} & 1 & 2 & 3 & 4 \\
\cline{2-6}
&  1 & 1,1 & 1,2 & 1,3 & 1,4 \\
& 2 & 2,1 & 2,2 & 2,3 & 2,6\\
& 3 & 3,1 & 3,2 & 3,3 & 3,8 \\
& 4 & 4,1 & 6,2 & 8,3 & 4,4\\
\end{tabular}
\caption{An example symmetric two-player game of four strategies \citep{Jordan10sw}.
Exploring strategies in the sequence (1,2,3,4) yields increasing regrets until the last step.}
 \label{tab:game_with_increasing_regret}
\end{table}

As the example suggests, it will be difficult to \textit{guarantee} progress throughout the strategy exploration process.
That does not mean, however, that exploration decisions are arbitrary.
Indeed, it can make quite a difference how one selects among strategies to add.
To compare alternative exploration policies, we generally evaluate their performance in expectation, with respect to distributions of games and stochastic elements of the EGTA process with automated strategy generation.

One natural approach to strategy exploration is the \term{double oracle} (DO) method of \citet{mcmahan03}.
DO was originally defined for two-player games,%
\footnote{The original paper likewise defined and analyzed DO for zero-sum games, but it can be applied without modification in the general-sum case.}
though the idea readily 
generalizes to $\numPlayers$ players.%
\footnote{Current usage convention retains the ``double'' part of the name for any $\numPlayers$.}
Let $X^k=(X_1^k,\dotsc,X_\numPlayers^k)$ denote the restricted strategy sets corresponding to iteration~$k$, with $\sigma^{*k}$ a NE profile for $\game_{\downarrow X^k}$.
$\BR_i$ is a best-response oracle for player~$i$.
DO augments the strategy sets for the next iteration with the best responses to the current iteration's NE: $X_i^{k+1}=X_i^k\cup \{ \BR_i(\sigma_{-i}^{*k})\}$.
If the best responses are already contained in the strategy sets, $\BR_i(\sigma_{-i}^{*k})\in X_i^k$, a NE for the full game has been found, and the process can terminate.
More typically, DO proceeds until the gains fall below a threshold, or until a time or iteration limit has been reached.

% \amy{what i would say here, or perhaps below when we mention PSRO is: it requires 3 subroutines:
% 1. a method for finding best responses (e.g., RL);
% 2. a method for evaluating policies (easy in zero-sum case, since the game has a value) --- necessary for convergence check;
% 3. a (tractable) meta-solver that outputs eqa, or similar, to be used as training targets in the next iteration}

The first application of automated strategy generation in EGTA was the use of genetic optimization by \citet{Phelps06} for bidding in a continuous double auction (CDA)\@.
% \amy{note to self: they did not compute BRs. they just used GAs or similar to automatically generate new strategies}
\citet{Schvartzman09,Schvartzman10} demonstrated the use of RL (non-deep: with tile-coded Q-functions) for optimization,
% \amy{i do not understand from the explanation about how ``optimization'' is a step in DO. what are we optimizing? clarify that this is the BR step.} 
also in CDA trading environments.
The latter studies were essentially instances of DO using RL to approximate the best-response oracle.
% \amy{ah, this makes sense!}
Across several trading games, these approaches were shown to produce new CDA strategies exceeding the performance of hand-coded predecessors.
Automated strategy generation using local search methods also played a role in EGTA applications to protocol compliance \citep{Wellman13kd} and credit network formation \citep{Dandekar15}.

\subsection{PSRO: Policy-Space Response Oracles}
\label{sec:psro}

\citet{Lanctot17} introduced \term{policy-space response oracles} (PSRO) as a general procedure combining deep RL with EGTA\@. 
Pseudocode for the PSRO algorithm, expressed in the terminology and notation of this survey, is presented as Alg.~\ref{alg:psro}.
The algorithm maintains an empirical game, $\hat{\game}=\langle \Players, (X_i), (\hat{u}_i)\rangle$, which models the game $\Gamma_{\downarrow{X}}$ over the strategy space $X$ as of the current PSRO iteration.
The empirical payoff matrix $\hat{u}$ can be computed and updated each iteration using any method for estimating or learning empirical games.
From $\hat{\game}$, PSRO derives a target profile $\sigma$.
For each player~$i$, it then uses deep RL to train a policy $\pi_i'$ that optimally responds to $\sigma_{-i}$.
As the \term{training target} $\sigma_{-i}$ is a mixed profile, the method operates by sampling during training. 
This deep RL computation represents the \term{response oracle} that gives PSRO its name.

\begin{algorithm}[htb]
 \SetKwInOut{Input}{input}\SetKwInOut{Output}{output}
\Input{initial strategy sets $X_1,\dotsc,X_\numPlayers$}
Estimate empirical payoff functions $\hat{u}_i$ by simulating  profiles in $X = X_1\times\cdots\times X_\numPlayers$ \;
Initialize training targets $\sigma_i \gets \textsc{Uniform}(X_i)$ \;
\While{within iteration limit}{
  \For{player $i \in \Players$}{
    \For{many episodes}{
      Sample $\pi_{-i} \sim \sigma_{-i}$ \;
      Train policy $\pi_i'$ with other agents playing $\pi_{-i}$ \; % \in \mbox{BR}_\epsilon(\sigma_{-i})$ \label{alg:psro:BR} \;
    }
    $X_i \gets X_i \cup \{ \pi_i' \}$ \;
  }
  Extend $\hat{u}_i$ to cover newly added strategies  \;
  Compute new training target $\sigma\gets\mathit{MSS}(\hat{\game})$ \;
}
Extract solution from final $\hat{\game}$ \;
\vspace{5mm}
\caption{PSRO pseudocode \citep{Lanctot17}.}
\label{alg:psro}
\end{algorithm}

% \amy{i'm missing something here. in the Train oracle step, what is $\rho$, and why does it need to be sampled, as we have already sampled $\pi_{-i}$ from $\sigma_{-i}$? also, how do you aggregate $\pi_i'$'s across episodes? maybe you don't? maybe you just add all the $\pi_i$'s you discover.}

% \begin{figure}[ht]
% 	\centering
% 	\includegraphics[width=0.75\textwidth]{figs/PSRO}
% 	\caption{PSRO pseudocode \citep{Lanctot17}, with annotations on key points discussed here.}\label{fig:psro}
% \end{figure}

% \begin{figure}[ht]
% 	\centering
% 	\includegraphics[width=0.75\textwidth]{figs/gpsro_overview_v3}
% 	\caption{PSRO phases explained,.}\label{fig:psro_phases}
% \end{figure}

PSRO's signal innovation was introducing the concept of \term{meta-strategy solvers} (MSSs), a generalized approach to selecting training targets.
The MSS selects an other-agent profile to train against, which in this framework is the essential driver of strategy exploration.
Double oracle as described above can be viewed as a special case of PSRO where the training target is the other-agent profile in a Nash equilibrium (i.e., the MSS is an NE-finding algorithm).
Though DO is often effective, there is ample evidence that best-response to NE is not always the best approach to strategy exploration.
\citet{Jordan10sw} demonstrate this for a simple auction game, where even adding random strategies could provide substantial speedups.
More generally, \citet{Lanctot17} argued that best-responding to Nash overfits to the current equilibrium strategies, and thus tends to produce results that may not be generally effective across the  relevant strategy space. 

This was indeed the motivation for employing a generalized MSS concept in PSRO, maintaining the principle of best-response but allowing the target to vary.
\citet{Lanctot17} propose several alternative MSSs, for example, their \term{projected replicator dynamics} method constrains the training target to include every strategy generated so far with at least some minimum probability.
This ensures that a broad set of opponent strategies are encountered 
during training.
It also limits the change in training targets from one round to the next, which serves to inhibit thrashing, in which an agent trained in one round throws away what the previous agent learned,
due to drastic changes in the training-target mixed strategy.
% \amy{discuss tradeoff: learn too slow vs. jump past solution}

The MSS abstraction is quite flexible, allowing PSRO to cover some classic game-solving approaches as special cases.
For example, selecting a uniform distribution over strategies $X^e$ as MSS essentially reproduces the  fictitious play algorithm \citep{brown1951iterative}.%
\footnote{To see this, recall that FP is defined by best response of each player to the others' distribution of play over prior iterations.
If we view the strategy introduced on each PSRO iteration as a play, the uniform distribution over these is exactly what FP would respond to.}
An MSS that simply extracts the most recent strategy corresponds to \textit{iterated best response}, which in a symmetric context is also termed \textit{self play} \citep{silver18}.

Any solution concept can readily be adopted as an MSS\@.
\citet{MullerORTPLHMLH20} employ an MSS based on \alpharank, and \citet{MarrisMLTG21} define MSSs that correspond to correlated equilibrium concepts or employ solution refinement criteria.
Combinations are also possible; for example \citet{wang19sywsjf} employed a mixture of NE and uniform, which essentially randomizes over whether to apply DO or FP on a given PSRO iteration.

\citet{balduzzi2019open} introduced an MSS specifically for zero-sum games, called \textit{rectified Nash}. 
Rectified Nash includes in the training target the subset of opponent strategies supported by the current equilibrium that the latest strategy beats or ties.
The idea behind this approach is to expand the scope of existing strategies by building on their strengths.
\citet{Dinh22} proposed a MSS for two-player zero-sum games that applies online learning to the empirical game and outputs the online profile as a best-response target. 

Whereas the MSS abstraction provides significant flexibility in the choice of training target, there may be some benefit to broadening the concept of best-responding to a fixed target. 
\citet{Wright19} suggest a \textit{history-aware} approach, where the BR to the primary training target is adjusted to improve performance with respect to previous targets.
\citet{Perez-Nieves21} propose maximizing a weighted combination of response quality and contribution of diversity to the current empirical game.
\citet{Yao23} likewise define a diversity metric and apply it as a regularizer during the best-response computation.

% \amy{1-2 sentences about correlated PSRO. the most interesting things to mention, i think, are: 1. heuristics for choosing among eqa; max welfare doesn't work so well; max Gini welfare works better, empirically. (why?) 2. there is not a single BR; there are multiple to consider, one per action, since each action induces its own training target. assuming we do not want to double the number of actions each iteration, which of these BRs should we add to the strategy set?}

\subsection{Frontiers in Strategy Exploration}

PSRO is actively being exercised and extended by several research groups \citep{Bighashdel24}.
Developments include computational enhancements, for example to parallelize best-response training \citep{McAleerLFB20} or adaptively optimize hyperparameters \citep{Li24lywhca}.
New MSS ideas are generated on a regular basis, and as yet there is no definitive understanding of which MSS is the best to employ for a given game environment.
In general, we cannot even be sure that exploration with a given MSS will produce progress from iteration to iteration (recall the example of Table~\ref{tab:game_with_increasing_regret}).
An exception is to best-respond to the profile in the empirical game $\hat{\game}_{\downarrow X}$ that minimizes full-game regret, what \citet{Jordan10sw} termed the \term{minimum regret constrained profile} (MRCP):
\begin{displaymath}
    \text{MRCP} (\game, X) = \arg \min_{\sigma \in \Delta(X)} \regret^\game (\sigma). 
%\amy{$= \arg \min_{\sigma \in \Delta(X)} \max_{i} \max_{\tau_i \in \Delta(S_i)} u_i (\tau_i, \sigma_{-i}) - (\sigma_i, \sigma_{-i})$}
\end{displaymath}
% where $X$ is the restricted strategy space of the empirical game.
\citet{McAleer22} proposed using MRCP as an MSS for two-player zero-sum games, observing that this would ensure an \textit{anytime} property of monotone improvement.
\citet{wang2022evaluating} likewise considered MRCP for general-sum games, addressing its computational challenges and evaluating it as an MSS for strategy exploration.
Despite the anytime property, these authors found that MCRP may not perform as well as alternatives.
% perhaps because MRCP is not well-behaved; in general, it may possess many local minima
An explicit regularization approach proposed by \citet{Wang23w}, which balances equilibrium and MRCP (i.e., trades off regret in the empirical game for reduced full-game regret), seems to offer robust performance compared to other MSSs.

In the absence of broad theoretical characterization of MSS effectiveness, the literature relies on computational experimentation over a variety of game environments. 
Conducting these experiments presents subtle issues.
As \citet{wang2022evaluating} point out, the object of evaluation is the series of restricted strategy sets produced in the exploration process.
These authors propose a consistency condition, which mandates that in comparing trajectories of strategy sets generated by alternative MSSs, the same solver should be employed.
For example, to compare DO and FP (implemented by Nash and uniform MSSs, respectively), one should fix a particular solver.
When feasible to compute, MRCP can be an appropriate solver for evaluation.
If instead one uses Nash for DO and uniform for FP (as in some prior literature),
there is a confound between the role of the solver in identifying solutions and in guiding exploration.

Typical evaluation of strategy exploration focuses on how quickly we can construct an empirical game that contains an approximate equilibrium of the full game.
For games with multiple equilibria, we generally care about \textit{which} equilibria are found, and may wish to cover a diverse set of equilibria.
\citet{Wang24w} show that varying the response objective (i.e., beyond maximization of own utility) can effectively direct exploration toward solutions with desired qualities.

Other recent ideas provide potential new directions for strategy exploration.
\citet{Smith23} investigate opportunities for \textit{strategic knowledge transfer}, whereby products of response learning can be reused or repurposed in subsequent related response computation.
One of their proposed methods, \textit{mixed-opponents}, uses the value functions underlying mixed strategies produced by an MSS to construct pure-strategy response targets representing qualitatively different but plausibly relevant behavior.
\citet{Li23combining} propose enhancing response policy generation with AlphaZero-style tree search \citep{silver18}, producing policies with runtime performance beyond the policy networks represented in the empirical game.
Dyna-PSRO \citep{Smith24w} co-learns a world model (environment transition dynamics and rewards) along with the empirical game, to gain the most leverage from experience accrued in simulation during game estimation and best-response computation.

\section{Applications}
\label{sec:applications}

The foregoing sections have cited numerous works that advanced the methodology of EGTA, many of which were driven by demands of particular applications. 
In this section we focus on the application areas, selectively outlining a few in which EGTA has contributed domain insights.

\paragraph{Recreational Games}
Like for AI in general, game-playing has been a driving application for EGTA advances \citep{TuylsPLHELSG20}.
Early studies applied EGTA to games played among heuristic strategies for Texas Hold'em poker \citep{PonsenRCDT08,PonsenTKR09}, and Leduc poker has served as a common benchmark for EGTA with RL \citep{Lanctot17}.
Recent breakthroughs in game-playing have featured empirical game-theoretic reasoning; for example, the development of AlphaStar \citep{Vinyals19} included a ``Nash league'' that tracked equilibria of candidate policies generated over many iterations.%
\footnote{\url{https://www.deepmind.com/blog/} \\ \url{alphastar-mastering-the-real-time-strategy-game-starcraft-ii}}
Empirical game versions of a wide variety of recreational games were also employed by \citet{Czarnecki20} to understand common strategic landscapes.

\paragraph{Economics and Finance}
The earliest EGTA developments (see Section~\ref{sec:history}) were motivated by games involving bidding in auctions, and other economic applications. 
Some agent-based finance studies not explicitly labeled EGTA essentially took this approach in estimating game models from simulation data \citep{Zhan07}.
\citet{Wellman20} surveys economic applications of EGTA, including several examples where such methods produced new insights for canonical auction games beyond analytic tractability.
Systematic EGTA studies established the centrality of price prediction in bidding heuristics for complex auctions \citep{Wellman08omr,Wellman17sg}, for example.
Other economic domains addressed by EGTA include financial and environmental regulation
\citep{Cheng17,Cheng19}, blockchain mechanisms \citep{Wu24}, simulated economies \citep{Dwarakanath24}, bank interest-rate risk \citep{Zhao23pv}, formation of credit networks \citep{Dandekar15}, adoption and use of payment mechanisms
\citep{Cheng16,Mayo21fw}, fraud detection \citep{Mayo24gw}, and management of debt in financial networks \citep{Mayo21w,Zhou24wvbscw}.

EGTA has been applied perhaps most extensively to strategic scenarios regarding trading in financial markets \citep{Wah16,Wah17}.
Early on, empirical game modeling shed light on heuristic strategies from the agent-based finance literature for trading in continuous double auctions \citep{KaisersTTP08,Schvartzman09}.
Financial market applications have also considered specialized domains like prediction markets \citep{Wah16lp} and markets for exchange-traded funds \citep{Shearer21bbw}, and issues like order priority rules \citep{Qi22v} and strategic choice between market mechanisms \citep{Wah15hw}.
EGTA studies have investigated the implications of market manipulation \citep{Liu22spoof,Wang21hyw}, including the prospect for automated learning of manipulative strategies \citep{Shearer23rw}.

\paragraph{Other Applications}
\citet{Wellman19wn} surveyed applications of empirical game-theoretic techniques to problems in cyber-security.
For example, \citet{Hutchins24} employed EGTA to rank network defense strategies.
Other domains subjected to EGTA treatment include pursuit-evasion games \citep{Li23wzxca}, social dilemmas \citep{Leibo17,Phelps16,Pretorius20,Willis23}, software development \citep{Gavidia-Calderon20shb2}, space debris removal \citep{Klima16}, and team formation \citep{Yang21}.

\section{Empirical Mechanism Design}
\label{sec:md}

The problem of \term{mechanism design} is to specify rules of interaction (i.e., the \textit{mechanism}) for a set of agents, based on design objectives.
The mechanism together with agent preferences define a game, and so in a sense the mechanism designer is specifying a game. 
Evaluating a mechanism essentially requires solving the game, so that the properties of the solution can be quantified.
With such an evaluator on hand, a designer can---in principle---search for mechanisms that optimize their desiderata.
The use of algorithms to design economic mechanisms has been termed \term{automated mechanism design} (AMD) \citep{Conitzer03a}. 
AMD formulates the design as an optimization problem, with specified objectives and constraints that enforce equilibrium behavior. 

% This mechanism design problem is a very complex constrained optimization problem, because it is a search over mechanisms whose properties depend on their participants' collective behavior.
In many real-world applications, such as developing tax or climate policies, the game induced by a mechanism is not available in analytic form.
If we can simulate agent strategies interacting through a mechanism, then reasoning about the game induced by a particular design is a form of EGTA.
Thus, the techniques described here can be used to characterize the strategic behavior induced by a mechanism.
In such cases, we refer to the mechanism design problem as \term{empirical mechanism design} (EMD).%
\footnote{\citet{Phelps10mp} define a related approach termed \textit{evolutionary mechanism design}, which likewise employs simulation over heuristic strategy sets but emphasizes evolutionary search methods and stability concepts.
The recent \textit{differentiable economics} approach of \citet{Duetting24} leverages machine learning methods to solve AMD problems where mechanism operations and equilibrium constraints can be expressed in neural networks.
}

In the first study framed explicitly as EMD, \citet{Vorobeychik06kw} used EGTA to evaluate candidate designs to fix a pathology observed in the inaugural TAC supply chain management tournament. 
Specifically, they investigated whether changing a storage cost parameter would be sufficient to deter excessive procurement of supplies. 
Their approach was simply to construct empirical game models for a discrete set of parameter settings over a specified interval. 
They found that while increasing storage costs did indeed decrease procurement levels, no settings in the range considered reasonable were sufficient to remove the pathology.

\citet{jordan10wb} took a similar approach to EMD in their study of the TAC Ad Auctions (TAC/AA) game. 
Taking the perspective of the search publisher, they examined the effect of auction parameters, such as reserve price, on publisher revenue.
For each candidate setting, they solved an empirical game, essentially re-equilibrating the play among the top agents from the TAC/AA tournament.
In another auction-related application, \citet{Brinkman17} used EMD to determine optimal clearing intervals for call markets.

The preceding studies evaluated a fixed set of candidate mechanisms.
\citet{Vorobeychik12} proposed an approach based on black-box optimization, where candidates are generated by a stochastic search process.
This work also incorporated constraints on mechanism properties (e.g., individual rationality). 
\citet{viqueira19cmg} likewise employed a black-box technique, specifically Bayesian optimization, to search for revenue-maximizing reserve prices in a simultaneous auction scenario based on the TAC Ad Exchange game \citep{Tao15}.
Each candidate vector of reserve prices (one per auction) defines as empirical game, a model of which they constructed using some of the sampling methods described in Section~\ref{sec:statistical}, assuming plausible bidding heuristics.
In another distinct approach, \citet{Zhang23facmh} show how to reformulate mechanism design problems as two-player zero-sum games, amenable to solution using PSRO.

% \amy{maybe Salesforce's AI economist work, or David's newer work on learning indirect mechanisms}

\section{Conclusion}

More than twenty years of research in empirical game-theoretic analysis has produced a large body of concepts, representations, algorithms, and application experience.
The enterprise started with the motivating idea that empirical methods such as simulation, machine learning, and statistics could broaden the practical scope of game-theoretic reasoning, beyond what is feasible through deductive analytic techniques alone.
The literature surveyed here demonstrates the extent of this broadening.

The core idea of EGTA is to induce a game model from simulation data.
Distinguishing the object of deductive reasoning---the \textit{empirical game}---from the source of knowledge about the game (e.g., an agent-based simulation model) has the practical effect of decoupling descriptive complexity from game-theoretic reasoning complexity. 
We can simulate a complicated world to produce an empirical game that is as simple or complex as we can computationally afford. 
Any simplification invariably sacrifices fidelity, but the EGTA perspective affords a smoother trade-off than we can typically achieve with analytic modeling.

In reviewing the ideas and techniques introduced over the course of EGTA's development as a methodology, we aim to provide the reader with an understanding of the state-of-art, as well as a structure for extending the EGTA toolbox.
Many of the components presented here correspond to well-defined subproblems, for example what we have labeled \textit{game model learning} (Section~\ref{sec:learning}), \textit{strategy exploration} (Section~\ref{sec:explore}), or \textit{dynamic sampling} (Section~\ref{sec:dynamic-sample}).
Progress on subproblems may be easier to evaluate than entire game reasoning frameworks. 
Other EGTA advances may operate at the interfaces, or could have cross-cutting effects on multiple subproblems.

For example, most EGTA techniques developed to date assume an empirical game model in normal form.
Some works have started to exploit structure in empirical game models, for example, symmetry or interaction sparsity, to support representational scalability \citep{Li20w}.
Capturing extensive-form (tree) structure in an empirical game model \citep{Konicki22,Konicki25,McAleer21} can also provide advantages in representation and reasoning, including the ability to consider refined solution concepts based on this structure.
In principle, EGTA could be applied with respect to any special game class or representation, adopting any solution concept deemed appropriate.
For example, recent work has shown how to conduct EGTA for \term{mean-field games} \citep{Muller22,Wang23mean} and team games \citep{McAleer23}.
In all these works, we see that extending EGTA beyond normal-form---and moreover taking full advantage of the special game features---entails further innovations in game model induction, strategy exploration, or other elements of the EGTA process.

Another path of future work should further develop tools for understanding the applicability of EGTA results. 
Solving an empirical game gives us a solution for that literal game model, which is a limited representation of the game we actually care about.
Statistical techniques like those presented in Section~\ref{sec:statistical} can inform probabilistic statements relating empirical-game solutions to the game without sampling error (i.e., what we have called the \textit{true} game). 
The true game, however, is typically defined over a restricted strategy space compared to the \textit{base} game.
We typically lack strong theoretical connections between restricted-game and base-game solutions (except in the asymptotic limit), and so our level of confidence generally relies on experimental evidence.
Moreover, even the base game may be a simplified version of the \textit{underlying} game of interest, limited by the fidelity of simulation modeling.
In many contexts, our interest in strategic analysis is not actually for any particular game instance, but rather in a class of strategic situations of which the underlying game is representative. 
Ultimately, we seek a more precise understanding of how insights from EGTA results (or any form of game-theoretic reasoning) may bear on general strategic situations beyond specific games analyzed.

The development of EGTA methodology has coincided with a significant increase in the adoption of game-theoretic principles throughout AI and computer science, as well as major advances in machine learning methods.
These ML advances have contributed to improved EGTA, as well as to game-theoretic reasoning approaches that do not necessarily employ empirical game models. 
We expect that the use of empirical methods for analyzing games will continue to evolve rapidly, further expanding the scope of principled strategic reasoning.

\section*{Acknowledgments}

Karl Tuyls conducted the work for this survey while at Google Deepmind, Paris, France.
We thank students and colleagues at the University of Michigan, Google DeepMind, and Brown University for comments and discussions on the content of this survey.
We are particularly grateful to Daniel Hennes and Bhaskar Mishra for assistance with computational experiments reported herein.

% Amy Greenwald is funded by ONR ...

% The authors would like to thank Bhaskar Mishra for his contribution to Section~\ref{sec:statistical}.

\bibliographystyle{plainnat}
\bibliography{egta}

\end{document}